\documentclass[twocolumn]{aastex7}
\usepackage{caption}
\usepackage{float}
\usepackage{subcaption} 
\usepackage{amsmath}
\usepackage{graphicx}
\usepackage{booktabs}
\usepackage{multirow}
\usepackage{longtable}
\usepackage{booktabs}
\usepackage{hyperref}
\usepackage{mhchem}
\usepackage{siunitx}
\newcommand{\Ms}{M$_{\odot}$}

\newcommand      \grays       {$\gamma$-rays}
\newcommand      \gray       {$\gamma$-ray}
\newcommand{\Ni}{$^{56}$Ni}

\newcommand{\Co}{$^{56}$Co}
\newcommand{\Fe}{$^{56}$Fe}
\newcommand{\mic}{$\mu$m}

\usepackage[normalem]{ulem}
\usepackage{longtable}
\shorttitle{Dust in Type Iax SNe}
\shortauthors{Kumar et al. }

\begin{document}
\title{Type Iax supernovae as a source of iron-rich silicate dust}

\author[0009-0004-0041-9271]{Aman Kumar}
\affiliation{Indian Institute of Astrophysics, 100 Feet Rd, Koramangala, Bengaluru, Karnataka 560034, India}
\affiliation{Academy of Scientific and Innovative Research (AcSIR), Ghaziabad, Uttar Pradesh, 201002, India}
\email{aman.kumar@iiap.res.in}

\author[0000-0002-9820-679X]{Arkaprabha Sarangi}
\affiliation{Indian Institute of Astrophysics, 
100 Feet Rd, Koramangala, Bengaluru, Karnataka 560034, India}
\email{arkaprabha.sarangi@iiap.res.in}


\begin{abstract}

We model the formation of dust in the ejecta of Type Iax supernovae (SNe), which is a low-luminosity subclass of Type Ia SNe. A non-equilibrium chemical kinetic approach is adopted to trace the synthesis of molecules, molecular clusters, and dust grains in the ejecta of thermonuclear SNe. We find that Type Iax SNe provide conditions conducive to the formation of several O-rich dust species in the ejecta. Particularly, iron-rich silicates of chemical type \ce{FeSiO3}, \ce{Fe2SiO4}, and \ce{MgFeSiO4} are found to form in abundance, suggesting that the ejecta of low-luminosity thermonuclear SNe can be a site where a large fraction of iron is locked up in dust, unlike other stellar sources.  
The final mass of dust formed in the ejecta ranges between $10^{-5}$ and $10^{-4}$ \Ms, where most of the dust forms between 1000 and 2000 days post-explosion. Apart from Fe-rich silicates, Mg-silicates, and silicon carbide are also formed in the ejecta of Type Iax SNe. When compared to the dust budget of typical Type Ia SNe, we find that the expected dust-to-ejecta mass ratio is 1 or 2 orders of magnitude larger in Type Iax SNe. We conclude that the ejecta of typical Type Ia SNe form a negligible amount of dust, in agreement with observation, while the low-luminosity subclass Type Iax SNe are potential producers of iron-rich silicates.

\end{abstract}

\section{Background}

Thermonuclear SNe have a white dwarf (WD) progenitor in a binary system, but they have a number of explosion channels, giving us a wide variety of peculiar SNe \citep{Hoyle_1960,liu2023}. A Type Iax SN is a subclass of Type Ia thermonuclear SN, earlier termed as `peculiar Ia', that is fainter and less energetic than the more common Type Ia. SN 2002cx was the first peculiar Type Ia SN observed by \cite{Foley2013, Li2003}, so they are also termed as `02cx-like'. Unlike their Type Ia cousins, Type Iax SNe don't entirely destroy the white dwarf during the explosion, instead leaving behind a stellar remnant. They are also termed as failed SNe, which leave behind a `zombie' star \citep{Foley2014,Kawabata2018, Kawabata2021}. This bound remnant might be affecting the light curve \citep{Foley2016, Magee2016}. They are characterized by their lower luminosity, lower ejecta velocities, and a lack of a secondary maximum in their near-infrared light curves\citep{Magee_2025,kasen_2006}. The classification is based on observational properties, including their unique spectral features of a prominent FeIII line and weak SiII and SII lines \citep{Branch2004} and light curves that are fainter and evolve more slowly than typical Type Ia SNe \citep{Foley2013}. Even among Type Iax there is a wide range in the peak absolute brightness ($M_v = -14$ to $-18$ mag) \citep{Singh_2018, Foley2013, singh_2025, kwok_2025}.

\cite{Foley2013, Li2001} estimate that Type Iax SNe have a 30\% occurrence rate as compared to Type Ia in a selected sample space,which is further constrained to $\sim$5\% by \cite{Dimitriadis2025} using Zwicky Transient Facility SN Ia Data. There are around 5--10 Type Iax SNe classified detections per year. Type Iax SNe are estimated to originate from a single degenerate channel. In a Type Ia single degenerate channel explosion, a deflagration subsonic wave originates within the WD core, which pre-expands in the WD, and spontaneously the subsonic wave converts into a supersonic wave and explode \citep{ Woosley_1986,  Hoeflich_1995, Ropke_2007, Ropke_2007b, Seithenzahl_2013}. This is known as the Delayed-Detonation model, the detonation burns down the WD constituents to \Ni, and the pre-expansion due to the deflagration wave quenches the burning, so that we are left with some intermediate-mass elements (IMEs) and iron-group elements (IGEs) like Si and Mg. This model resembles the observed abundances and high \Ni~ content of Type Ia SN ejecta. In Type Iax SNe, the lower ejecta velocity, low \Ni~ content, and no secondary maximum in the infra-red light curve, which suggests well-mixed ejecta, can be explained by a pure deflagration model \citep{Liu2015, Barna_2018, Zang2019, Gamezo_2003, Garcia_2005, Ropke_2006a, Jordan_2012a, Ma_2013, Long_2014, Fink_2014, Lach_2022a, kwok_2025}. In the pure deflagration model, the subsonic deflagration wave never goes into the supersonic phase, and depending upon its energy, it ejects some of the outer parts of the WD, which also explains the wide range in peak absolute brightness of Type Iax SN. Deflagration quenches the nuclear burning of C/O WD, which leaves us with less \Ni \  content with well-mixed high abundances of C, O, and other IME like Mg, Si and S. The centrally peaked emission lines detected by \texttt{JWST} in early nebular spectra of Type Iax SN 2024pxl and SN 2024vjm, also indicate the mixing of metals among all the radial layers \citep{kwok_2025}. 

Cosmic dust is made of clusters of different sizes of diverse chemical nature. Dust formation requires moderately high temperatures, in the range of 1000--2000~K, and high gas density with abundances of the refractory elements \citep{sarangi2018book}. Core-collapse SNe, outflows from AGB stars, mass ejections from LBV and Wolf-Rayet stars are responsible for the dust formations \citep{dwe11}. SNe are one of the primary dust factories in galaxies. The ejecta of a Core-collapse SN provide the necessary environmental conditions and chemical compositions for dust production. Especially Type II SNe, with a massive, slow-moving ejecta, are observationally confirmed to produce large masses of dust \citep{matsuura_2015, shahbandeh_2023, clayton_2025}.



Type Ia SN ejecta is dominated by \Ni~ which will further decay into \Fe, they produce around 0.6 \Ms~ of \Ni~ \citep{Nozawa_2011}.On the other side, Core-collapse SNe have much smaller \Ni~ mass, around 0.055 \Ms  \citep{seitenzahl_2014, Rodriguez_2023}. Type Ia SNe, being highly energetic, do not show signs of dust formation \citep{gom12b}. Interestingly, Type Iax SNe being less energetic can be good candidates for dust production, and having significant \Ni, mixed microscopically in the entire ejecta, can provide a pathway for the synthesis of iron-rich dust. 
\begin{longtable}[c]{ccccccc}
\caption{ The deflagration models used in this study, Mass of ejecta  (M$_{ej}$), Bound remnant mass (M$_b$), synthesized \Ni ~mass (M$_{^{56}Ni}$) are in \Ms \ and ejecta kinetic energy (E$_{kin,ej}$) is in $10^{50}$erg unit, are from \cite{Fink_2014}. The early light curves and densities of a few Type Iax SNe match these models in the mentioned references.}
\label{tab: Model}\\
\hline
Model & M$_{ej}$ & M$_b$ & M$_{^{56}Ni}$ & E$_{kin,ej}$ & SN & Ref. \\ \hline
\endfirsthead
\endhead
\hline
\endfoot
\endlastfoot
\multirow{3}{*}{\texttt{N1def}} & \multirow{3}{*}{0.0843} & \multirow{3}{*}{1.32} & \multirow{3}{*}{0.0345} & \multirow{3}{*}{0.149} & SN~2019muj, & \citep{Barna2021}, \\
 &  &  &  &  & SN~2014dt & \citep{CN2023} \\
 &  &  &  &  & SN~2024pxl &\citep{kwok_2025} \\\hline
\multirow{2}{*}{\texttt{N3def}} & \multirow{2}{*}{0.195} & \multirow{2}{*}{1.21} & \multirow{2}{*}{0.073} & \multirow{2}{*}{0.439} & SN~2002cx, & \multirow{2}{*}{\citep{Barna_2018},} \\
 &  &  &  &  & SN~2015H &  \\ \hline 

\multirow{3}{*}{\texttt{N5def}} & \multirow{3}{*}{0.372} & \multirow{3}{*}{1.03} & \multirow{3}{*}{0.158} & \multirow{3}{*}{1.35} & SN~2020udy &\citep{Mag2023}\\
  &  &  &  &  & SN~2005hk & \citep{Barna_2018}\\
    &  &  &  &  & SN~2020rea & \citep{MSingh2022}\\\hline
\multirow{2}{*}{\texttt{N10def}} & \multirow{2}{*}{0.478} & \multirow{2}{*}{0.926} & \multirow{2}{*}{0.183} & \multirow{2}{*}{1.95} & SN~2011ay & \multirow{2}{*}{\citep{Barna_2018}} \\
 &  &  &  &  & SN~2012Z &  \\ \hline
\texttt{N20def} & 0.859 & 0.545 & 0.264 & 3.75 & - & - \\
\texttt{N40def} & 1.21 & 0.19 & 0.335 & 5.22 & - & - \\
\texttt{N100def} & 1.31 & 0.090 & 0.355 & 6.11 & - & - \\ \hline
\end{longtable}
\textit{Herschel} PACS and SPIRE photometry at 70--500 \mic, of Type Ia remnants Tycho and Kepler finds dust masses of $\sim$ 3 $\times$ 10$^{-3}$~\Ms\ \citep{gom12b}, at a temperature of $\sim$ 82~K. The IR emission was found to be spatially consistent with the X-ray emitting region and the circumstellar region swept up by the SN blast-wave. Very recently, a large mass of dust ($\sim$ 10$^{-2}$ \Ms) was reported in Type Ia-CSM SN~2018evt, where the dust was estimated to form in the circumstellar environment \citep{wang_2024}. Evidence of new dust formation in the ejecta of Type Ia is yet to be confirmed by observations. Initial theoretical models have predicted new dust formation in the ejecta of Type Ia SNe, with significant dust masses, ranging between 10$^{-4}$ -- 0.1 \Ms\ \citep{Nozawa_2011}. However, the approach adopted in those models may oversimplify the chemical evolution of the ejecta \citep{don85}. 

\textit{Spitzer} observations found strong mid-IR emission in nearby (19 Mpc) type Iax SN~2014dt, in 3.6 and 4.5 \mic\ fluxes \citep{fox_14dt_2016}. A dust mass of about 10$^{-5}$ \Ms\ at $\sim$~700~K was estimated by a single carbon dust component fit \citep{fox_14dt_2016}. The SN remnant 1181 was estimated to have the origin in Type Iax SN Pa~30, which shows dust emission in the far-IR through IRAS and AKARI imaging \citep{lykou_2023,cunningham_2024}; however, it is difficult to distinguish between dust produced in the ejecta and in the swept-up interstellar/circumstellar matter. Interestingly, based on the X-ray properties, the SN remnant Sagittarius A East in our galactic center is reported to be of Type Iax origin, and also hosts large masses of warm dust (0.02 \Ms) \citep{zhou_2021}. Even though all the dust may not have condensed in the ejecta alone, it projects a strong case for Type Iax SNe to be dust producers in galaxies. In this paper, we will explore the possibility of dust production in Type Iax SNe and also compare our results with Type Ia SNe.

\begin{figure*}[htbp] 
    \centering
    \begin{subfigure}{0.475\textwidth}
        \centering
        \includegraphics[width=0.968\linewidth]{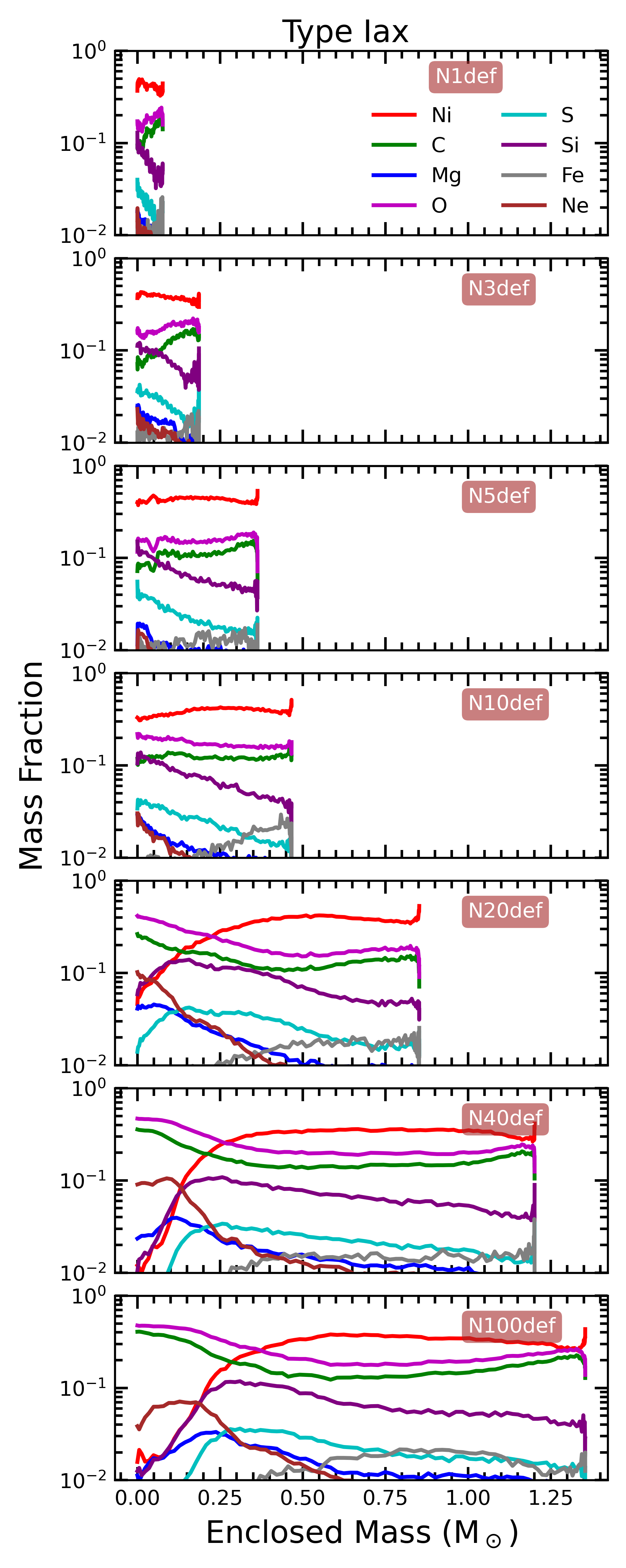}
        \label{fig: abun_def}
    \end{subfigure}
    \hfill
    \begin{subfigure}{0.485\textwidth}
        \centering
        \includegraphics[width=0.967\linewidth]{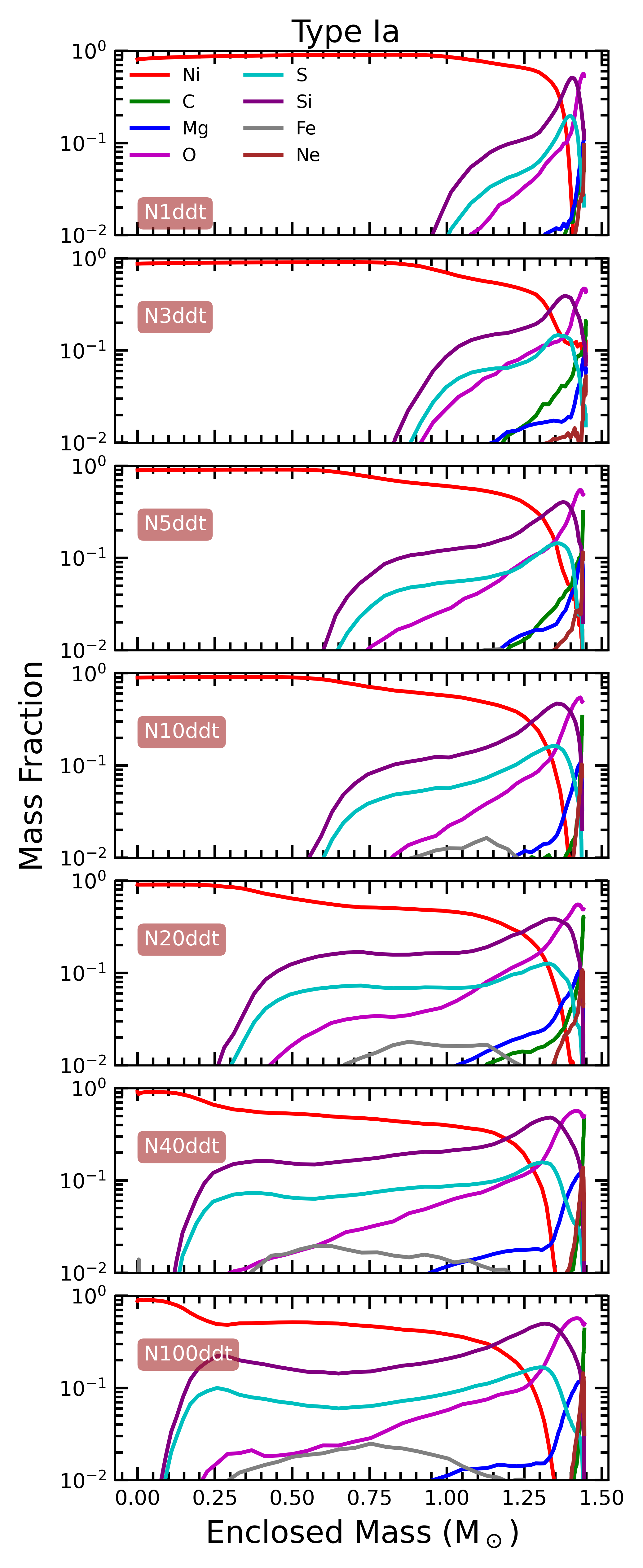}
        \label{fig: abun_ddt}
    \end{subfigure}
    \caption{\textit{Left}: Abundance profile of deflagration models (Type Iax) in enclosed mass coordinate\citep{Fink_2014}. \textit{Right:}Abundance profile of delayed detonation models (Type Ia) in enclosed mass coordinate \citep{Seithenzahl_2013}. Both models use a similar 1.4 \Ms\  initial mass of WD. Type Iax ejecta is increasing in mass for higher configuration models due to more powerful deflagration waves. Type Ia ejecta as initial deflagration ignition sites are increasing the ejecta is giving elements other than \Ni~ up to more inner regions. }
    \label{fig: abundance}

\end{figure*}
\section{Rationale of this work}
SN Type Iax are a subclass of thermonuclear SN explosions.
 Similar to Type Ia, they are rich in IME and IGE \citep{Li_2018}. Also, the light curve of Type Iax SNe is powered by \grays~ produced by radioactive decay of $^{56}Ni~\rightarrow~^{56}Co~\rightarrow~^{56}Fe$ \citep{Colgate_1969, Hillebrandt_2013, Arnett_1982, Barna_2018}. Unlike typical Type Ia explosions, which are energetic, fast expanding (10,000-20,000 kms$^{-1}$) \citep{Srivastav_2023}, have secondary maximum in their near-infrared light curves\citep{Elias_1981}, and completely disrupt the progenitor white dwarf\citep{Bersten_2017,liu2023}, Type Iax explosions are less energetic, exhibit lower ejecta velocities (2,000–7,000 kms\(^{-1}\))\citep{Jha_2017,Jha2019}, and may leave behind a bound remnant\citep{Foley2014}. These factors lead to slower expansion and higher ejecta densities at a time when the temperature is lower than the ionization temperature, allowing atoms and molecules to collide more frequently and enhancing dust nucleation rates. The presence of significant unburned carbon and oxygen is expected to form the necessary dust-precursor molecules in the chemically mixed ejecta, providing the important ingredients for the formation of iron-rich dust species.
  

The paper progresses from the physical properties of Type Iax ejecta to the formation of dust and its astrophysical context. The following section \ref{Ejecta} introduces the ejecta properties of Type Iax SNe, beginning with a review of explosion models and progenitor channels from the literature, and summarizing the key input parameters such as abundance profiles, temperature, densities of the system at 100 days after explosion, along with their \Ni ~yield. Section \ref{chem} focuses on the chemistry of the ejecta, describing the reaction network, modifications implemented in this study, the species considered, and the reactions that are found to be the most important for dust formation. Section \ref{molecule} addresses molecule formation, identifying key molecules produced in the ejecta and their relation to the dust species. Section \ref{dust} presents the dust budget, detailing dust types, quantities, and formation timescales. Section \ref{compare} describes the dust formation scenario in Type Ia and presents its dust mass evolution. Section \ref{discussion} provides a broader discussion, including the ``missing iron in the ISM" problem \citep{Dwek_2016,Psaradaki_2023,Pinto_2013}, the sensitivity of results to key parameters, and the need for further observational constraints. Finally, Section \ref{summary} summarizes the main findings and highlights their significance for understanding dust production in thermonuclear explosions.


\section{Ejecta of Type Iax and Ia SNe}\label{Ejecta}

In this work, we adopt the three-dimensional pure deflagration model of \cite{Fink_2014} as the mechanism for Type Iax SN explosion.
We use \cite{Seithenzahl_2013} delayed detonation models for Type Ia SNe, which explore a wide range of ignition configurations within Chandrasekhar mass C/O WDs.

\subsection{Various progenitor channels}

\cite{Fink_2014} models consider different numbers and geometries of ignition kernels, leading to variations in explosion strength, ejecta mass, and the mass of synthesized \Ni~ mentioned in table \ref{tab: Model}. The ignition kernels are the origin site of the deflagration wave. The lower-kernel configurations (\texttt{N1def,N3def,N5def}) produce weaker explosions, with smaller \Ni\ masses and a more massive bound remnants. The light curves predicted by several of these models and model densities show good agreement with those observed for certain known Type Iax SNe SN~2014dt \citep{CN2023}, SN~2019muj \citep{Barna2021}, SN~2005hk SN~2002cx, SN~2015H, SN~2011ay, SN~2012Z \citep{Barna_2018}, SN~2020udy \citep{Mag2023} and SN~2020rea \citep{MSingh2022}. The energetics and abundances from these models serve as our initial conditions, allowing us to investigate how differences in ignition channels influence the chemical composition and physical conditions relevant for molecule and dust formation. Higher configuration models (\texttt{N150def, N200def, N300def, N1600def}) are also available; however, they don't align with any of the current Type Iax observations, and therefore are not significant for our study. 


\cite{Seithenzahl_2013} uses a similar type of ignition kernel geometries for delayed-detonation models. The Lower-kernel configurations (N1ddt, N3ddt, N5ddt) have fewer deflagration wave origin sites, which leads to less mixing. Following detonation, this results in more \Ni\ masses, while other elements are only present in the outer regions. The higher configuration models(\texttt{N20ddt,  N40ddt, N100ddt}) have more deflagration ignition kernels, which will give well-mixed elemental composition up to the more inner regions, giving less \Ni~ abundance. \texttt{N100ddt} model among other models produces 0.6 \Ms\ of \Ni, which is similar to the estimated Type Ia \Ni~ content from observations \citep{Stritzinger_2006}.

\subsection{Abundances}

\begin{figure}[htbp]
    \centering   
    \includegraphics[]{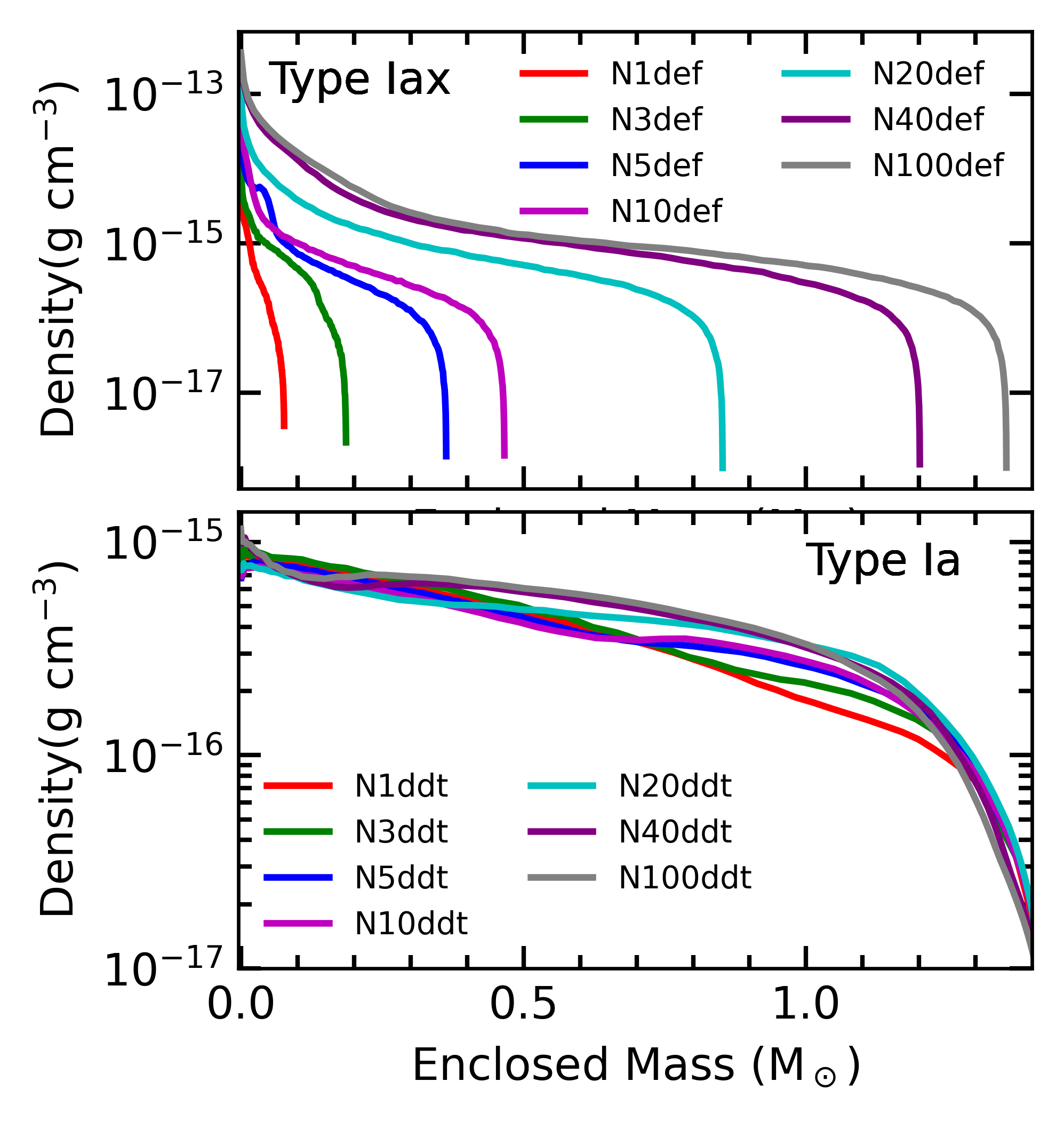}
\caption{\textit{Top}: Density profile of \cite{Fink_2014} deflagration models of Type Iax in enclosed mass coordinate at 100 days. \textit{Bottom}: Density profile of \cite{Seithenzahl_2013} DDT models of Type Ia in enclosed mass coordinate at 100 days.}
\label{fig: density}
\end{figure}

Figure \ref{fig: abundance} shows the abundance profiles as a function of enclosed mass for a set of deflagration (\texttt{N1def} to \texttt{N100def}) models representing Type Iax SNe and delayed-detonation (DDT) (\texttt{N1ddt} to \texttt{N100ddt}) models representing Type Ia SNe. The source data of \cite{Fink_2014, Seithenzahl_2013} \footnote{HESMA Hydro models: \url{https://hesma.h-its.org/hydro/}} provide abundances in radial coordinates at 100s post-explosion. We derive the abundance distribution as a function of mass coordinates by integrating over the density profile (Fig.: \ref{fig: density}).

The ejecta is characterized by a chemically mixed composition in the pure deflagration models. While \Ni~ is the most abundant species, the ejecta also contains significant amounts of O, C, and Si. The large abundance of unburnt C/O exceeds the mass of the Si/S, which is a consequence of the subsonic and turbulent nature of the deflagration flame, allowing the white dwarf to expand and quench the burning before completion. This turbulent process thoroughly mixes the burned and unburned material. \\
In contrast, in the DDT models, the ejecta is more dominated by \Ni, with contributions from S, Si, and O. The explosion begins similarly to a Type Iax, with a subsonic deflagration flame that pre-expands the entire white dwarf, reducing its overall density. At some point, this deflagration spontaneously transitions into a detonation. A detonation is a supersonic shockwave that rips through the pre-expanded star at immense speed. The highest density material at the center burns to iron elements, while the lower density material further out burns to intermediate elements, creating a structure where elements are less mixed with a core of Ni/Fe and outer layers of Si/S. The powerful detonation is efficient at converting large portions of the WD's C/O into Si/S in intermediate-density layers.
The dominating \Ni~ in both cases will eventually decay into \Co~ and then into \Fe, emitting highly energetic \grays. Large Fe abundances along the entire ejecta help to form iron-rich dust. 


\begin{figure}[htbp]
    \centering   
    \includegraphics[width=0.45\textwidth]{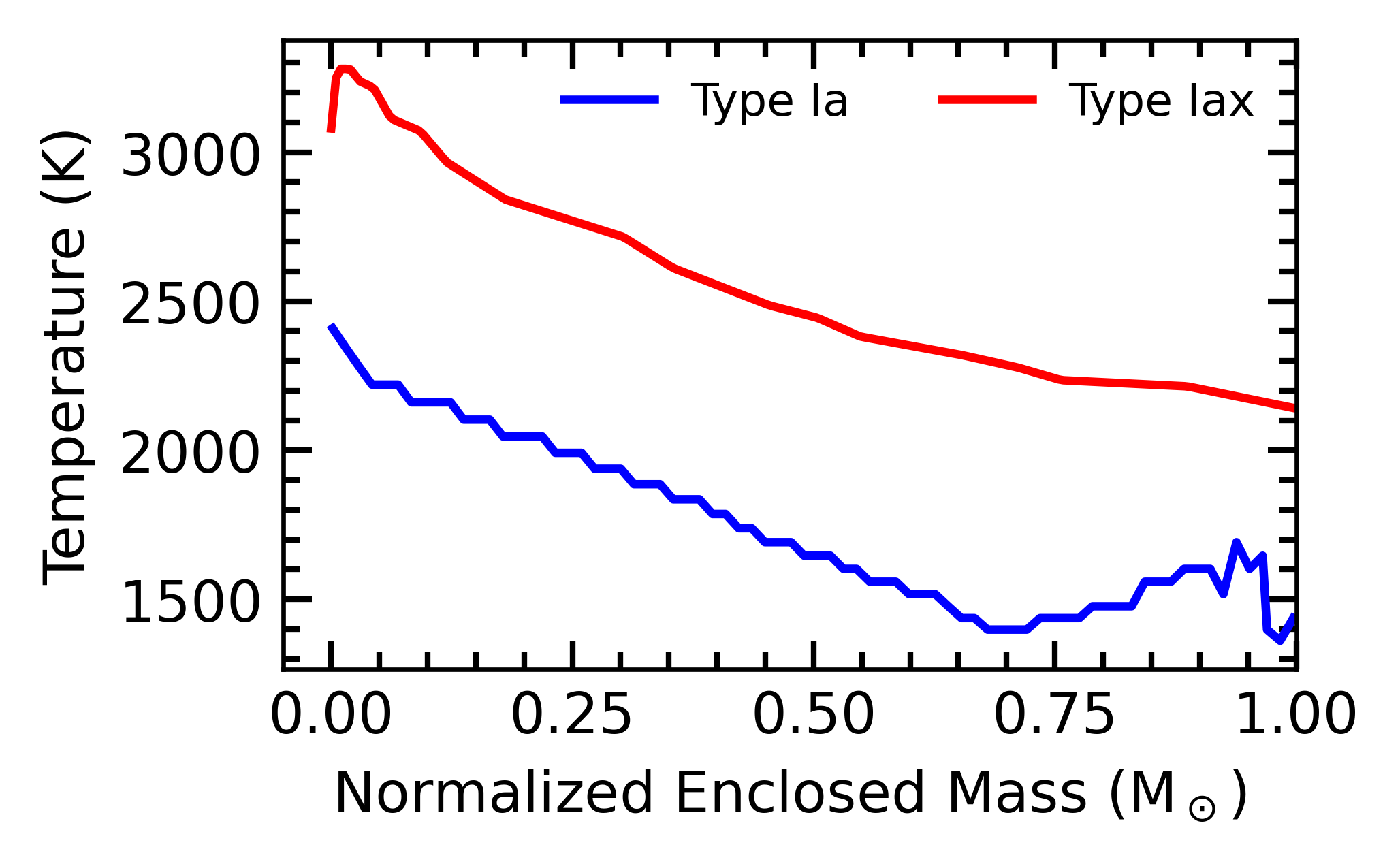}
\caption{ The Temperature profiles at 100 days used in this study. Type Ia temperature profile is taken from \cite{Nozawa_2011} and the hotter profile of Type Iax is taken from \cite{jack_2011}. The enclosed mass is normalized to the ejecta mass of the SN Ia ($M_{ejecta}$ = 1.38 $M_{\odot}$) used in \cite{Nozawa_2011}. Both temperature profiles follows a power law with exponent -0.89 with time.}
\label{fig: temp}
\end{figure}

\begin{figure*}[h] 
    \centering
    \begin{subfigure}{0.48\textwidth}
        \centering
        \includegraphics[width=\linewidth]{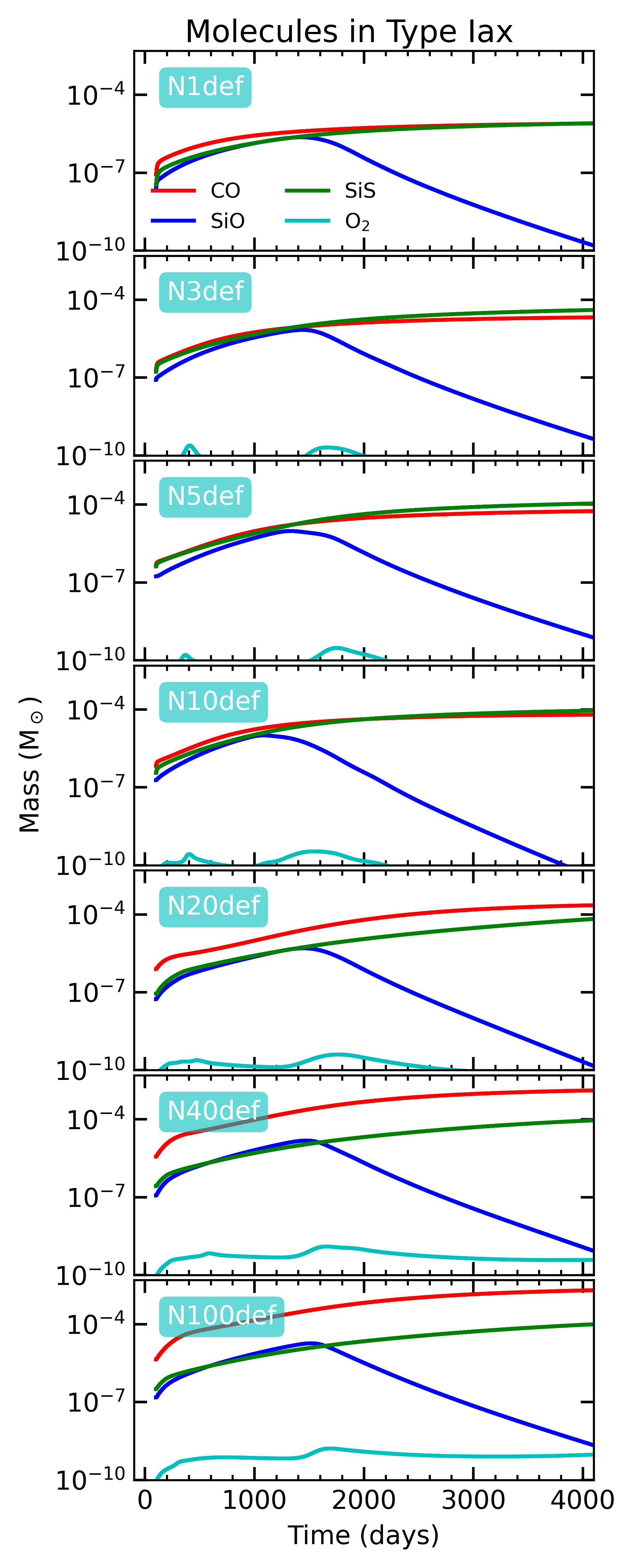}
        \label{fig: molecules_def}
    \end{subfigure}
    \hfill
    \begin{subfigure}{0.48\textwidth}
        \centering
        \includegraphics[width=\linewidth]{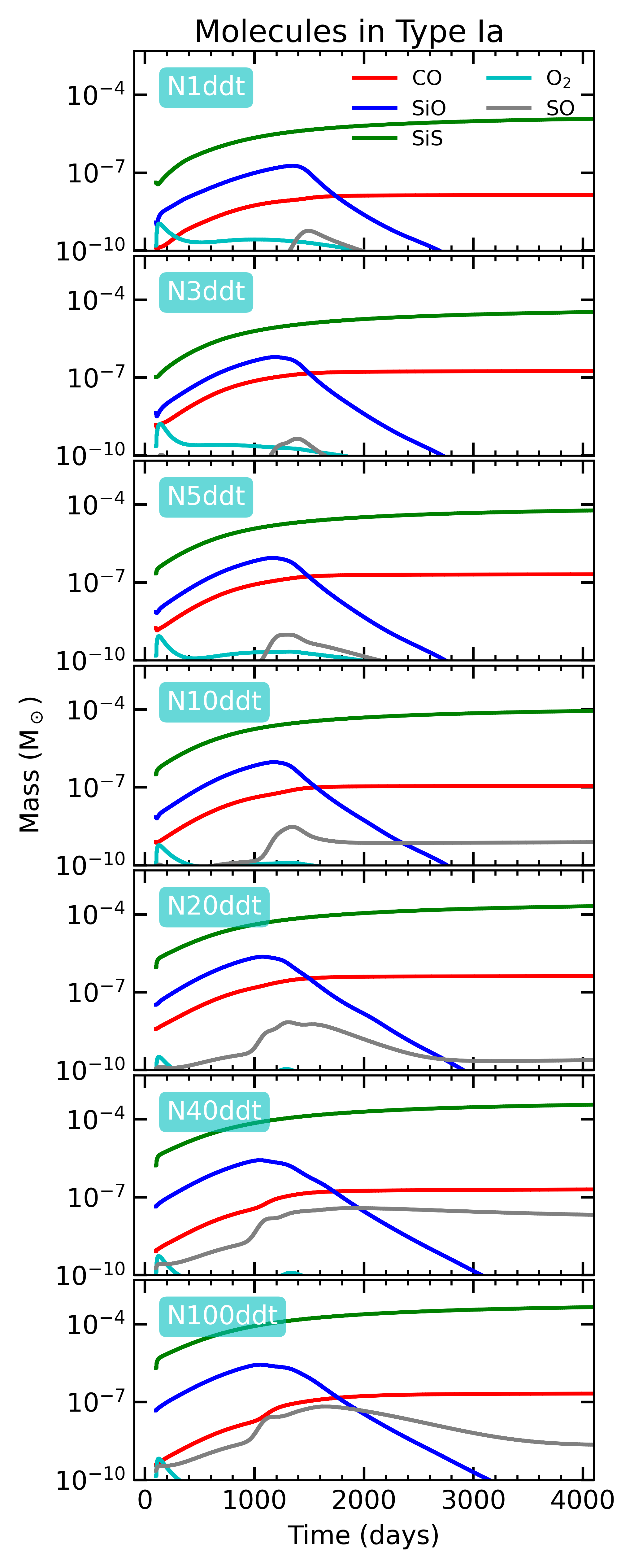}
        \label{fig: molecules_ddt}
    \end{subfigure}
    \caption{Comparison of molecules produced in Type Iax and Type Ia SN ejecta.}
    \label{fig: molecules}
\end{figure*}

\subsection{Gas Temperature}

\cite{Nozawa_2011} obtained temperature distributions for Type Ia gas ejecta using the deflagration models of \cite{Nomoto_1984b,Thielemann_1986} for 100 and 300 days after explosion. These temperature profiles follow a power law with exponent -0.89 with time.  For all DDT models, the ejecta mass is similar, and we scaled the normalized mass in fig \ref{fig: temp} to 1.4 \Ms\ ejecta mass. 
There is no model available for the temperature profile of Type Iax SN ejecta. We expect that due to lower velocities and a bound remnant, the initial temperatures of Type Iax ejecta should be higher than those of Type Ia SN. \cite{Banhidi2025} reported photospheric temperature of 4475~K  for SN~2022xlp at 90 days, \cite{MSingh2022} reported 5500 K for SN~2020udy at ~100-125 days, and $\sim$4500~K for SN~2020rea at 140 days. Additionally, \cite{Kawabata2018} reported photospheric temperature around 3700~K for Type Iax SN~2014dt at 132 days, and remained constant until 410 days, after the explosion. However, at relatively early epochs ($\sim$100 days), the ejecta of Type Iax supernovae are expected to be optically thin due to their low ejecta masses \citep{kwok_2025}, so the gas temperature is expected to drop. Hence, the evolution of photospheric temperature should not be a tracer for the evolution of gas temperatures. In our study for Type Iax SN ejecta, we used the theoretical radial temperature profiles of Type Ia from \cite{jack_2011} at 50 days (shown in Fig.~\ref{fig: temp}) and scale according to the ejecta mass. Similar to Type Ia, we assume the temporal profiles to follow a power-law exponent of -0.89. Our model temperature for the inner ejecta at 100 days is comparable to the estimated photospheric temperature for SN~2014dt by \cite{Kawabata2018}. 

\subsection{Gas Density}

We use densities provided by \cite{Fink_2014} at 100s post-explosion, in velocity space. Figure \ref{fig: density} presents the density profiles as a function of enclosed mass, for deflagration and DDT models at 100 days. All deflagration models exhibit a steep density drop in the innermost regions, followed by a more gradual decline toward the outer layers. Lower-energy explosions (e.g., \texttt{N1def, N3def, N5def}) produce a small amount of ejecta with a sharp decline in the outer regions, while higher-energy models (e.g. \texttt{N40def, N100def}) display more extended density gradients.\\
The DDT models of \cite{Seithenzahl_2013} have lower densities since the star was pre-expanded and exploded by detonation. All models have similar densities, with slight differences in intermediate regions due to variations in the initial deflagration wave configurations.

\section{Ejecta chemistry}\label{chem}
The chemical evolution of the SN ejecta is a complex interplay of various physical and chemical processes. We describe our chemical kinetic model, building upon previous work \citep{sar13, sarangi_2022b} while incorporating new species and reactions relevant to the unique conditions of Type Iax SNe.

Our model is based on the well-established Non-Equilibrium Chemistry Solver Algorithm (NECSA), which has been successfully applied to study the chemistry of Type II SN ejecta \citep{sar13,sar15,sarangi2018book, Sarangi_2022}. The formation of dust and molecules is based on simultaneous phases of nucleation and condensation \citep{cherchneff2009,cherchneff2010, sar15}. The chemical network accounting for all possible reactions involving atoms, molecules, and molecular clusters present in the gas is modelled by \cite{sar13}. A comprehensive list of chemical reactions is presented in \cite{cherchneff2009} and \cite{sar13}, encompassing both thermal and nonthermal processes. In this paper, we have significantly expanded the chemical network to account for a more comprehensive set of reactions involving iron and its compounds. We have added new species, including iron oxides, Mg-Fe silicates (Pyroxenes and Olivines); Table \ref{tab: network} lists all the new reactions. The nucleation proceeds by consecutive phases of Mg/Fe addition and oxidation by single O addition, \ce{O$_2$} and \ce{SO}. Due to the lack of documentation, some of the rates are estimated using similar reactions with known reaction rates. Our final chemical network consists of 109 species and 504 reactions. It includes possible thermal and non-thermal processes constrained by the temperature and density conditions.
\begin{figure*}[htbp]
    \centering   
    \includegraphics[]{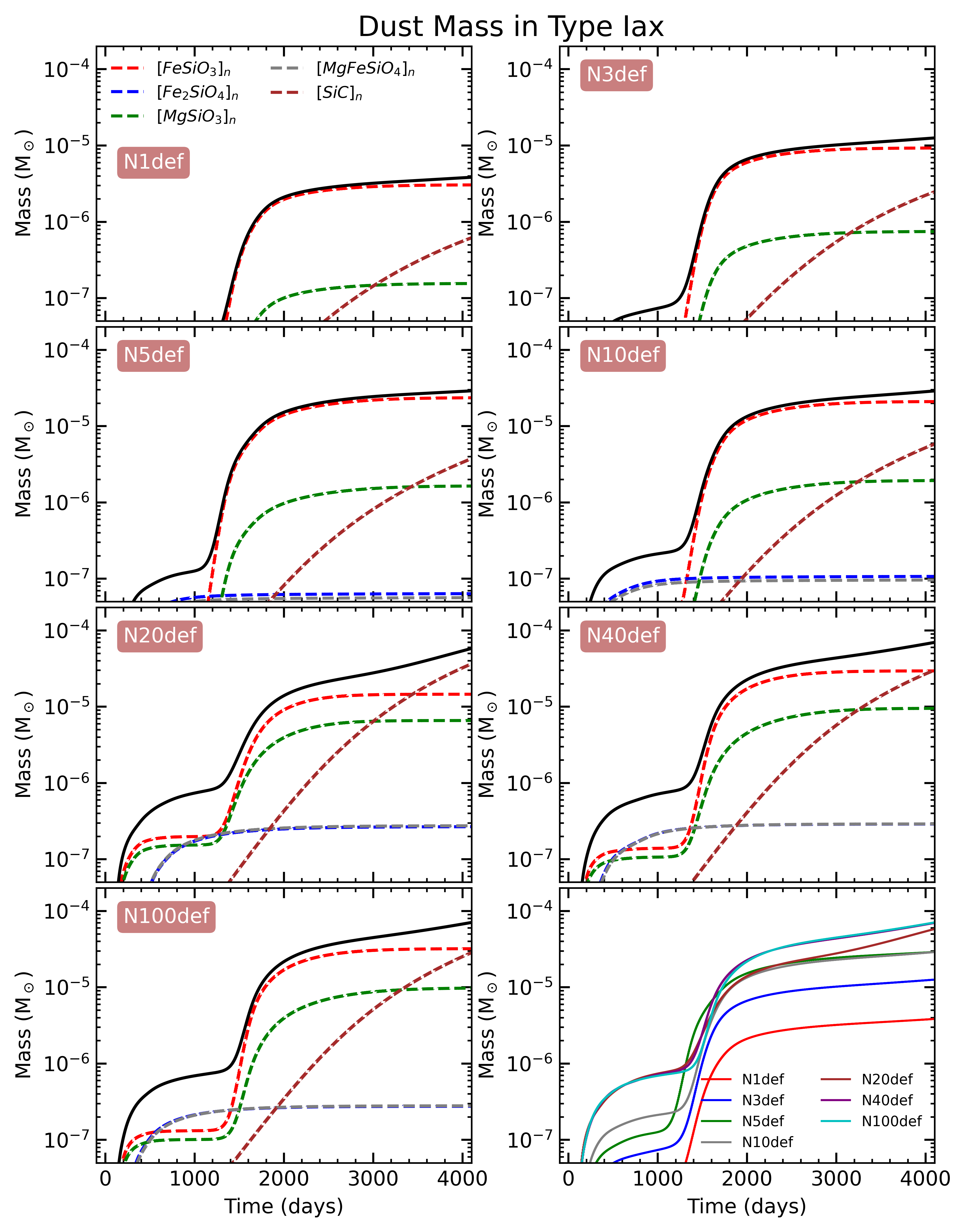}
\caption{Dust mass evolution in Type Iax SN ejecta for different deflagration models. Each panel shows the time evolution of individual dust species produced in dashed colored lines with the total dust mass in a solid black line. The bottom-right panel compares total dust masses across all models.}
\label{fig: defmodel}
\end{figure*}
\begin{figure*}[htbp]
    \centering   
    \includegraphics[]{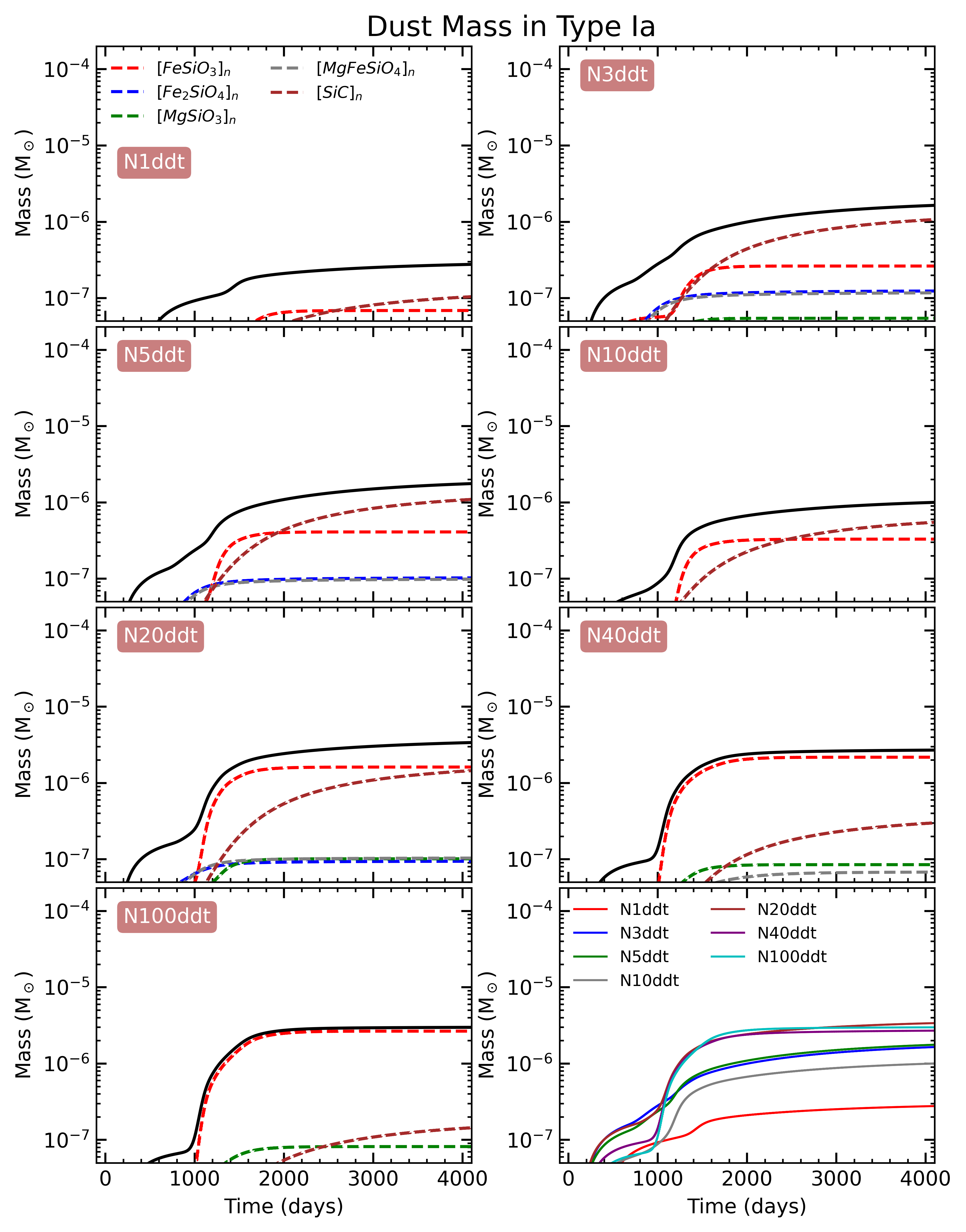}
\caption{Dust mass evolution in Type Ia SN ejecta for different delayed-detonation models. Each panel shows the time evolution of individual dust species produced in dashed colored lines with the total dust mass in a solid black line. The bottom-right panel compares total dust masses across all models \citep{Seithenzahl_2013}.}
\label{fig: ddtmodel}
\end{figure*}

Nonthermal processes are mostly influenced by radioactive \Ni \  which will further decay to \Co \ and then to \Fe\ , producing highly energetic \grays\ , that degrade to UV photons and Compton electrons (CE). They disrupt and delay the formation of both precursor molecules and dust grains. Type Iax SNe have a relatively smaller amount of \Ni \  as mentioned in table \ref{tab: Model}, when compared to Type Ia SNe. The reactions with fast CE depend upon the amount of \Ni\ produced and the ejecta mass. For deflagration models \texttt{N1def, N10def, N100def}, the rates are given in table \ref{tab: rates} in Arrhenius form, calculated using the prescription of \citep{cherchneff2009}. However, Type Ia SN produces a significant amount of \Ni\ (around 0.6 \Ms\ ). Theoretical DDT models produce \Ni\ masses ranging from 1.11 to 0.6 \Ms\ . Table \ref{tab: rates} presents the CE rates for the \texttt{N100ddt} model with the lowest \Ni\ mass among all DDT models. 

\section{Molecules in the ejecta of Type Iax compared to Ia}\label{molecule}
We followed the evolution of the species from day 100 to 4100, post-explosion. The synthesis of diatomic molecules, namely \ce{SiO}, \ce{CO}, \ce{O$_2$}, \ce{SO}, and \ce{SiS}, is important in the ejecta in terms of their abundances and epochs of formation shown in Figure \ref{fig: molecules} (Left-panel for Type Iax and Right-panel for Type Ia). We begin our analysis of the ejecta chemistry from day 100 after the explosion. The conditions of the ejecta before that are not suitable for the synthesis of neutral molecules or clusters.

In Type Iax SNe, for all the models, \ce{CO} abundance increases as early as 200 days and becomes almost saturated by 2000 days, post-explosion. There is no further change in the abundances of \ce{CO} at late times since CO does not act as a precursor molecule to any dust species \citep{sar13}.  
\ce{SiO} is the most important precursor molecule in dust production. The epoch of \ce{SiO} depletion is similar to the epoch of rise in dust masses, highlighting that the formation of \ce{SiO} molecules in the ejecta leads to rapid conversion into small dust clusters. \ce{O$_2$} gets quickly depleted to form CO and SiO molecules, and continues to act as an oxidizing agent for O-rich clusters of silicates. The zones rich in Si and S and scarce in oxygen prefer the formation of \ce{SiS} molecules. \ce{SO} synthesized in the ejecta also has an oxidizing effect on the small clusters, similar to that of O$_2$. Despite large abundances of Fe atoms, the abundances of \ce{FeO} molecules are not significant. 


The ejecta of the DDT models(Type Ia SNe), are rich in \ce{Si}, \ce{S}, which explains the relatively much higher production of \ce{SiS} than the \ce{CO} molecules, and the production of \ce{SO} molecules also increases in higher configuration models, which is missing in the deflagration models.

\section{Dust budget of Type Iax SNe}\label{dust}


As discussed in section \ref{chem}, we introduced a chemical network (table \ref{tab: network})for Fe-Mg silicates and iron oxides based on nucleation. For all of these models, there was no significant iron oxide production in the molecular phase.  Pyroxenes like Ferrosilite (\ce{FeSiO3}) and Enstatite(\ce{MgSiO3}) are the most abundant dust species followed by Olivines like Fayalite(\ce{Fe2SiO4}) and Forsterite(\ce{MgFeSiO4}) species contributing to total dust mass, as shown in Figure \ref{fig: defmodel}. The formation of Fe-rich silicates is not proposed for Type II SNe \citep{sar13}, but the unique abundance distribution in Type Iax SNe leads to the formation of such silicates. 


For Type Iax, the total ejecta in each model was divided into multiple zones of the same size, and then the total dust mass was calculated by summing over all zones. Olivines like \ce{Fe2SiO4} and \ce{MgFeSiO4} species form early in small masses at 300 days in the innermost zones. Later, at 1500 days, \ce{FeSiO3} and \ce{MgSiO3} appear in the innermost regions, leading to a sudden increase in dust masses. After this, we are still left with a lot of elements in free gas form, and free Si and C combine to form \ce{SiC} at later times and become equally significant in higher configuration models. 

The formation of different species and epochs of formation depends on the physical conditions of the environment and local chemistry.
Figure \ref{fig: defmodel} and table \ref{tab: dust mass} highlight the individual species contribution in each deflagration model of Type Iax SNe, along with a comparison between the dust production in different models.
\texttt{N1def} and \texttt{N3def} models are low energy, have low ejecta mass, forming dust of masses around \SI{3.7e-06}{M_{\odot}} and \SI{1.2e-05}{M_{\odot}} at 4000 days after explosion, respectively. \texttt{N20def, N40def} and \texttt{N100def} are high-energy models with small bound remnant mass, have high ejecta mass giving dust mass of \SI{5.4e-05}{M_{\odot}} to \SI{6.7e-05}{M_{\odot}} at 4000 days after explosion.
Intermediate models \texttt{N5def} and \texttt{N10def} produce dust masses of \SI{2.8e-05}{M_{\odot}}; however, the time of formation is earlier than the other cases.\\
All models vary in dust composition, depending on their ejecta mass, elemental composition, and \Ni ~content. Figure \ref{fig: velocity} (left-panel) describes the dust formation zones as a function of time, in velocity space. As we see in the Figure \ref{fig: comparison}, the dust-to-gas mass ratio of all the deflagration models is almost the same, i.e. 4 to \SI{8e-05}{M_{\odot}} at 4000 days (table \ref{tab: compare}), suggesting that Type Iax SNe are efficient dust producers. Specifically, \texttt{N3def, N5def} and \texttt{N10def} models provide the best conditions for dust production, and there are some observed SN candidates of these models as shown in table \ref{tab: Model}.
\begin{longtable*}{cccccccc}
\caption{Mass evolution of individual dust species in time along with the total dust masses in all deflagration models and one scaled model for SN 2012Z. The masses are in \Ms. }
\label{tab: dust mass}\\
\hline
Days & \ce{[FeSiO3]_n} & \ce{[MgSiO3]_n} & \ce{[Fe2SiO4]_n} & \ce{[Mg2SiO4]_n} & \ce{[MgFeSiO4]_n} & \ce{[SiC]_n} & Total Dust Mass \\ \hline
\endfirsthead
\multicolumn{8}{c}%
{{\bfseries Table \thetable\ (continue)}} \\
\hline
Days & \ce{[FeSiO_3]_n} & \ce{[MgSiO_3]_n} & \ce{[Fe_2SiO_4]_n} & \ce{[Mg_2SiO_4]_n} & \ce{[MgFeSiO_4]_n} & \ce{[SiC]_n} & Total Dust Mass \\ \hline
\endhead
\hline
\endfoot
\endlastfoot
\multicolumn{8}{c}{\texttt{N1def}}                                                  \\ \hline
300  & \SI{2.5e-10}{}&\SI{ 1.9E-10 }{}&\SI{ 3.2E-09 }{}&\SI{ 1.1E-10 }{}&\SI{ 2.8E-09 }{}&\SI{ 4.3E-11 }{}&\SI{ 6.6E-09 }{} \\
500  & \SI{3.4e-10}{}&\SI{ 2.5E-10 }{}&\SI{ 5.9E-09 }{}&\SI{ 2.1E-10 }{}&\SI{ 5.3E-09 }{}&\SI{ 1.0E-10 }{}&\SI{ 1.2E-08}{}  \\
800  & \SI{3.7E-10}{}&\SI{ 2.8E-10 }{}&\SI{ 8.0E-09 }{}&\SI{ 2.8E-10 }{}&\SI{ 7.1E-09 }{}&\SI{ 3.2E-10 }{}&\SI{ 1.6E-08}{}  \\
1000 & \SI{4.7E-10}{}&\SI{ 2.9E-10 }{}&\SI{ 8.9E-09 }{}&\SI{ 3.1E-10 }{}&\SI{ 7.9E-09 }{}&\SI{ 6.5E-10 }{}&\SI{ 1.9E-08 }{} \\
1500 & \SI{3.0E-07}{}&\SI{ 1.6E-08 }{}&\SI{ 1.0E-08 }{}&\SI{ 3.3E-10 }{}&\SI{ 8.9E-09 }{}&\SI{ 3.6E-09 }{}&\SI{ 3.4E-07 }{} \\
2000 &\SI{2.0E-06}{}&\SI{ 1.0E-07 }{}&\SI{ 1.1E-08 }{}&\SI{ 3.5E-10 }{}&\SI{ 9.3E-09 }{}&\SI{ 1.6E-08 }{}&\SI{ 2.1E-06 }{} \\
2500 & \SI{2.6E-06}{}&\SI{ 1.3E-07 }{}&\SI{ 1.1E-08 }{}&\SI{ 3.5E-10 }{}&\SI{ 9.5E-09 }{}&\SI{ 5.5E-08 }{}&\SI{ 2.8E-06 }{} \\ 
3000& \SI{2.9E-06}{}& \SI{ 1.4E-07 }{}& \SI{ 1.1E-08 }{}&\SI{ 3.6E-10 }{}&\SI{ 9.7E-09 }{}&\SI{ 1.4E-07 }{}& \SI{ 3.2E-06 }{}\\ 
4000 & \SI{3.0E-06}{}& \SI{ 1.5E-07 }{}&\SI{ 1.1E-08 }{}& \SI{ 3.7E-10 }{}& \SI{ 9.9E-09 }{}& \SI{ 5.5E-07 }{}&\SI{ 3.7E-06 }{} \\ \hline
\multicolumn{8}{c}{\texttt{N3def}}                                                  \\ \hline
300  & \SI{2.3E-09 }{}&\SI{ 1.7E-09 }{}&\SI{ 1.2E-08 }{}&\SI{ 6.1E-10 }{}&\SI{ 1.1E-08 }{}&\SI{ 1.3E-10 }{}&\SI{ 2.7E-08}{}  \\
500  & \SI{3.0E-09 }{}&\SI{ 2.3E-09 }{}&\SI{ 2.2E-08 }{}&\SI{ 1.1E-09 }{}&\SI{ 2.0E-08 }{}&\SI{ 3.3E-10 }{}&\SI{ 4.9E-08 }{} \\
800  &\SI{ 3.4E-09 }{}&\SI{ 2.6E-09 }{}&\SI{ 3.0E-08 }{}&\SI{ 1.5E-09 }{}&\SI{ 2.7E-08 }{}&\SI{ 1.0E-09 }{}&\SI{ 6.5E-08 }{} \\
1000 &\SI{ 3.6E-09 }{}&\SI{ 2.6E-09 }{}&\SI{ 3.3E-08 }{}&\SI{ 1.7E-09 }{}&\SI{ 3.0E-08 }{}&\SI{ 2.1E-09 }{}&\SI{ 7.3E-08}{}  \\
1500 & \SI{8.3E-07 }{}&\SI{ 7.1E-08 }{}&\SI{ 4.0E-08 }{}&\SI{ 1.9E-09 }{}&\SI{ 3.6E-08 }{}&\SI{ 1.2E-08 }{}&\SI{ 9.9E-07 }{} \\
2000 & \SI{6.0E-06 }{}&\SI{ 4.8E-07 }{}&\SI{ 4.2E-08 }{}&\SI{ 2.0E-09 }{}&\SI{ 3.8E-08 }{}&\SI{ 5.4E-08 }{}&\SI{ 6.6E-06 }{} \\
2500 & \SI{8.0E-06 }{}&\SI{ 6.4E-07 }{}&\SI{ 4.4E-08 }{}&\SI{ 2.1E-09 }{}&\SI{ 4.0E-08 }{}&\SI{ 1.9E-07 }{}&\SI{ 8.9E-06}{} \\
3000 & \SI{8.0E-06 }{}&\SI{ 6.4E-07 }{}& \SI{ 4.4E-08 }{}&\SI{ 2.1E-09 }{}&\SI{ 4.0E-08 }{}& \SI{ 1.9E-07 }{}&\SI{ 1.0E-05}{} \\
4000 &  \SI{8.0E-06 }{}&\SI{ 6.4E-07 }{}& \SI{ 4.4E-08 }{}& \SI{ 2.1E-09 }{}& \SI{ 4.0E-08 }{}&\SI{ 1.9E-07 }{}& \SI{ 1.2E-05}{} \\ \hline
\multicolumn{8}{c}{\texttt{N5def}}                                                  \\ \hline
300  & \SI{5.4E-09 }{}&\SI{ 4.1E-09 }{}&\SI{ 1.9E-08 }{}&\SI{ 6.2E-10 }{}&\SI{ 1.7E-08 }{}&\SI{ 2.1E-10 }{}&\SI{ 4.7E-08}{}  \\
500  & \SI{6.6E-09 }{}&\SI{ 5.0E-09 }{}&\SI{ 3.6E-08 }{}&\SI{ 1.2E-09 }{}&\SI{ 3.1E-08 }{}&\SI{ 5.0E-10 }{}&\SI{ 8.1E-08 }{} \\
800  & \SI{6.9E-09 }{}&\SI{ 5.3E-09 }{}&\SI{ 5.2E-08 }{}&\SI{ 1.7E-09 }{}&\SI{ 4.6E-08 }{}&\SI{ 1.5E-09 }{}&\SI{ 1.1E-07}{}  \\
1000 & \SI{8.7E-09 }{}&\SI{ 5.4E-09 }{}&\SI{ 5.7E-08 }{}&\SI{ 1.9E-09 }{}&\SI{ 5.0E-08 }{}&\SI{ 3.0E-09 }{}&\SI{ 1.3E-07}{}  \\
1500 & \SI{3.8E-06 }{}&\SI{ 3.0E-07 }{}&\SI{ 6.1E-08 }{}&\SI{ 2.0E-09 }{}&\SI{ 5.4E-08 }{}&\SI{ 1.7E-08 }{}&\SI{ 4.3E-06}{}  \\
2000 & \SI{1.4E-05 }{}&\SI{ 9.7E-07 }{}&\SI{ 6.2E-08 }{}&\SI{ 2.1E-09 }{}&\SI{ 5.5E-08 }{}&\SI{ 8.1E-08 }{}&\SI{ 1.5E-05}{}  \\
2500 & \SI{1.9E-05 }{}&\SI{ 1.3E-06 }{}&\SI{ 6.3E-08 }{}&\SI{ 2.1E-09 }{}&\SI{ 5.5E-08 }{}&\SI{ 3.0E-07 }{}&\SI{ 2.1E-05}{}\\
3000 & \SI{2.2E-05 }{}&\SI{ 1.5E-06 }{}&\SI{ 6.3E-08 }{}&\SI{2.1E-09 }{}&\SI{ 5.6E-08 }{}&\SI{ 8.1E-07 }{}&\SI{ 2.4E-05}{} \\
4000 & \SI{2.3E-05 }{}& \SI{ 1.6E-06 }{}& \SI{ 6.4E-08 }{}&\SI{ 2.2E-09 }{}&\SI{ 5.6E-08 }{}& \SI{ 3.3E-06 }{}&\SI{ 2.8E-05}{} \\ \hline
\multicolumn{8}{c}{\texttt{N10def}}                                                 \\ \hline
300  & \SI{1.2E-08 }{}&\SI{ 9.0E-09 }{}&\SI{ 2.7E-08 }{}&\SI{ 1.1E-09 }{}&\SI{ 2.5E-08 }{}&\SI{ 3.1E-10 }{}&\SI{ 7.5E-08}{}  \\
500  & \SI{1.4E-08 }{}&\SI{ 1.1E-08 }{}&\SI{ 5.7E-08 }{}&\SI{ 2.5E-09 }{}&\SI{ 5.2E-08 }{}&\SI{ 7.1E-10 }{}&\SI{ 1.4E-07}{}  \\
800  & \SI{1.5E-08 }{}&\SI{ 1.2E-08 }{}&\SI{ 8.2E-08 }{}&\SI{ 3.6E-08 }{}&\SI{ 7.4E-08 }{}&\SI{ 2.1E-09 }{}&\SI{ 1.9E-07 }{} \\
1000 & \SI{1.5E-08 }{}&\SI{ 1.2E-08 }{}&\SI{ 9.3E-08 }{}&\SI{ 4.0E-08 }{}&\SI{ 8.3E-08 }{}&\SI{ 4.4E-09 }{}&\SI{ 2.1E-07}{}  \\
1500 & \SI{1.2E-06 }{}&\SI{ 1.4E-07 }{}&\SI{ 1.0E-07 }{}&\SI{ 4.4E-08 }{}&\SI{ 9.1E-08 }{}&\SI{ 2.5E-08 }{}&\SI{ 1.5E-06}{}  \\
2000 & \SI{1.2E-05 }{}&\SI{ 1.0E-06 }{}&\SI{ 1.0E-07 }{}&\SI{ 4.4E-08 }{}&\SI{ 9.4E-08 }{}&\SI{ 1.2E-07 }{}&\SI{ 1.4E-05}{}  \\
2500 & \SI{1.7E-05 }{}&\SI{ 1.6E-06 }{}&\SI{ 1.0E-07 }{}&\SI{ 4.5E-08 }{}&\SI{ 9.5E-08 }{}&\SI{ 4.4E-07 }{}&\SI{ 1.9E-05}{} \\
3000 &\SI{1.9E-05 }{}&\SI{ 1.8E-06 }{}& \SI{ 1.1E-07 }{}&\SI{ 4.5E-08 }{}&\SI{ 9.5E-08 }{}& \SI{ 1.2E-06 }{}& \SI{ 2.3E-05}{}\\
4000 &  \SI{2.1E-05 }{}& \SI{ 1.9E-06 }{}& \SI{ 1.1E-07 }{}&\SI{ 4.6E-08 }{}& \SI{ 9.6E-08 }{}&\SI{ 5.2E-06 }{}& \SI{ 2.8E-05}{}\\ \hline
\multicolumn{8}{c}{\texttt{N20def}}                                                 \\ \hline
300  & \SI{1.2E-07 }{}&\SI{ 9.4E-08 }{}&\SI{ 1.8E-09 }{}&\SI{ 5.9E-09 }{}&\SI{ 1.8E-09 }{}&\SI{ 5.1E-10 }{}&\SI{ 2.2E-07}{}  \\
500  & \SI{1.8E-07 }{}&\SI{ 1.4E-07 }{}&\SI{ 4.4E-08 }{}&\SI{ 1.8E-08 }{}&\SI{ 4.4E-08 }{}&\SI{ 1.5E-09 }{}&\SI{ 4.2E-07}{}  \\
800  & \SI{1.9E-07 }{}&\SI{ 1.5E-07 }{}&\SI{ 1.3E-07 }{}&\SI{ 2.4E-08 }{}&\SI{ 1.3E-07 }{}&\SI{ 5.4E-09 }{}&\SI{ 6.4E-07}{}  \\
1000 & \SI{2.0E-07 }{}&\SI{ 1.5E-07 }{}&\SI{ 1.7E-07 }{}&\SI{ 3.3E-08 }{}&\SI{ 1.7E-07 }{}&\SI{ 1.7E-08 }{}&\SI{ 7.3E-07}{}  \\
1500 & \SI{1.6E-06 }{}&\SI{ 7.5E-07 }{}&\SI{ 2.3E-07 }{}&\SI{ 3.7E-08 }{}&\SI{ 2.3E-07 }{}&\SI{ 7.5E-08 }{}&\SI{ 2.4E-06}{}  \\
2000 & \SI{9.0E-06 }{}&\SI{ 3.9E-06 }{}&\SI{ 2.5E-07 }{}&\SI{ 3.8E-08 }{}&\SI{ 2.6E-07 }{}&\SI{ 4.3E-07 }{}&\SI{ 1.4E-05 }{} \\
2500 & \SI{1.3E-05 }{}&\SI{ 5.8E-06 }{}&\SI{ 2.6E-07 }{}&\SI{ 3.8E-08 }{}&\SI{ 2.7E-07 }{}&\SI{ 1.9E-06 }{}&\SI{ 2.2E-05 }{}\\
3000 & \SI{1.4E-05 }{}& \SI{ 6.4E-06 }{}& \SI{ 2.6E-07 }{}& \SI{ 3.9E-08 }{}& \SI{ 2.7E-07 }{}& \SI{ 6.3E-06 }{}& \SI{ 2.8E-05 }{} \\
4000 & \SI{1.5E-05 }{}& \SI{ 6.6E-06 }{}& \SI{ 2.7E-07 }{}& \SI{ 3.9E-08 }{}& \SI{ 2.8E-07 }{}& \SI{ 3.2E-05 }{}& \SI{ 5.4E-05 }{} \\ \hline
\multicolumn{8}{c}{\texttt{N40def}}                                                 \\ \hline
300  & \SI{9.6E-08 }{}&\SI{ 7.3E-08 }{}&\SI{ 2.7E-08 }{}&\SI{ 1.7E-09 }{}&\SI{ 2.5E-08 }{}&\SI{ 7.1E-10 }{}&\SI{ 2.2E-07 }{} \\
500  & \SI{1.3E-07 }{}&\SI{ 9.7E-08 }{}&\SI{ 1.0E-07 }{}&\SI{ 8.9E-09 }{}&\SI{ 9.7E-08 }{}&\SI{ 1.7E-09 }{}&\SI{ 4.3E-07 }{} \\
800  & \SI{1.4E-07 }{}&\SI{ 1.1E-07 }{}&\SI{ 1.7E-07 }{}&\SI{ 2.0E-08 }{}&\SI{ 1.7E-07 }{}&\SI{ 5.9E-09 }{}&\SI{ 6.2E-07 }{} \\
1000 & \SI{1.4E-07 }{}&\SI{ 1.1E-07 }{}&\SI{ 2.2E-07 }{}&\SI{ 2.6E-08 }{}&\SI{ 2.2E-07 }{}&\SI{ 1.2E-08 }{}&\SI{ 7.2E-07 }{} \\
1500 & \SI{1.3E-06 }{}&\SI{ 5.4E-07 }{}&\SI{ 2.6E-07 }{}&\SI{ 3.3E-08 }{}&\SI{ 2.6E-07 }{}&\SI{ 7.5E-08 }{}&\SI{ 2.5E-06 }{} \\
2000 & \SI{1.7E-05 }{}&\SI{ 4.5E-06 }{}&\SI{ 2.8E-07 }{}&\SI{ 3.5E-08 }{}&\SI{ 2.8E-07 }{}&\SI{ 4.1E-07 }{}&\SI{ 2.3E-05}{}  \\
2500 & \SI{2.5E-05 }{}&\SI{ 7.3E-06 }{}&\SI{ 2.8E-07 }{}&\SI{ 3.7E-08 }{}&\SI{ 2.9E-07 }{}&\SI{ 1.8E-06 }{}&\SI{ 3.5E-05}{}\\
3000 &  \SI{2.8E-05 }{}&\SI{ 8.8E-06 }{}& \SI{ 2.8E-07 }{}& \SI{ 3.7E-08 }{}&\SI{ 2.9E-07 }{}& \SI{ 5.7E-06 }{}& \SI{4.4E-05}{} \\
4000 &  \SI{2.9E-05 }{}&\SI{ 9.4E-06 }{}& \SI{ 2.9E-07 }{}& \SI{ 3.8E-08 }{}& \SI{ 2.9E-07 }{}& \SI{ 2.7E-05 }{}&\SI{ 6.6E-05}{}\\ \hline
\multicolumn{8}{c}{\texttt{N100def}}                                                \\ \hline
300  &\SI{ 9.6E-08 }{}&\SI{ 7.3E-08 }{}&\SI{ 3.2E-08 }{}&\SI{ 1.5E-09 }{}&\SI{ 3.0E-08 }{}&\SI{ 7.1E-10 }{}&\SI{ 2.3E-07 }{} \\
500  & \SI{1.2E-07 }{}&\SI{ 9.4E-08 }{}&\SI{ 1.0E-07 }{}&\SI{ 8.6E-09 }{}&\SI{ 9.9E-08 }{}&\SI{ 1.7E-09 }{}&\SI{ 4.2E-07 }{} \\
800  & \SI{1.3E-07 }{}&\SI{ 1.0E-07 }{}&\SI{ 1.8E-07 }{}&\SI{ 2.0E-08 }{}&\SI{ 1.8E-07 }{}&\SI{ 5.3E-09 }{}&\SI{ 6.1E-07 }{} \\
1000 & \SI{1.3E-07 }{}&\SI{ 1.0E-07 }{}&\SI{ 2.1E-07 }{}&\SI{ 2.5E-08 }{}&\SI{ 2.1E-07 }{}&\SI{ 1.1E-08 }{}&\SI{ 6.9E-07}{}  \\
1500 & \SI{7.6E-07 }{}&\SI{ 2.4E-07 }{}&\SI{ 2.5E-07 }{}&\SI{ 3.2E-08 }{}&\SI{ 2.5E-07 }{}&\SI{ 6.4E-08 }{}&\SI{ 1.6E-06 }{} \\
2000 & \SI{1.6E-05 }{}&\SI{ 3.9E-06 }{}&\SI{ 2.6E-07 }{}&\SI{ 3.5E-08 }{}&\SI{ 2.6E-07 }{}&\SI{ 3.4E-07 }{}&\SI{ 2.2E-05 }{} \\
2500 & \SI{2.7E-05 }{}&\SI{ 7.0E-06 }{}&\SI{ 2.7E-07 }{}&\SI{ 3.6E-08 }{}&\SI{ 2.7E-07 }{}&\SI{ 1.5E-06 }{}&\SI{ 3.6E-05}{} \\
3000 & \SI{3.0E-05 }{}&\SI{ 8.7E-06 }{}& \SI{ 2.7E-07 }{}& \SI{ 3.7E-08 }{}& \SI{ 2.7E-07 }{}& \SI{ 5.1E-06 }{}& \SI{ 4.5E-05}{}\\
4000 & \SI{3.1E-05 }{}& \SI{ 9.7E-06 }{}& \SI{ 2.7E-07 }{}& \SI{ 3.7E-08 }{}& \SI{ 2.8E-07 }{}& \SI{ 2.5E-05 }{}& \SI{ 6.7E-05}{}\\ \hline
\multicolumn{8}{c}{SN~2012Z}                                                  \\ \hline
300  & \SI{8.67e-08}{}&\SI{ 6.6E-08 }{}&\SI{ 2.6E-09 }{}&\SI{ 8.5E-10 }{}&\SI{ 3.2E-09 }{}&\SI{ 6.8E-10 }{}&\SI{ 1.6E-07 }{} \\
500  & \SI{1.4e-07}{}&\SI{ 1.1E-07 }{}&\SI{ 9.1E-09 }{}&\SI{ 3.3E-09 }{}&\SI{ 1.2E-08 }{}&\SI{ 1.8E-09 }{}&\SI{ 2.7E-07}{}  \\
800  & \SI{1.7E-07}{}&\SI{ 1.3E-07 }{}&\SI{ 4.8E-08}{}&\SI{ 1.2E-08 }{}&\SI{ 5.5E-08 }{}&\SI{ 5.0E-09 }{}&\SI{ 4.2E-07}{}  \\
1000 & \SI{1.8E-07}{}&\SI{ 1.4E-07 }{}&\SI{ 7.3E-08 }{}&\SI{ 1.7E-08 }{}&\SI{ 8.3E-08 }{}&\SI{ 9.2E-09 }{}&\SI{ 4.9E-07 }{} \\
1500 & \SI{4.9E-06}{}&\SI{ 3.2E-06 }{}&\SI{ 1.1E-07 }{}&\SI{ 2.5E-08 }{}&\SI{ 1.2E-07 }{}&\SI{ 4.1E-08 }{}&\SI{ 8.3E-06 }{} \\
2000 &\SI{3.4E-05}{}&\SI{ 1.8E-05 }{}&\SI{ 1.2E-07 }{}&\SI{ 2.8E-08 }{}&\SI{ 1.3E-07 }{}&\SI{ 1.7E-07 }{}&\SI{ 5.2E-05 }{} \\
2500 & \SI{5.8E-05}{}&\SI{ 2.9E-05 }{}&\SI{ 1.2E-07 }{}&\SI{ 3.0E-08 }{}&\SI{ 1.4E-07 }{}&\SI{ 6.8E-07 }{}&\SI{8.9E-05 }{} \\
3000 & \SI{7.3E-05}{}& \SI{ 3.7E-05 }{}& \SI{ 1.3E-07 }{}& \SI{ 3.1E-08 }{}& \SI{ 1.4E-07 }{}& \SI{ 2.5E-06 }{}&\SI{1.1E-04 }{}\\ 
4000 &  \SI{8.1E-05}{}& \SI{ 4.2E-05 }{}& \SI{ 1.3E-07 }{}& \SI{ 3.1E-08 }{}& \SI{ 1.5E-07 }{}& \SI{ 2.3E-05 }{}& \SI{1.5E-04 }{}\\ \hline
\end{longtable*}

\begin{figure*}[htbp] 
    \centering
    \begin{subfigure}{0.48\textwidth}
        \centering
        \includegraphics[width=\linewidth]{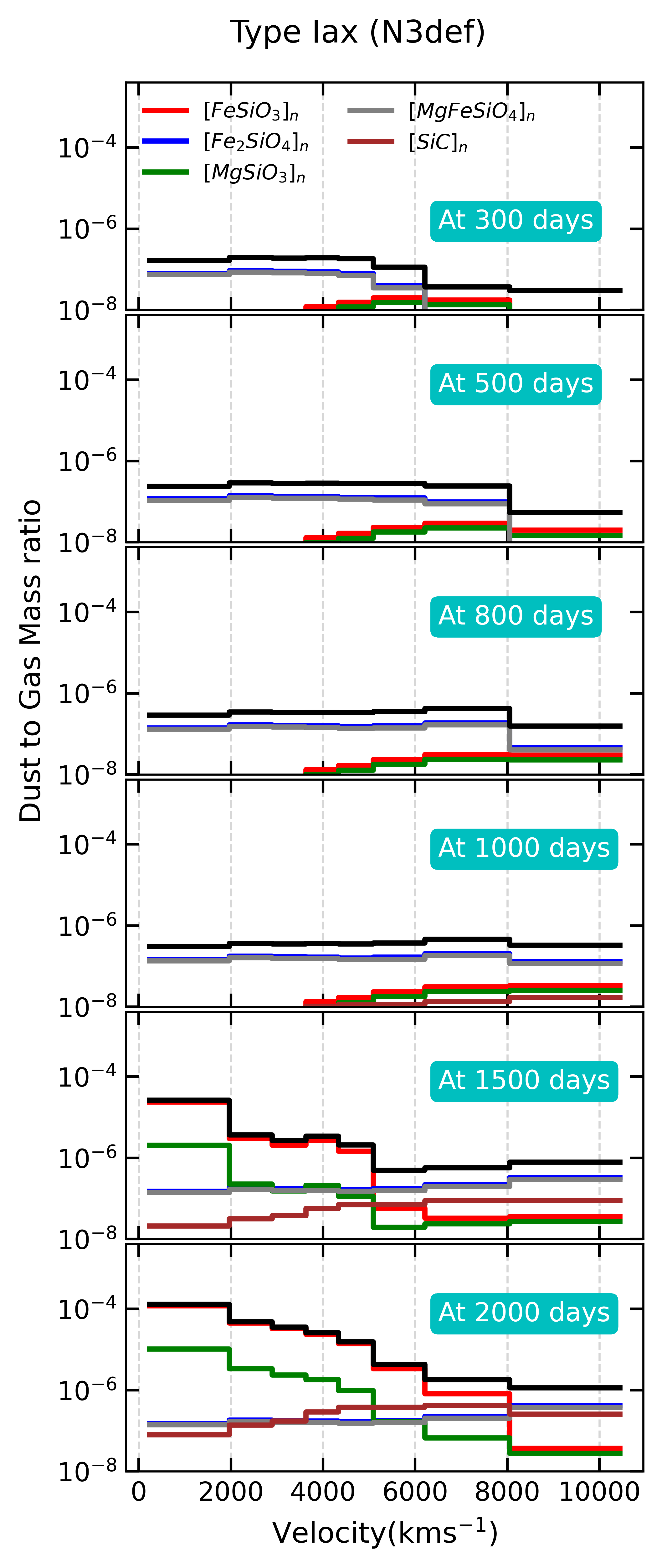}
        \label{fig: defvelocity}
    \end{subfigure}
    \hfill
    \begin{subfigure}{0.48\textwidth}
        \centering
        \includegraphics[width=\linewidth]{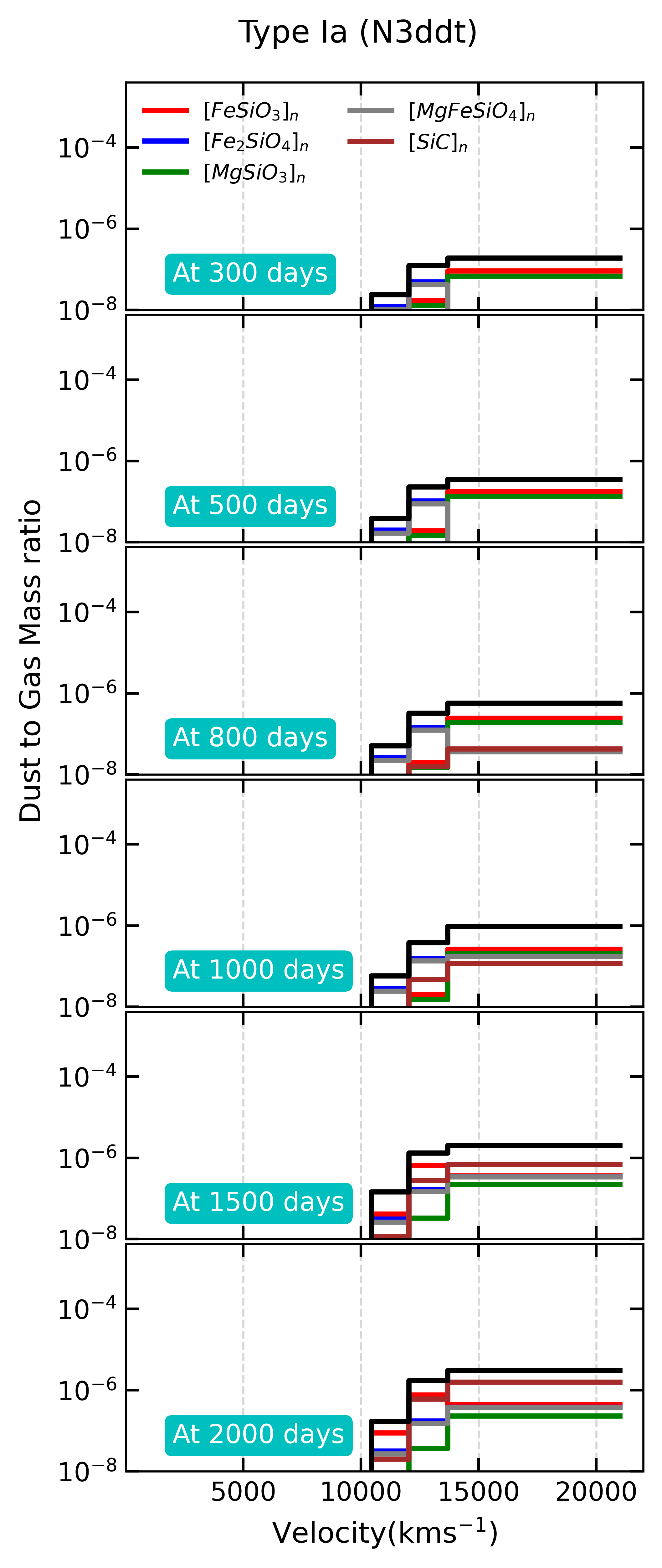}
        \label{fig: velocity}
    \end{subfigure}
    \caption{Comparison of dust to gas mass ratio in Type Iax and Type Ia \texttt{N3} model in velocity coordinate. The plot shows that the \texttt{N3def} model (Type Iax) is much more efficient in dust production than the \texttt{N3ddt} model (Type Ia). The comparison is also highlighting the reason for it, that in the Type Iax model, the dust formation is more in the inner high-density regions, while for the Type Ia, dust formation is only possible in the outer low-density regions.}
    \label{fig: velocity}
\end{figure*}

\subsection{Dust masses in Type Iax SN~2012Z}
\label{sec_SN2012Z}
\begin{figure}[htbp]
    \centering   
    \includegraphics[]{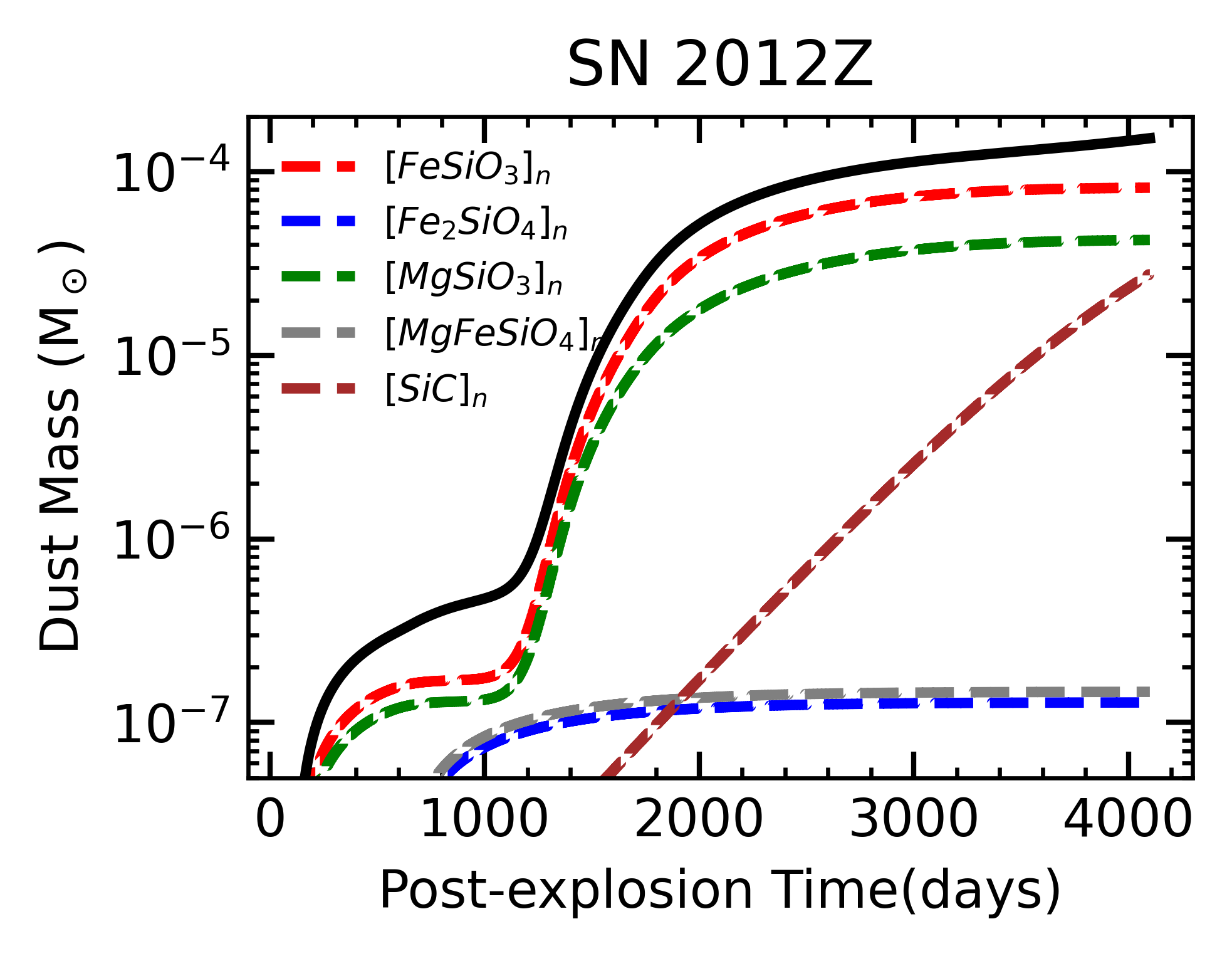}
\caption{Dust mass evolution of each dust species is shown in dashed lines with total dust mass in a solid black line for the \texttt{N10def} model scaled to SN~2012Z ejecta mass. }
\label{fig: obs}
\end{figure} 
 The deflagration models we have used are theoretical simulations and can differ from the actual observational results. The early light curves of these models are comparable to some observed sources, as mentioned in table \ref{tab: Model}. However, \cite{MSingh2024, Singh_2018} reported much larger ejecta mass as compared to the model ejecta masses. \cite{Mag2023} showed that SN~2020udy observations are in good agreement with the \texttt{N5def} model, but they also found the slow light curve evolution of SN~2020udy compared to the \texttt{N5def} model. This can be due to more radiation trapping, suggesting deflagration models are underestimating the ejecta mass. Similarly \cite{MSingh2022}  reported underestimation of ejecta mass by models for SN~2012Z and SN~2020rea.\\
\cite{MSingh2022} reported slow light curve evolution of SN~2012Z as compared to \texttt{N10def} model with \Ni \  mass of 0.12 \Ms\ and ejecta mass,$M_{ej}=1.09$ \Ms, much higher than the deflagration models. These quantities differ from the model; we scaled the \texttt{N10def} model to these parameters, keeping the elemental abundances relative to \Ni \  conserved. Also scaling the \texttt{N10def} model's density and temperature profile mentioned in Figure \ref{fig: density} and \ref{fig: temp}, total dust budget at 4000 days is \SI{1.47e-04}{M_{\odot}}. Profiles of each dust species are mentioned in Table \ref{tab: dust mass} and Figure \ref{fig: obs}. \\
This result shows that Type Iax ejecta, based on observational constraints, will result in even larger dust masses than those estimated from deflagration models. Further development of theoretical models consistent with observations are required to constrain the results.

\section{Comparison with standard Type Ia ejecta}\label{compare}

\begin{figure}[htbp]
    \centering   
    \includegraphics[]{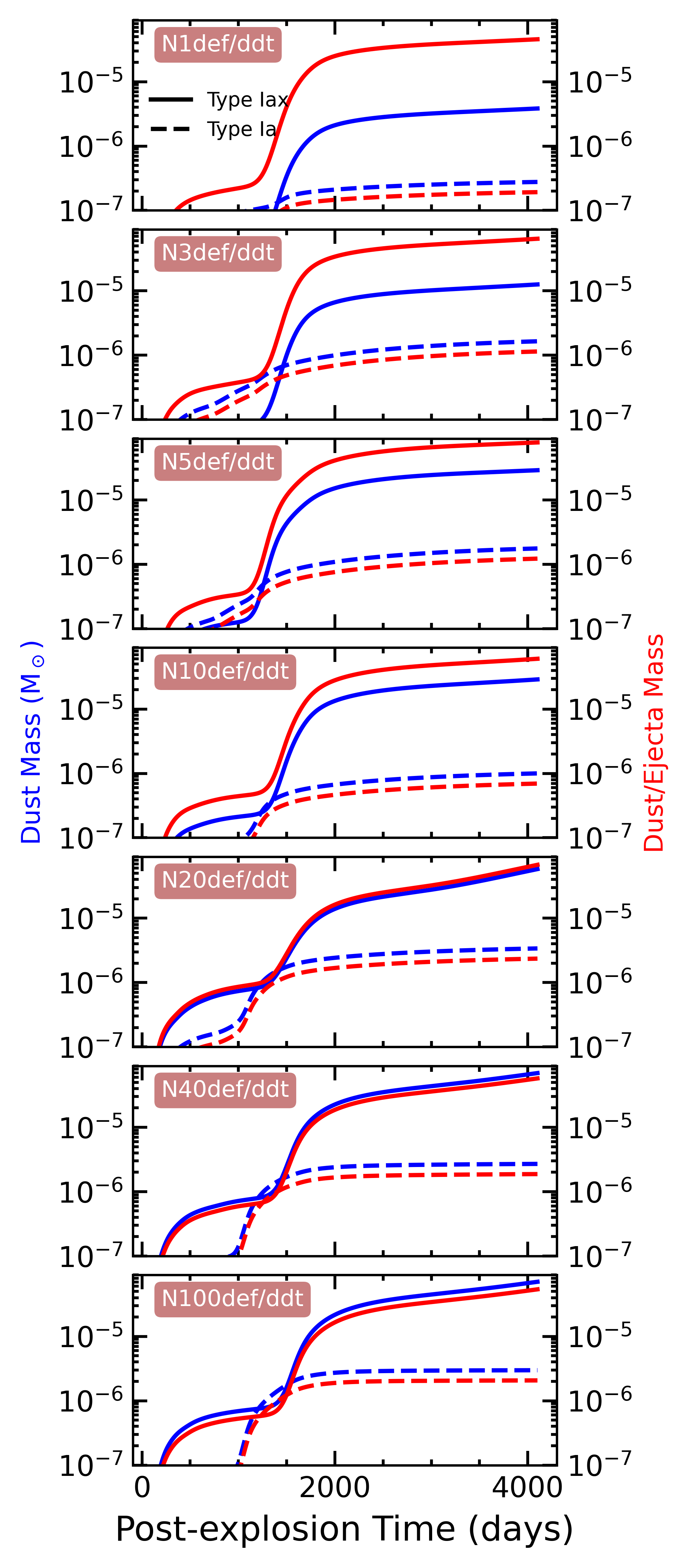}
\caption{Comparison of dust production efficiency of deflagration and DDT models of similar ignition channels. Solid lines are production outcomes from deflagration (Type Iax) models, and similarly dashed lines are for DDT (Type Ia) models. The blue lines are for the dust masses produced, and the red lines are for the dust-to-gas mass ratio along post-explosion time. }
\label{fig: comparison}
\end{figure}
As discussed earlier in section \ref{Ejecta}, the detonation reduces the important oxidizing agent O but enhances Si abundance, in addition to higher energies leading to lower gas densities. These conditions do not favor dust production in Type Ia SNe. Figure \ref{fig: velocity} highlights the contribution from different velocity zones in dust production for \texttt{N3} model of both deflagration and DDT models. In Type Iax SNe, the higher abundances of IGEs and IMEs in the dense inner regions promote efficient dust production. In contrast, in Type Ia SNe, these elements are pushed outward into lower-density regions, which is less favorable for dust formation. This effect is further compounded by the CE reactions, since Type Ia produces larger masses of \Ni.
\begin{longtable}{ccc|ccc}
\caption{Quantitative comparison of dust mass production in Type Iax SNe (deflagration) and Type Ia SNe (DDT) models at 4000 days. The dust masses are in \Ms\ . }
\label{tab: compare}\\
\hline
\multicolumn{3}{c|}{Type Iax}              & \multicolumn{3}{c}{Type Ia}               \\ \hline
\endfirsthead
\endhead
\hline
\endfoot
\endlastfoot
Def. Model & Dust Mass     & Dust/Gas Mass & DDT Model & Dust Mass     & Dust/Gas Mass \\ \hline
\texttt{N1def}      & \SI{3.7e-6}{} & \SI{4.4e-5}{} & \texttt{N1ddt}     & \SI{2.7e-7}{} & \SI{1.9e-7}{} \\
\texttt{N3def}      & \SI{1.2e-5}{} & \SI{6.2e-5}{} & \texttt{N3ddt}     & \SI{1.6e-6}{} & \SI{1.2e-6}{} \\
\texttt{N5def}      & \SI{2.8e-5}{} & \SI{7.5e-5}{} & \texttt{N5ddt}     & \SI{1.7e-6}{} & \SI{1.2e-6}{} \\
\texttt{N10def}     & \SI{2.8e-5}{} & \SI{5.8e-5}{} & \texttt{N10ddt}    & \SI{9.9e-7}{} & \SI{7.1e-7}{} \\
\texttt{N20def}     & \SI{5.4e-5}{} & \SI{6.2e-5}{} & \texttt{N20ddt}    & \SI{3.3e-6}{} & \SI{2.4e-6}{} \\
\texttt{N40def}     & \SI{6.6e-5}{} & \SI{5.5e-5}{} & \texttt{N40ddt}    & \SI{2.7e-6}{} & \SI{1.9e-6}{} \\
\texttt{N100def}    & \SI{6.7e-5}{} & \SI{5.1e-5}{} & \texttt{N100ddt}   & \SI{2.9e-6}{} & \SI{2.1e-6}{} \\ \hline
\end{longtable}
\texttt{N1ddt} model yields dust masses of around \SI{2.7e-7}{M_{\odot}} at 4000 days. As we go further for higher configurations (\texttt{N40ddt, N100ddt}) models, the initial deflagration waves become stronger and give us more mixed layers and less \Ni~ content, hence the dust mass increases. Due to low temperature, the time of dust formation is earlier for Type Iax, but even the \texttt{N100ddt} model with high abundances and low \Ni~ mass gives a dust mass of around \SI{2.9e-6}{M_{\odot}} at 4000 days, while for other models it ranges from \SI{2.7e-07}{M_{\odot}} to \SI{2.7e-06}{M_{\odot}} which is far less than the deflagration counterpart. Figure \ref{fig: ddtmodel} gives us the dust masses produced in all DDT models. Initial models are rich in \ce{SiC} since \ce{Si} abundance is higher, and as we go further, \ce{FeSiO3} becomes a significant contributor.

Figure \ref{fig: comparison} highlights the difference between the dust production efficiency of Type Ia and Type Iax SNe. The dust-to-gas mass ratio for Type Iax is about one or two orders of magnitude larger than the same model of Type Ia, at 4000 days as shown in table \ref{tab: compare}. The chemical mixing and incomplete burning due to the pure deflagration are the key factors that make Type Iax SNe potential sources of stellar dust, with a unique abundance of Fe-rich silicates. On the other hand, the models predict a negligible mass of dust to form in standard Type Ia SNe. 

\section{Implications}\label{discussion}
\cite{Dwek_2016,Psaradaki_2023,Pinto_2013} found that most of the iron in ISM is depleted from gas, and almost $90\%$ of it is locked in dust. As mentioned earlier, Type II  SNe are confirmed dust producers in galaxies, but they form too little Fe and cannot account for iron-rich dust. However, Type Ia SN are rich in Fe, but there is no observation yet that can confirm dust production in their ejecta. In this case, Type Iax SNe can provide a pathway for iron-rich dust production in SN environments.  

Our results highlight the possibility of dust production in Type Iax. Even though the dust masses are only around \SI{5.0e-5}{M_{\odot}}, the unique chemical composition in thermonuclear SNe makes this special SN subclass interesting to study the dust budget. Our results are highly sensitive to ejecta mass, the relative abundances of the elements, \Ni \  content, alongside temperatures and density of the environment. As mentioned in Section \ref{sec_SN2012Z}, models might underestimate the ejecta mass, and a model scaled to observed parameters gives us higher dust masses.

Our results suggest that silicon carbide dust should form in Type Iax SNe at late times, post 2000 days. In the pre-solar grains in meteorites, SiC is one of the most commonly found dust species \citep{fok_2024, hoppe_2024, hoppe_2019, hoppe_2018, hoppe_2010, amari_2014, zin07, nit08}, attributed to origins in AGB stars as well as core-collapse SNe based on their isotopic ratios. Interestingly, SiC-X grains found in meteorites are reported to have formed in the ejecta at least 2000 days post-explosion in core-collapse SNe \citep{liu_2018, ott_2019}. However, in thermonuclear SNe, the abundances of $^{13}$C and $^{17}$O are very low; those isotopic ratios do not match the reported ratios in presolar grains. 

\begin{figure}[t]
    \centering   
    \includegraphics[]{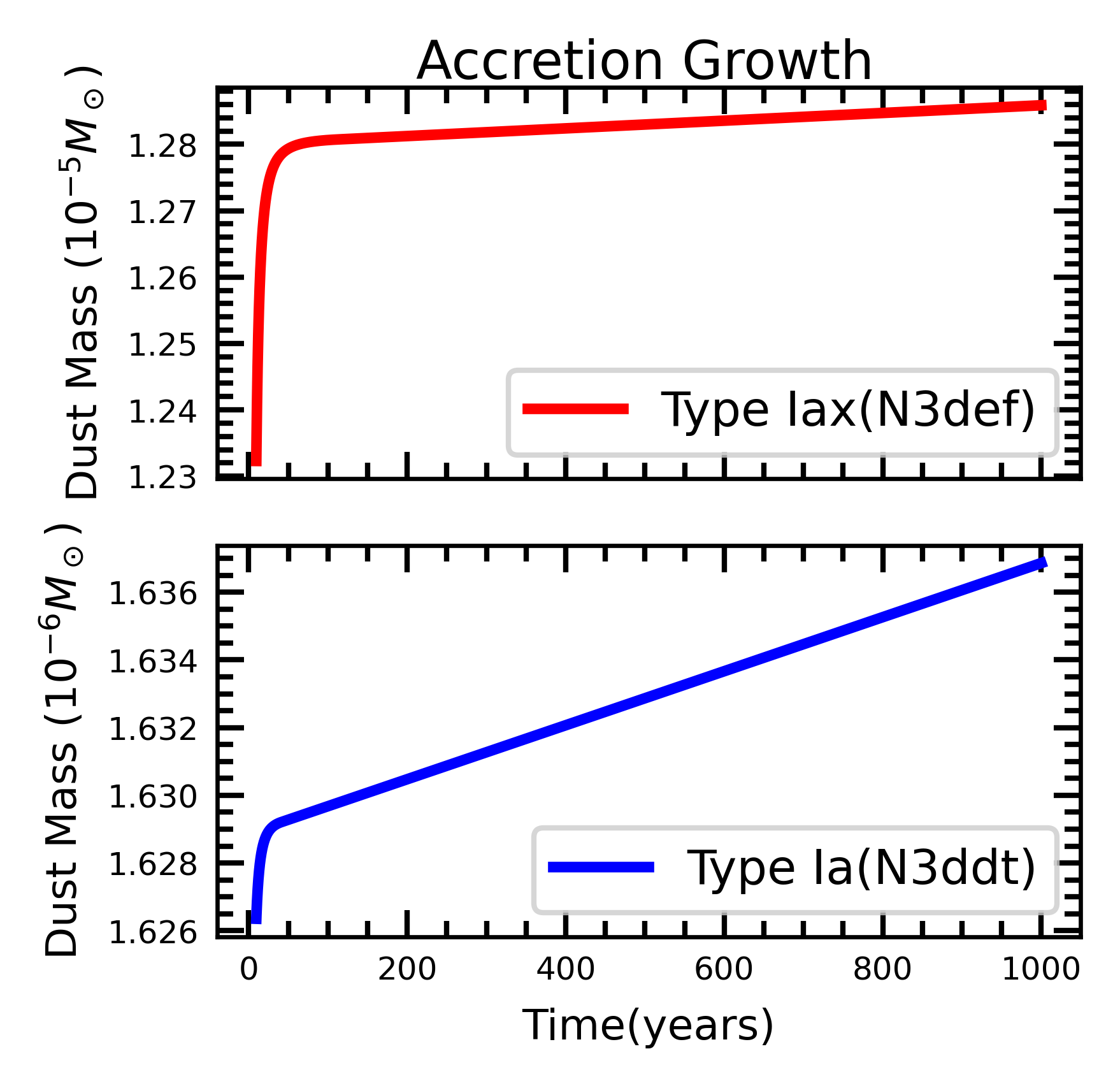}
\caption{ The dust mass growth in Type Iax (\texttt{N3def}) and Type Ia (\texttt{N3ddt}) ejecta due to accretion on surface of a $0.01\mu m$ \ce{FeSiO3} grain starting from about 7 years until 1000 years. }
\label{fig: acc}
\end{figure}

The model will improve if we have better observational constraints on ejecta mass, dust mass, and gas temperatures. Late-time follow-up observations around 5--6 years after explosions to look for possible IR emission from dust is required to verify this hypothesis. Since only a small part of the ejecta is involved in dust production, after 4000 days, many elements are still present in gas. They may accrete onto the surfaces of already produced dust grains. We have used a standard prescription of grain growth via accretion \citep{dwe11, sar2019}, taking $\mu$m grains of \ce{FeSiO3} and a sticking coefficient of 0.5 as initial conditions, for both cases of Type Iax (\texttt{N3def}) and Type Ia (\texttt{N3ddt}) SN ejecta.  The results are shown in Figure \ref{fig: acc} for possible grain growth up to 1000 years post-explosion. However, within that period, we did not find any significant change in dust masses due to accretion of metals on the surface. 

Spatially resolved galactic SN remnants, such as Pa~30, provide the opportunity to trace the location and composition of dust at late times. Remnant Pa~30 was reported to originate from a subluminous thermonuclear explosion, likely to be a Type Iax SN \citep{cunningham_2024}. It is bright in the far-IR wavelengths, revealing the presence of cold dust \citep{lykou_2023}. However, the IRAS 12 \mic\ resulted in non-detection. The high-resolution spectrum expected from the ongoing \texttt{JWST (GO 9111)} program will enable us to understand the abundance distribution in the nebula, the location of warm dust from its mid-IR emission, and decode whether there is significant dust formation in the ejecta of such SNe.  

\section{Summary and Conclusion}\label{summary}

Our findings suggest that the low-luminosity subclass of Type Ia -- the Type Iax SNe are efficient dust producers. We compare dust production in Type Ia and Type Iax SNe, showing the clear advantage of Type Iax SNe in producing dust in the ejecta. The main highlights for these results are

\begin{itemize}

    \item Favorable Conditions: The less energetic, pure deflagration explosion in Type Iax SNe leads to several conditions that are ideal for dust formation. The ejecta from Type Iax SNe are slower (2,000-7,000 $km~s^{-1}$) compared to Type Ia ejecta (10,000–20,000 $km~s^{-1}$). This slower expansion allows the ejecta to maintain a higher density for a longer period. The higher density promotes more frequent collisions between atoms and molecules, which enhances the rate of dust nucleation.
    \item \Ni~ Content: Type Iax SNe produce significantly less radioactive \Ni~ compared to Type Ia SNe. The decay of \Ni~ releases high-energy gamma rays, which can destroy newly formed dust and precursor molecules. The lower \Ni~ mass in Type Iax SNe allows the ejecta to cool more quickly and reduces this destruction.
    \item Precursor Materials: The incomplete burning in a pure deflagration leaves behind large amounts of unburned carbon (C) and oxygen (O), in addition to IMEs and IGEs. These are the crucial ingredients for forming precursor molecules like CO and SiO, which then lead to the formation of iron-rich dust species such as silicates.
    \item Dust Masses: At 4,000 days after the explosion, the dust mass produced by Type Iax SNe models ranged from \SI{3.7e-6}{}to \SI{6.7e-5}{M_{\odot}}. In contrast, Type Ia SNe models produced much less dust, ranging from \SI{2.7e-7}{}  to \SI{2.9e-6}{M_{\odot}}. The dust-to-gas mass ratio, a measure of production efficiency, was also much higher for Type Iax SNe, at about 4 to \SI{8e-5}{}, while for Type Ia SNe, it was only about 0.2 to \SI{2.4e-6}{}. Model scaled to reported SN~2012Z ejecta mass producing dust mass of \SI{1.5e-4}{M_{\odot}}  and dust-to-gas mass ratio of \SI{1.4e-4}{} much efficient than the \texttt{N100ddt}, model consistent with Type Ia ejecta.
    \item State-of-the-art model: Our model accounts for a comprehensive non-equilibrium chemical evolution of Type Ia~SNe. The small to negligible dust masses we predict are in agreement with non-detections of dust emission in a typical Type Ia SN ejecta \citep{gom12b}. We propose that due to the use of classical nucleation theory, the previous model of Type Ia SNe by \cite{Nozawa_2011} largely overestimates the dust masses. 
    \item Chemical network: The chemical pathway that leads to the synthesis of Fe-rich silicates in SNe is developed in this study, and can be used for any other stellar or SN environments. This result in a larger context can shed some light on how iron is integrated in dust grains, and so is depleted from the ISM \citep{Dwek_2016}. 
    
\end{itemize}

\section{Acknowledgments}
The authors gratefully acknowledge the support of the Department of Science and Technology (DST), and the Council of Scientific \& Industrial Research (CSIR),  Government of India. We also extend our sincere thanks to Dr.~Barnabas Barna, Dr.~Mridweeka Singh, Prof.~Saurabh Jha, Prof.~D.~K.~Sahu, and Mr.~Hrishav Das for very insightful discussions.

\bibliographystyle{aasjournalv7}
\bibliography{Bibliography,Bibliography_sarangi}

@article{ jack_2011,
	author = {{Jack, D.} and {Hauschildt, P. H.} and {Baron, E.}},
	title = {Theoretical light curves of type Ia supernovae},
	DOI= "10.1051/0004-6361/201014778",
	url= "https://doi.org/10.1051/0004-6361/201014778",
	journal = {A\&A},
	year = 2011,
	volume = 528,
	pages = "A141",
	month = "",
}

@article{ Sarangi_2022,
	author = {{Sarangi, Arkaprabha}},
	title = {Formation, distribution, and IR emission of dust in the clumpy ejecta of Type II-P core-collapse supernovae, in isotropic and anisotropic scenarios},
	DOI= "10.1051/0004-6361/202244391",
	url= "https://doi.org/10.1051/0004-6361/202244391",
	journal = {A\&A},
	year = 2022,
	volume = 668,
	pages = "A57",
}

@article{Barna_2018,
   title={Type Iax supernovae as a few-parameter family},
   volume={480},
   ISSN={1365-2966},
   url={http://dx.doi.org/10.1093/mnras/sty2065},
   DOI={10.1093/mnras/sty2065},
   number={3},
   journal={Monthly Notices of the Royal Astronomical Society},
   publisher={Oxford University Press (OUP)},
   author={Barna, Barnabás and Szalai, Tamás and Kerzendorf, Wolfgang E and Kromer, Markus and Sim, Stuart A and Magee, Mark R and Leibundgut, Bruno},
   year={2018},
   month=aug, pages={3609–3627} }

@ARTICLE{Singh_2018,
       author = {{Singh}, Mridweeka and {Misra}, Kuntal and {Sahu}, D.~K. and {Dastidar}, Raya and {Gangopadhyay}, Anjasha and {Bose}, Subhash and {Srivastav}, Shubham and {Anupama}, G.~C. and {Chakradhari}, N.~K. and {Kumar}, Brajesh and {Kumar}, Brijesh and {Pandey}, S.~B.},
        title = "{Exploring the optical behaviour of a Type Iax supernova SN 2014dt}",
      journal = {\mnras},
         year = 2018,
        month = feb,
       volume = {474},
       number = {2},
        pages = {2551-2563},
          doi = {10.1093/mnras/stx2916},
archivePrefix = {arXiv},
       eprint = {1711.00292},
 primaryClass = {astro-ph.HE},
       adsurl = {https://ui.adsabs.harvard.edu/abs/2018MNRAS.474.2551S}
}

@ARTICLE{Fink_2014,
       author = {{Fink}, Michael and {Kromer}, Markus and {Seitenzahl}, Ivo R. and {Ciaraldi-Schoolmann}, Franco and {R{\"o}pke}, Friedrich K. and {Sim}, Stuart A. and {Pakmor}, R{\"u}diger and {Ruiter}, Ashley J. and {Hillebrandt}, Wolfgang},
        title = "{Three-dimensional pure deflagration models with nucleosynthesis and synthetic observables for Type Ia supernovae}",
      journal = {\mnras},
         year = 2014,
        month = feb,
       volume = {438},
       number = {2},
        pages = {1762-1783},
          doi = {10.1093/mnras/stt2315},
archivePrefix = {arXiv},
       eprint = {1308.3257},
 primaryClass = {astro-ph.SR},
       adsurl = {https://ui.adsabs.harvard.edu/abs/2014MNRAS.438.1762F}
}

@ARTICLE{Nozawa_2011,
       author = {{Nozawa}, Takaya and {Maeda}, Keiichi and {Kozasa}, Takashi and {Tanaka}, Masaomi and {Nomoto}, Ken'ichi and {Umeda}, Hideyuki},
        title = "{Formation of Dust in the Ejecta of Type Ia Supernovae}",
      journal = {\apj},
         year = 2011,
        month = jul,
       volume = {736},
       number = {1},
          eid = {45},
        pages = {45},
          doi = {10.1088/0004-637X/736/1/45},
archivePrefix = {arXiv},
       eprint = {1105.0973},
 primaryClass = {astro-ph.SR},
       adsurl = {https://ui.adsabs.harvard.edu/abs/2011ApJ...736...45N}
}

@ARTICLE{Seithenzahl_2013,
       author = {{Seitenzahl}, Ivo R. and {Ciaraldi-Schoolmann}, Franco and {R{\"o}pke}, Friedrich K. and {Fink}, Michael and {Hillebrandt}, Wolfgang and {Kromer}, Markus and {Pakmor}, R{\"u}diger and {Ruiter}, Ashley J. and {Sim}, Stuart A. and {Taubenberger}, Stefan},
        title = "{Three-dimensional delayed-detonation models with nucleosynthesis for Type Ia supernovae}",
      journal = {\mnras},
         year = 2013,
        month = feb,
       volume = {429},
       number = {2},
        pages = {1156-1172},
          doi = {10.1093/mnras/sts402},
archivePrefix = {arXiv},
       eprint = {1211.3015},
 primaryClass = {astro-ph.SR},
       adsurl = {https://ui.adsabs.harvard.edu/abs/2013MNRAS.429.1156S}
}

@article{cherchneff2010,
	adsurl = {http://adsabs.harvard.edu/abs/2010ApJ...713....1C},
	archiveprefix = {arXiv},
	author = {{Cherchneff}, I. and {Dwek}, E.},
	doi = {10.1088/0004-637X/713/1/1},
	eprint = {1002.3060},
	journal = {The Astrophysical Journal},
	month = apr,
	pages = {1-24},
	primaryclass = {astro-ph.SR},
	title = {{The Chemistry of Population III Supernova Ejecta. II. The Nucleation of Molecular Clusters as a Diagnostic for Dust in the Early Universe}},
	volume = 713,
	year = 2010}

@article{sar13,
	adsurl = {http://adsabs.harvard.edu/abs/2013ApJ...776..107S},
	archiveprefix = {arXiv},
	author = {{Sarangi}, A. and {Cherchneff}, I.},
	doi = {10.1088/0004-637X/776/2/107},
	eid = {107},
	eprint = {1309.5887},
	journal = {The Astrophysical Journal},
	month = oct,
	pages = {107},
	primaryclass = {astro-ph.SR},
	title = {{The Chemically Controlled Synthesis of Dust in Type II-P Supernovae}},
	volume = 776,
	year = 2013}

@inbook{Jha_2017,
   title={Type Iax Supernovae},
   ISBN={9783319218465},
   url={http://dx.doi.org/10.1007/978-3-319-21846-5_42},
   DOI={10.1007/978-3-319-21846-5_42},
   booktitle={Handbook of Supernovae},
   publisher={Springer International Publishing},
   author={Jha, Saurabh W.},
   year={2017},
   pages={375–401} }

@ARTICLE{Liu2015,
       author = {{Liu}, Zheng-Wei and {Stancliffe}, Richard J. and {Abate}, C. and {Wang}, B.},
        title = "{Pre-explosion Companion Stars in Type Iax Supernovae}",
      journal = {\apj},
         year = 2015,
        month = aug,
       volume = {808},
       number = {2},
          eid = {138},
        pages = {138},
          doi = {10.1088/0004-637X/808/2/138},
archivePrefix = {arXiv},
       eprint = {1506.04903},
 primaryClass = {astro-ph.SR},
       adsurl = {https://ui.adsabs.harvard.edu/abs/2015ApJ...808..138L}
}

@ARTICLE{Zang2019,
       author = {{Zhang}, Michael and {Fuller}, Jim and {Schwab}, Josiah and {Foley}, Ryan J.},
        title = "{The Long-term Evolution and Appearance of Type Iax Postgenitor Stars}",
      journal = {\apj},
         year = 2019,
        month = feb,
       volume = {872},
       number = {1},
          eid = {29},
        pages = {29},
          doi = {10.3847/1538-4357/aafb34},
archivePrefix = {arXiv},
       eprint = {1812.08793},
 primaryClass = {astro-ph.SR},
       adsurl = {https://ui.adsabs.harvard.edu/abs/2019ApJ...872...29Z}
}

@ARTICLE{CN2023,
       author = {{Camacho-Neves}, Yssavo and {Jha}, Saurabh W. and {Barna}, Barnabas and {Dai}, Mi and {Filippenko}, Alexei V. and {Foley}, Ryan J. and {Hosseinzadeh}, Griffin and {Howell}, D. Andrew and {Johansson}, Joel and {Kelly}, Patrick L. and {Kerzendorf}, Wolfgang E. and {Kwok}, Lindsey A. and {Larison}, Conor and {Magee}, Mark R. and {McCully}, Curtis and {O'Brien}, John T. and {Pan}, Yen-Chen and {Pandya}, Viraj and {Singhal}, Jaladh and {Stahl}, Benjamin E. and {Szalai}, Tam{\'a}s and {Wieber}, Meredith and {Williamson}, Marc},
        title = "{Over 500 Days in the Life of the Photosphere of the Type Iax Supernova SN 2014dt}",
      journal = {\apj},
         year = 2023,
        month = jul,
       volume = {951},
       number = {1},
          eid = {67},
        pages = {67},
          doi = {10.3847/1538-4357/acd558},
archivePrefix = {arXiv},
       eprint = {2302.03105},
 primaryClass = {astro-ph.HE},
       adsurl = {https://ui.adsabs.harvard.edu/abs/2023ApJ...951...67C}
}

@ARTICLE{Barna2021,
       author = {{Barna}, Barnab{\'a}s and {Szalai}, Tam{\'a}s and {Jha}, Saurabh W. and {Camacho-Neves}, Yssavo and {Kwok}, Lindsey and {Foley}, Ryan J. and {Kilpatrick}, Charles D. and {Coulter}, David A. and {Dimitriadis}, Georgios and {Rest}, Armin and {Rojas-Bravo}, C{\'e}sar and {Siebert}, Matthew R. and {Brown}, Peter J. and {Burke}, Jamison and {Padilla Gonzalez}, Estefania and {Hiramatsu}, Daichi and {Howell}, D. Andrew and {McCully}, Curtis and {Pellegrino}, Craig and {Dobson}, Matthew and {Smartt}, Stephen J. and {Swift}, Jonathan J. and {Stacey}, Holland and {Rahman}, Mohammed and {Sand}, David J. and {Andrews}, Jennifer and {Wyatt}, Samuel and {Hsiao}, Eric Y. and {Anderson}, Joseph P. and {Chen}, Ting-Wan and {Della Valle}, Massimo and {Galbany}, Llu{\'\i}s and {Gromadzki}, Mariusz and {Inserra}, Cosimo and {Lyman}, Joe and {Magee}, Mark and {Maguire}, Kate and {M{\"u}ller-Bravo}, Tom{\'a}s E. and {Nicholl}, Matt and {Srivastav}, Shubham and {Williams}, Steven C.},
        title = "{SN 2019muj - a well-observed Type Iax supernova that bridges the luminosity gap of the class}",
      journal = {\mnras},
         year = 2021,
        month = feb,
       volume = {501},
       number = {1},
        pages = {1078-1099},
          doi = {10.1093/mnras/staa3543},
archivePrefix = {arXiv},
       eprint = {2011.03068},
 primaryClass = {astro-ph.HE},
       adsurl = {https://ui.adsabs.harvard.edu/abs/2021MNRAS.501.1078B}
}

@ARTICLE{MSingh2024,
       author = {{Singh}, Mridweeka and {Sahu}, Devendra K. and {Barna}, Barnab{\'a}s and {Gangopadhyay}, Anjasha and {Dastidar}, Raya and {Teja}, Rishabh Singh and {Misra}, Kuntal and {Howell}, D. Andrew and {Wang}, Xiaofeng and {Mo}, Jun and {Yan}, Shengyu and {Hiramatsu}, Daichi and {Pellegrino}, Craig and {Anupama}, G.~C. and {Joshi}, Arti and {Bostroem}, K. Azalee and {Burke}, Jamison and {McCully}, Curtis and {Subramanian V}, Rama and {Li}, Gaici and {Xi}, Gaobo and {Li}, Xin and {Li}, Zhitong and {Srivastav}, Shubham and {Im}, Hyobin and {Dutta}, Anirban},
        title = "{SN 2020udy: A New Piece of the Homogeneous Bright Group in the Diverse Iax Subclass}",
      journal = {\apj},
         year = 2024,
        month = apr,
       volume = {965},
       number = {1},
          eid = {73},
        pages = {73},
          doi = {10.3847/1538-4357/ad2618},
archivePrefix = {arXiv},
       eprint = {2401.07107},
 primaryClass = {astro-ph.HE},
       adsurl = {https://ui.adsabs.harvard.edu/abs/2024ApJ...965...73S}
}

@ARTICLE{Cherchneff2009,
       author = {{Cherchneff}, Isabelle and {Dwek}, Eli},
        title = "{The Chemistry of Population III Supernova Ejecta. I. Formation of Molecules in the Early Universe}",
      journal = {\apj},
         year = 2009,
        month = sep,
       volume = {703},
       number = {1},
        pages = {642-661},
          doi = {10.1088/0004-637X/703/1/642},
archivePrefix = {arXiv},
       eprint = {0907.3621},
 primaryClass = {astro-ph.SR},
       adsurl = {https://ui.adsabs.harvard.edu/abs/2009ApJ...703..642C}
}

@INCOLLECTION{Bersten_2017,
       author = {{Bersten}, Melina C. and {Mazzali}, Paolo A.},
        title = "{Light Curves of Type I Supernovae}",
    booktitle = {Handbook of Supernovae},
         year = 2017,
       editor = {{Alsabti}, Athem W. and {Murdin}, Paul},
        pages = {723},
          doi = {10.1007/978-3-319-21846-5_25},
       adsurl = {https://ui.adsabs.harvard.edu/abs/2017hsn..book..723B}
}

@article{sarangi2018book,
	adsurl = {http://adsabs.harvard.edu/abs/2018SSRv..214...63S},
	author = {{Sarangi}, A. and {Matsuura}, M. and {Micelotta}, E.~R.},
	doi = {10.1007/s11214-018-0492-7},
	eid = {63},
	journal = {\ssr},
	month = apr,
	pages = {63},
	title = {{Dust in Supernovae and Supernova Remnants I: Formation Scenarios}},
	volume = 214,
	year = 2018}

@ARTICLE{Foley2013,
       author = {{Foley}, Ryan J. and {Challis}, P.~J. and {Chornock}, R. and {Ganeshalingam}, M. and {Li}, W. and {Marion}, G.~H. and {Morrell}, N.~I. and {Pignata}, G. and {Stritzinger}, M.~D. and {Silverman}, J.~M. and {Wang}, X. and {Anderson}, J.~P. and {Filippenko}, A.~V. and {Freedman}, W.~L. and {Hamuy}, M. and {Jha}, S.~W. and {Kirshner}, R.~P. and {McCully}, C. and {Persson}, S.~E. and {Phillips}, M.~M. and {Reichart}, D.~E. and {Soderberg}, A.~M.},
        title = "{Type Iax Supernovae: A New Class of Stellar Explosion}",
      journal = {\apj},
         year = 2013,
        month = apr,
       volume = {767},
       number = {1},
          eid = {57},
        pages = {57},
          doi = {10.1088/0004-637X/767/1/57},
archivePrefix = {arXiv},
       eprint = {1212.2209},
 primaryClass = {astro-ph.SR},
       adsurl = {https://ui.adsabs.harvard.edu/abs/2013ApJ...767...57F}
}

@ARTICLE{Dwek_2016,
       author = {{Dwek}, Eli},
        title = "{Iron: A Key Element for Understanding the Origin and Evolution of Interstellar Dust}",
      journal = {\apj},
         year = 2016,
        month = jul,
       volume = {825},
       number = {2},
          eid = {136},
        pages = {136},
          doi = {10.3847/0004-637X/825/2/136},
archivePrefix = {arXiv},
       eprint = {1605.01957},
 primaryClass = {astro-ph.GA},
       adsurl = {https://ui.adsabs.harvard.edu/abs/2016ApJ...825..136D}
}

@article{sar15,
	adsurl = {http://adsabs.harvard.edu/abs/2015A%26A...575A..95S},
	archiveprefix = {arXiv},
	author = {{Sarangi}, A. and {Cherchneff}, I.},
	doi = {10.1051/0004-6361/201424969},
	eid = {A95},
	eprint = {1412.5522},
	journal = {\aap},
	month = mar,
	pages = {A95},
	primaryclass = {astro-ph.SR},
	title = {{Condensation of dust in the ejecta of Type II-P supernovae}},
	volume = 575,
	year = 2015}

@article{gom12b,
	adsurl = {http://adsabs.harvard.edu/abs/2012MNRAS.420.3557G},
	archiveprefix = {arXiv},
	author = {{Gomez}, H.~L. and {Clark}, C.~J.~R. and {Nozawa}, T. and {Krause}, O. and {Gomez}, E.~L. and {Matsuura}, M. and {Barlow}, M.~J. and {Besel}, M.-A. and {Dunne}, L. and {Gear}, W.~K. and {Hargrave}, P. and {Henning}, T. and {Ivison}, R.~J. and {Sibthorpe}, B. and {Swinyard}, B.~M. and {Wesson}, R.},
	doi = {10.1111/j.1365-2966.2011.20272.x},
	eprint = {1111.6627},
	journal = {\mnras},
	month = mar,
	pages = {3557-3573},
	primaryclass = {astro-ph.GA},
	title = {{Dust in historical Galactic Type Ia supernova remnants with Herschel}},
	volume = 420,
	year = 2012}

@ARTICLE{sar2019,
       author = {{Sarangi}, Arkaprabha and {Dwek}, Eli and {Kazanas}, Demos},
        title = "{Dust Formation in AGN Winds}",
      journal = {\apj},
         year = 2019,
        month = nov,
       volume = {885},
       number = {2},
          eid = {126},
        pages = {126},
          doi = {10.3847/1538-4357/ab46a9},
archivePrefix = {arXiv},
       eprint = {1909.10426},
 primaryClass = {astro-ph.GA},
       adsurl = {https://ui.adsabs.harvard.edu/abs/2019ApJ...885..126S}
}

@ARTICLE{Mag2023,
       author = {{Maguire}, Kate and {Magee}, Mark R. and {Leloudas}, Giorgos and {Miller}, Adam A. and {Dimitriadis}, Georgios and {Pursiainen}, Miika and {Bulla}, Mattia and {De}, Kishalay and {Gal-Yam}, Avishay and {Perley}, Daniel A. and {Fremling}, Christoffer and {Karambelkar}, Viraj R. and {Nordin}, Jakob and {Reusch}, Simeon and {Schulze}, Steve and {Sollerman}, Jesper and {Terreran}, Giacomo and {Yang(杨轶)}, Yi and {Bellm}, Eric C. and {Groom}, Steven L. and {Kasliwal}, Mansi M. and {Kulkarni}, Shrinivas R. and {Lacroix}, Leander and {Masci}, Frank J. and {Purdum}, Josiah N. and {Sharma}, Yashvi and {Smith}, Roger},
        title = "{SN 2020udy: an SN Iax with strict limits on interaction consistent with a helium-star companion}",
      journal = {\mnras},
         year = 2023,
        month = oct,
       volume = {525},
       number = {1},
        pages = {1210-1228},
          doi = {10.1093/mnras/stad2316},
archivePrefix = {arXiv},
       eprint = {2304.12361},
 primaryClass = {astro-ph.HE},
       adsurl = {https://ui.adsabs.harvard.edu/abs/2023MNRAS.525.1210M}
}

@ARTICLE{MSingh2022,
       author = {{Singh}, Mridweeka and {Misra}, Kuntal and {Sahu}, Devendra K. and {Ailawadhi}, Bhavya and {Dutta}, Anirban and {Howell}, D. Andrew and {Anupama}, G.~C. and {Bostroem}, K. Azalee and {Burke}, Jamison and {Dastidar}, Raya and {Gangopadhyay}, Anjasha and {Hiramatsu}, Daichi and {Im}, Hyobin and {McCully}, Curtis and {Pellegrino}, Craig and {Srivastav}, Shubham and {Teja}, Rishabh Singh},
        title = "{Optical studies of a bright Type Iax supernova SN 2020rea}",
      journal = {\mnras},
         year = 2022,
        month = dec,
       volume = {517},
       number = {4},
        pages = {5617-5626},
          doi = {10.1093/mnras/stac3059},
archivePrefix = {arXiv},
       eprint = {2210.11752},
 primaryClass = {astro-ph.HE},
       adsurl = {https://ui.adsabs.harvard.edu/abs/2022MNRAS.517.5617S}
}

@article{Psaradaki_2023,
   title={Oxygen and iron in interstellar dust: An X-ray investigation},
   volume={670},
   ISSN={1432-0746},
   url={http://dx.doi.org/10.1051/0004-6361/202244110},
   DOI={10.1051/0004-6361/202244110},
   journal={Astronomy \& Astrophysics},
   publisher={EDP Sciences},
   author={Psaradaki, I. and Costantini, E. and Rogantini, D. and Mehdipour, M. and Corrales, L. and Zeegers, S. T. and de Groot, F. and den Herder, J. W. A. and Mutschke, H. and Trasobares, S. and de Vries, C. P. and Waters, L. B. F. M.},
   year={2023},
   month=feb, pages={A30} }

@article{Pinto_2013,
    author = "Pinto, C. and Kaastra, J. S. and Costantini, E. and de Vries, C.",
    title = "{ISM composition through X-ray spectroscopy of LMXBs}",
    eprint = "1301.1612",
    archivePrefix = "arXiv",
    primaryClass = "astro-ph.GA",
    doi = "10.1051/0004-6361/201220481",
    journal = "Astron. Astrophys.",
    volume = "551",
    pages = "A25",
    year = "2013"
}

@ARTICLE{zhou_2021,
       author = {{Zhou}, Ping and {Leung}, Shing-Chi and {Li}, Zhiyuan and {Nomoto}, Ken'ichi and {Vink}, Jacco and {Chen}, Yang},
        title = "{Chemical Abundances in Sgr A East: Evidence for a Type Iax Supernova Remnant}",
      journal = {\apj},
         year = 2021,
        month = feb,
       volume = {908},
       number = {1},
          eid = {31},
        pages = {31},
          doi = {10.3847/1538-4357/abbd45},
archivePrefix = {arXiv},
       eprint = {2006.15049},
 primaryClass = {astro-ph.HE},
       adsurl = {https://ui.adsabs.harvard.edu/abs/2021ApJ...908...31Z}
}

@article{Jha2019,
  author   = {Jha, Saurabh W. and Maguire, Kate and Sullivan, Mark},
  title    = {Observational properties of thermonuclear supernovae},
  journal  = {Nature Astronomy},
  year     = {2019},
  month    = {aug},
  volume   = {3},
  number   = {8},
  pages    = {706--716},
  [cite_start]abstract = {The explosive death of a star as a supernova is one of the most dramatic events in the Universe. [cite: 1] [cite_start]Supernovae have an outsized impact on many areas of astrophysics: they are major contributors to the chemical enrichment of the cosmos and significantly influence the formation of subsequent generations of stars and the evolution of galaxies. [cite: 2] [cite_start]Here we review the observational properties of thermonuclear supernovae—exploding white dwarf stars resulting from the stellar evolution of low-mass stars in close binary systems. [cite: 3] [cite_start]The best known objects in this class are type-Ia supernovae (SNe Ia), astrophysically important in their application as standardizable candles to measure cosmological distances and the primary source of iron group elements in the Universe. [cite: 4] [cite_start]Surprisingly, given their prominent role, SN Ia progenitor systems and explosion mechanisms are not fully understood; [cite: 5] [cite_start]the observations we describe here provide constraints on models, not always in consistent ways. [cite: 6] [cite_start]Recent advances in supernova discovery and follow-up have shown that the class of thermonuclear supernovae includes more than just SNe Ia, and we characterize that diversity in this review. [cite: 7]},
  issn     = {2397-3366},
  doi      = {10.1038/s41550-019-0858-0},
  url      = {https://doi.org/10.1038/s41550-019-0858-0}
}

@article{liu2023,
  title={Type Ia supernova explosions in binary systems: a review},
  author={Liu, Zheng-Wei and R{\"o}pke, Friedrich K and Han, Zhanwen},
  journal={Research in Astronomy and Astrophysics},
  volume={23},
  number={8},
  pages={082001},
  year={2023},
  publisher={IOP Publishing}
}

@ARTICLE{Li2003,
       author = {{Li}, Weidong and {Filippenko}, Alexei V. and {Chornock}, Ryan and {Berger}, Edo and {Berlind}, Perry and {Calkins}, Michael L. and {Challis}, Peter and {Fassnacht}, Chris and {Jha}, Saurabh and {Kirshner}, Robert P. and {Matheson}, Thomas and {Sargent}, Wallace L.~W. and {Simcoe}, Robert A. and {Smith}, Graeme H. and {Squires}, Gordon},
        title = "{SN 2002cx: The Most Peculiar Known Type Ia Supernova}",
      journal = {\pasp},
         year = 2003,
        month = apr,
       volume = {115},
       number = {806},
        pages = {453-473},
          doi = {10.1086/374200},
archivePrefix = {arXiv},
       eprint = {astro-ph/0301428},
 primaryClass = {astro-ph},
       adsurl = {https://ui.adsabs.harvard.edu/abs/2003PASP..115..453L}
}

@ARTICLE{Li2001,
       author = {{Li}, Weidong and {Filippenko}, Alexei V. and {Treffers}, Richard R. and {Riess}, Adam G. and {Hu}, Jingyao and {Qiu}, Yulei},
        title = "{A High Intrinsic Peculiarity Rate among Type IA Supernovae}",
      journal = {\apj},
         year = 2001,
        month = jan,
       volume = {546},
       number = {2},
        pages = {734-743},
          doi = {10.1086/318299},
archivePrefix = {arXiv},
       eprint = {astro-ph/0006292},
 primaryClass = {astro-ph},
       adsurl = {https://ui.adsabs.harvard.edu/abs/2001ApJ...546..734L}
}

@ARTICLE{Foley2014,
       author = {{Foley}, Ryan J. and {McCully}, Curtis and {Jha}, Saurabh W. and {Bildsten}, Lars and {Fong}, Wen-fai and {Narayan}, Gautham and {Rest}, Armin and {Stritzinger}, Maximilian D.},
        title = "{Possible Detection of the Stellar Donor or Remnant for the Type Iax Supernova 2008ha}",
      journal = {\apj},
         year = 2014,
        month = sep,
       volume = {792},
       number = {1},
          eid = {29},
        pages = {29},
          doi = {10.1088/0004-637X/792/1/29},
archivePrefix = {arXiv},
       eprint = {1408.1091},
 primaryClass = {astro-ph.HE},
       adsurl = {https://ui.adsabs.harvard.edu/abs/2014ApJ...792...29F}
}

@ARTICLE{Kawabata2018,
       author = {{Kawabata}, Miho and {Kawabata}, Koji S. and {Maeda}, Keiich and {Yamanaka}, Masayuki and {Nakaoka}, Tatsuya and {Takaki}, Katsutoshi and {Fukushima}, Daiki and {Kojiguchi}, Naoto and {Masumoto}, Kazunari and {Matsumoto}, Katsura and {Akitaya}, Hiroshi and {Itoh}, Ryosuke and {Kanda}, Yuka and {Moritani}, Yuki and {Takata}, Koji and {Uemura}, Makoto and {Ui}, Takahiro and {Yoshida}, Michitoshi and {Hattori}, Takashi and {Lee}, Chien-Hsiu and {Tominaga}, Nozomu and {Nomoto}, Ken'ichi},
        title = "{Extended optical/NIR observations of Type Iax supernova 2014dt: Possible signatures of a bound remnant}",
      journal = {\pasj},
         year = 2018,
        month = dec,
       volume = {70},
       number = {6},
          eid = {111},
        pages = {111},
          doi = {10.1093/pasj/psy116},
archivePrefix = {arXiv},
       eprint = {1810.00922},
 primaryClass = {astro-ph.HE},
       adsurl = {https://ui.adsabs.harvard.edu/abs/2018PASJ...70..111K}
}

@ARTICLE{Kawabata2021,
       author = {{Kawabata}, Miho and {Maeda}, Keiichi and {Yamanaka}, Masayuki and {Nakaoka}, Tatsuya and {Kawabata}, Koji S. and {Aoki}, Kentaro and {Anupama}, G.~C. and {Burgaz}, Umut and {Dutta}, Anirban and {Isogai}, Keisuke and {Kino}, Masaru and {Kojiguchi}, Naoto and {Kota}, Iida and {Kumar}, Brajesh and {Kuroda}, Daisuke and {Maehara}, Hiroyuki and {Matsubayashi}, Kazuya and {Morihana}, Kumiko and {Murata}, Katsuhiro L. and {Ohshima}, Tomohito and {Otsuka}, Masaaki and {Sahu}, Devendra K. and {Singh}, Avinash and {Sugitani}, Koji and {Takahashi}, Jun and {Takagi}, Kengo},
        title = "{Intermediate luminosity type Iax supernova 2019muj with narrow absorption lines: Long-lasting radiation associated with a possible bound remnant predicted by the weak deflagration model}",
      journal = {\pasj},
         year = 2021,
        month = oct,
       volume = {73},
       number = {5},
        pages = {1295-1314},
          doi = {10.1093/pasj/psab075},
archivePrefix = {arXiv},
       eprint = {2107.02822},
 primaryClass = {astro-ph.HE},
       adsurl = {https://ui.adsabs.harvard.edu/abs/2021PASJ...73.1295K}
}

@ARTICLE{Magee2016,
       author = {{Magee}, M.~R. and {Kotak}, R. and {Sim}, S.~A. and {Kromer}, M. and {Rabinowitz}, D. and {Smartt}, S.~J. and {Baltay}, C. and {Campbell}, H.~C. and {Chen}, T.-W. and {Fink}, M. and {Gal-Yam}, A. and {Galbany}, L. and {Hillebrandt}, W. and {Inserra}, C. and {Kankare}, E. and {Le Guillou}, L. and {Lyman}, J.~D. and {Maguire}, K. and {Pakmor}, R. and {R{\"o}pke}, F.~K. and {Ruiter}, A.~J. and {Seitenzahl}, I.~R. and {Sullivan}, M. and {Valenti}, S. and {Young}, D.~R.},
        title = "{The type Iax supernova, SN 2015H. A white dwarf deflagration candidate}",
      journal = {\aap},
         year = 2016,
        month = may,
       volume = {589},
          eid = {A89},
        pages = {A89},
          doi = {10.1051/0004-6361/201528036},
archivePrefix = {arXiv},
       eprint = {1603.04728},
 primaryClass = {astro-ph.HE},
       adsurl = {https://ui.adsabs.harvard.edu/abs/2016A&A...589A..89M}
}

@ARTICLE{Foley2016,
       author = {{Foley}, Ryan J. and {Jha}, Saurabh W. and {Pan}, Yen-Chen and {Zheng}, Wei Kang and {Bildsten}, Lars and {Filippenko}, Alexei V. and {Kasen}, Daniel},
        title = "{Late-time spectroscopy of Type Iax Supernovae}",
      journal = {\mnras},
         year = 2016,
        month = sep,
       volume = {461},
       number = {1},
        pages = {433-457},
          doi = {10.1093/mnras/stw1320},
archivePrefix = {arXiv},
       eprint = {1601.05955},
 primaryClass = {astro-ph.HE},
       adsurl = {https://ui.adsabs.harvard.edu/abs/2016MNRAS.461..433F}
}

@ARTICLE{Branch2004,
       author = {{Branch}, David and {Baron}, E. and {Thomas}, R.~C. and {Kasen}, D. and {Li}, Weidong and {Filippenko}, Alexei V.},
        title = "{Reading the Spectra of the Most Peculiar Type Ia Supernova 2002cx}",
      journal = {\pasp},
         year = 2004,
        month = oct,
       volume = {116},
       number = {824},
        pages = {903-908},
          doi = {10.1086/425081},
archivePrefix = {arXiv},
       eprint = {astro-ph/0408130},
 primaryClass = {astro-ph},
       adsurl = {https://ui.adsabs.harvard.edu/abs/2004PASP..116..903B}
}

@ARTICLE{Woosley_1986,
       author = {{Woosley}, S.~E. and {Taam}, R.~E. and {Weaver}, T.~A.},
        title = "{Models for Type I Supernova. I. Detonations in White Dwarfs}",
      journal = {\apj},
         year = 1986,
        month = feb,
       volume = {301},
        pages = {601},
          doi = {10.1086/163926},
       adsurl = {https://ui.adsabs.harvard.edu/abs/1986ApJ...301..601W}
}

@ARTICLE{Hoeflich_1995,
       author = {{Hoeflich}, P. and {Khokhlov}, A.~M. and {Wheeler}, J.~C.},
        title = "{Delayed Detonation Models for Normal and Subluminous Type IA Supernovae: Absolute Brightness, Light Curves, and Molecule Formation}",
      journal = {\apj},
         year = 1995,
        month = may,
       volume = {444},
        pages = {831},
          doi = {10.1086/175656},
       adsurl = {https://ui.adsabs.harvard.edu/abs/1995ApJ...444..831H}
}

@ARTICLE{Ropke_2007,
       author = {{R{\"o}pke}, F.~K. and {Niemeyer}, J.~C.},
        title = "{Delayed detonations in full-star models of type Ia supernova explosions}",
      journal = {\aap},
         year = 2007,
        month = mar,
       volume = {464},
       number = {2},
        pages = {683-686},
          doi = {10.1051/0004-6361:20066585},
archivePrefix = {arXiv},
       eprint = {astro-ph/0703378},
 primaryClass = {astro-ph},
       adsurl = {https://ui.adsabs.harvard.edu/abs/2007A&A...464..683R}
}

@ARTICLE{Ropke_2007b,
       author = {{R{\"o}pke}, F.~K. and {Woosley}, S.~E. and {Hillebrandt}, W.},
        title = "{Off-Center Ignition in Type Ia Supernovae. I. Initial Evolution and Implications for Delayed Detonation}",
      journal = {\apj},
         year = 2007,
        month = may,
       volume = {660},
       number = {2},
        pages = {1344-1356},
          doi = {10.1086/512769},
archivePrefix = {arXiv},
       eprint = {astro-ph/0609088},
 primaryClass = {astro-ph},
       adsurl = {https://ui.adsabs.harvard.edu/abs/2007ApJ...660.1344R}
}

@ARTICLE{Gamezo_2003,
       author = {{Gamezo}, Vadim N. and {Khokhlov}, Alexei M. and {Oran}, Elaine S. and {Chtchelkanova}, Almadena Y. and {Rosenberg}, Robert O.},
        title = "{Thermonuclear Supernovae: Simulations of the Deflagration Stage and Their Implications}",
      journal = {Science},
         year = 2003,
        month = jan,
       volume = {299},
       number = {5603},
        pages = {77-81},
          doi = {10.1126/science.299.5603.77},
archivePrefix = {arXiv},
       eprint = {astro-ph/0212054},
 primaryClass = {astro-ph},
       adsurl = {https://ui.adsabs.harvard.edu/abs/2003Sci...299...77G}
}

@ARTICLE{Garcia_2005,
       author = {{Garc{\'\i}a-Senz}, D. and {Bravo}, E.},
        title = "{Type Ia Supernova models arising from different distributions of igniting points}",
      journal = {\aap},
         year = 2005,
        month = feb,
       volume = {430},
        pages = {585-602},
          doi = {10.1051/0004-6361:20041628},
archivePrefix = {arXiv},
       eprint = {astro-ph/0409480},
 primaryClass = {astro-ph},
       adsurl = {https://ui.adsabs.harvard.edu/abs/2005A&A...430..585G}
}

@ARTICLE{Ropke_2006a,
       author = {{R{\"o}pke}, F.~K. and {Gieseler}, M. and {Reinecke}, M. and {Travaglio}, C. and {Hillebrandt}, W.},
        title = "{Type Ia supernova diversity in three-dimensional models}",
      journal = {\aap},
         year = 2006,
        month = jul,
       volume = {453},
       number = {1},
        pages = {203-217},
          doi = {10.1051/0004-6361:20053430},
archivePrefix = {arXiv},
       eprint = {astro-ph/0506107},
 primaryClass = {astro-ph},
       adsurl = {https://ui.adsabs.harvard.edu/abs/2006A&A...453..203R}
}

@ARTICLE{ Jordan_2012a,
       author = {{Jordan}, IV, George C. and {Perets}, Hagai B. and {Fisher}, Robert T. and {van Rossum}, Daniel R.},
        title = "{Failed-detonation Supernovae: Subluminous Low-velocity Ia Supernovae and their Kicked Remnant White Dwarfs with Iron-rich Cores}",
      journal = {\apjl},
         year = 2012,
        month = dec,
       volume = {761},
       number = {2},
          eid = {L23},
        pages = {L23},
          doi = {10.1088/2041-8205/761/2/L23},
archivePrefix = {arXiv},
       eprint = {1208.5069},
 primaryClass = {astro-ph.HE},
       adsurl = {https://ui.adsabs.harvard.edu/abs/2012ApJ...761L..23J}
}

@ARTICLE{Ma_2013,
       author = {{Ma}, H. and {Woosley}, S.~E. and {Malone}, C.~M. and {Almgren}, A. and {Bell}, J.},
        title = "{Carbon Deflagration in Type Ia Supernova. I. Centrally Ignited Models}",
      journal = {\apj},
         year = 2013,
        month = jul,
       volume = {771},
       number = {1},
          eid = {58},
        pages = {58},
          doi = {10.1088/0004-637X/771/1/58},
archivePrefix = {arXiv},
       eprint = {1305.2433},
 primaryClass = {astro-ph.HE},
       adsurl = {https://ui.adsabs.harvard.edu/abs/2013ApJ...771...58M}
}

@ARTICLE{Long_2014,
       author = {{Long}, Min and {Jordan}, IV, George C. and {van Rossum}, Daniel R. and {Diemer}, Benedikt and {Graziani}, Carlo and {Kessler}, Richard and {Meyer}, Bradley and {Rich}, Paul and {Lamb}, Don Q.},
        title = "{Three-dimensional Simulations of Pure Deflagration Models for Thermonuclear Supernovae}",
      journal = {\apj},
         year = 2014,
        month = jul,
       volume = {789},
       number = {2},
          eid = {103},
        pages = {103},
          doi = {10.1088/0004-637X/789/2/103},
archivePrefix = {arXiv},
       eprint = {1307.8221},
 primaryClass = {astro-ph.HE},
       adsurl = {https://ui.adsabs.harvard.edu/abs/2014ApJ...789..103L}
}

@ARTICLE{Lach_2022a,
       author = {{Lach}, F. and {Callan}, F.~P. and {Bubeck}, D. and {R{\"o}pke}, F.~K. and {Sim}, S.~A. and {Schrauth}, M. and {Ohlmann}, S.~T. and {Kromer}, M.},
        title = "{Type Iax supernovae from deflagrations in Chandrasekhar mass white dwarfs}",
      journal = {\aap},
         year = 2022,
        month = feb,
       volume = {658},
          eid = {A179},
        pages = {A179},
          doi = {10.1051/0004-6361/202141453},
archivePrefix = {arXiv},
       eprint = {2109.02926},
 primaryClass = {astro-ph.SR},
       adsurl = {https://ui.adsabs.harvard.edu/abs/2022A&A...658A.179L}
}

@ARTICLE{Hoyle_1960,
       author = {{Hoyle}, F. and {Fowler}, William A.},
        title = "{Nucleosynthesis in Supernovae.}",
      journal = {\apj},
         year = 1960,
        month = nov,
       volume = {132},
        pages = {565},
          doi = {10.1086/146963},
       adsurl = {https://ui.adsabs.harvard.edu/abs/1960ApJ...132..565H}
}

@ARTICLE{Magee_2025,
       author = {{Magee}, M.~R. and {Killestein}, T.~L. and {Pursiainen}, M. and {Godson}, B. and {Jarvis}, D. and {Jim{\'e}nez-Palau}, C. and {Lyman}, J.~D. and {Steeghs}, D. and {Warwick}, B. and {Anderson}, J.~P. and {Butterley}, T. and {Chen}, T.-W. and {Dhillon}, V.~S. and {Galbany}, L. and {Gonz{\'a}lez-Gait{\'a}n}, S. and {Gromadzki}, M. and {Inserra}, C. and {Kelsey}, L. and {Kumar}, A. and {Leloudas}, G. and {Mattila}, S. and {Moran}, S. and {M{\"u}ller-Bravo}, T.~E. and {Noysena}, K. and {Ramsay}, G. and {Srivastav}, S. and {Starling}, R. and {Wilson}, R.~W. and {Young}, D.~R. and {Ackley}, K. and {Breton}, R.~P. and {Casares Vel{\'a}zquez}, J. and {Dyer}, M.~J. and {Galloway}, D.~K. and {Kankare}, E. and {Kotak}, R. and {Nuttall}, L.~K. and {O'Neill}, D. and {Pessi}, P. and {Pollacco}, D. and {Ulaczyk}, K. and {Yaron}, O.},
        title = "{SN 2024bfu, SN 2025qe, and the early light curves of type Iax supernovae}",
      journal = {\mnras},
         year = 2025,
        month = nov,
       volume = {543},
       number = {4},
        pages = {3731-3753},
          doi = {10.1093/mnras/staf1675},
archivePrefix = {arXiv},
       eprint = {2506.02118},
 primaryClass = {astro-ph.HE},
       adsurl = {https://ui.adsabs.harvard.edu/abs/2025MNRAS.543.3731M}
}

@article{kasen_2006,
  title={Secondary maximum in the near-infrared light curves of type Ia supernovae},
  author={Kasen, Daniel},
  journal={The Astrophysical Journal},
  volume={649},
  number={2},
  pages={939},
  year={2006},
  publisher={IOP Publishing}
}

@article{Elias_1981,
  title={Infrared light curves of Type I supernovae},
  author={Elias, JH and Frogel, Jay A and Hackwell, JA and Persson, SE},
  journal={Astrophysical Journal, Part 2-Letters to the Editor, vol. 251, Dec. 1, 1981, p. L13-L16.},
  volume={251},
  pages={L13--L16},
  year={1981}
}

@ARTICLE{Li_2018,
       author = {{Li}, Linyi and {Wang}, Xiaofeng and {Zhang}, Jujia and {Arcavi}, Iair and {Zhang}, Tianmeng and {Rui}, Liming and {Hosseinzadeh}, Griffin and {Howell}, D. Andrew and {McCully}, Curtis and {Zhang}, Kaicheng and {Valenti}, Stefano and {Mo}, Jun and {Li}, Wenxiong and {Huang}, Fang and {Xiang}, Danfeng and {Wang}, Lifan and {Zhou}, Xu},
        title = "{Optical observations of the 2002cx-like supernova 2014ek and characterizations of SNe Iax}",
      journal = {\mnras},
         year = 2018,
        month = aug,
       volume = {478},
       number = {4},
        pages = {4575-4589},
          doi = {10.1093/mnras/sty1303},
archivePrefix = {arXiv},
       eprint = {1805.05810},
 primaryClass = {astro-ph.HE},
       adsurl = {https://ui.adsabs.harvard.edu/abs/2018MNRAS.478.4575L}
}

@ARTICLE{Srivastav_2023,
       author = {{Srivastav}, Shubham and {Smartt}, S.~J. and {Huber}, M.~E. and {Dimitriadis}, G. and {Chambers}, K.~C. and {Fulton}, Michael D. and {Moore}, Thomas and {Callan}, F.~P. and {Gillanders}, James H. and {Maguire}, K. and {Nicholl}, M. and {Shingles}, Luke J. and {Sim}, S.~A. and {Smith}, K.~W. and {Anderson}, J.~P. and {de Boer}, Thomas and {Chen}, Ting-Wan and {Gao}, Hua and {Young}, D.~R.},
        title = "{The Luminous Type Ia Supernova 2022ilv and Its Early Excess Emission}",
      journal = {\apjl},
         year = 2023,
        month = feb,
       volume = {943},
       number = {2},
          eid = {L20},
        pages = {L20},
          doi = {10.3847/2041-8213/acb2ce},
archivePrefix = {arXiv},
       eprint = {2211.10544},
 primaryClass = {astro-ph.HE},
       adsurl = {https://ui.adsabs.harvard.edu/abs/2023ApJ...943L..20S}
}

@ARTICLE{Colgate_1969,
       author = {{Colgate}, Stirling A. and {McKee}, Chester},
        title = "{Early Supernova Luminosity}",
      journal = {\apj},
         year = 1969,
        month = aug,
       volume = {157},
        pages = {623},
          doi = {10.1086/150102},
       adsurl = {https://ui.adsabs.harvard.edu/abs/1969ApJ...157..623C}
}

@ARTICLE{Hillebrandt_2013,
       author = {{Hillebrandt}, W. and {Kromer}, M. and {R{\"o}pke}, F.~K. and {Ruiter}, A.~J.},
        title = "{Towards an understanding of Type Ia supernovae from a synthesis of theory and observations}",
      journal = {Frontiers of Physics},
         year = 2013,
        month = apr,
       volume = {8},
       number = {2},
        pages = {116-143},
          doi = {10.1007/s11467-013-0303-2},
archivePrefix = {arXiv},
       eprint = {1302.6420},
 primaryClass = {astro-ph.CO},
       adsurl = {https://ui.adsabs.harvard.edu/abs/2013FrPhy...8..116H}
}

@ARTICLE{Arnett_1982,
       author = {{Arnett}, W.~D.},
        title = "{Type I supernovae. I - Analytic solutions for the early part of the light curve}",
      journal = {\apj},
         year = 1982,
        month = feb,
       volume = {253},
        pages = {785-797},
          doi = {10.1086/159681},
       adsurl = {https://ui.adsabs.harvard.edu/abs/1982ApJ...253..785A}
}

@ARTICLE{Stritzinger_2006,
       author = {{Stritzinger}, M. and {Mazzali}, P.~A. and {Sollerman}, J. and {Benetti}, S.},
        title = "{Consistent estimates of $^{56}$Ni yields for type Ia supernovae}",
      journal = {\aap},
         year = 2006,
        month = dec,
       volume = {460},
       number = {3},
        pages = {793-798},
          doi = {10.1051/0004-6361:20065514},
archivePrefix = {arXiv},
       eprint = {astro-ph/0609232},
 primaryClass = {astro-ph},
       adsurl = {https://ui.adsabs.harvard.edu/abs/2006A&A...460..793S}
}

@ARTICLE{Nomoto_1984b,
       author = {{Nomoto}, K. and {Thielemann}, F.-K. and {Yokoi}, K.},
        title = "{Accreting white dwarf models for type I supernovae. III. Carbon deflagration supernovae.}",
      journal = {\apj},
         year = 1984,
        month = nov,
       volume = {286},
        pages = {644-658},
          doi = {10.1086/162639},
       adsurl = {https://ui.adsabs.harvard.edu/abs/1984ApJ...286..644N}
}

@ARTICLE{Thielemann_1986,
       author = {{Thielemann}, F.-K. and {Nomoto}, K. and {Yokoi}, K.},
        title = "{Explosive nucleosynthesis in carbon deflagration models of Type I supernovae}",
      journal = {\aap},
         year = 1986,
        month = apr,
       volume = {158},
       number = {1-2},
        pages = {17-33},
       adsurl = {https://ui.adsabs.harvard.edu/abs/1986A&A...158...17T}
}

@ARTICLE{sarangi_2022b,
       author = {{Sarangi}, Arkaprabha},
        title = "{Formation, distribution and IR emission of dust in the clumpy ejecta of Type II-P core-collapse supernovae, in isotropic and anisotropic scenarios}",
      journal = {arXiv e-prints},
         year = 2022,
        month = sep,
          eid = {arXiv:2209.14896},
        pages = {arXiv:2209.14896},
archivePrefix = {arXiv},
       eprint = {2209.14896},
 primaryClass = {astro-ph.SR},
       adsurl = {https://ui.adsabs.harvard.edu/abs/2022arXiv220914896S}
}

@ARTICLE{kwok_2025,
       author = {{Kwok}, Lindsey A. and {Singh}, Mridweeka and {Jha}, Saurabh W. and {Blondin}, St{\'e}phane and {Dastidar}, Raya and {Larison}, Conor and {Miller}, Adam A. and {Andrews}, Jennifer E. and {Andrews}, Moira and {Anupama}, G.~C. and {Auchettl}, Katie and {B{\'a}nhidi}, Dominik and {Barna}, Barnabas and {Bostroem}, K. Azalee and {Brink}, Thomas G. and {Cartier}, R{\'e}gis and {Chen}, Ping and {Christy}, Collin T. and {Coulter}, David A. and {Covarrubias}, Sofia and {Davis}, Kyle W. and {Dickinson}, Connor B. and {Dong}, Yize and {Farah}, Joseph R. and {Filippenko}, Alexei V. and {Fl{\"o}rs}, Andreas and {Foley}, Ryan J. and {Franz}, Noah and {Fremling}, Christoffer and {Galbany}, Llu{\'\i}s and {Gangopadhyay}, Anjasha and {Garg}, Aarna and {Garnavich}, Peter and {Gates}, Elinor L. and {Graur}, Or and {Gordon}, Alexa C. and {Hiramatsu}, Daichi and {Hoang}, Emily and {Howell}, D. Andrew and {Hsu}, Brian and {Johansson}, Joel and {Joshi}, Arti and {Kahinga}, Lordrick A. and {Kaur}, Ravjit and {Kumar}, Sahana and {Kumnurdmanee}, Piramon and {Kuncarayakti}, Hanindyo and {LeBaron}, Natalie and {Liu}, Chang and {Maeda}, Keiichi and {Maguire}, Kate and {McCully}, Curtis and {Mehta}, Darshana and {Menotti}, Luca M. and {Metevier}, Anne J. and {Misra}, Kuntal and {Murphey}, C. Tanner and {Newsome}, Megan and {Padilla Gonzalez}, Estefania and {Patra}, Kishore C. and {Pearson}, Jeniveve and {Piro}, Anthony L. and {Polin}, Abigail and {Ravi}, Aravind P. and {Rest}, Armin and {Rehemtulla}, Nabeel and {Meza Retamal}, Nicolas and {Robinson}, Olivia M. and {Rojas-Bravo}, C{\'e}sar and {Sahu}, Devendra K. and {Sand}, David J. and {Schmidt}, Brian P. and {Schulze}, Steve and {Schwab}, Michaela and {Shrestha}, Manisha and {Siebert}, Matthew R. and {Simha}, Sunil and {Smith}, Nathan and {Sollerman}, Jesper and {Subrayan}, Bhagya M. and {Szalai}, Tam{\'a}s and {Taggart}, Kirsty and {Teja}, Rishabh Singh and {Temim}, Tea and {Terwel}, Jacco H. and {Tinyanont}, Samaporn and {Valenti}, Stefano and {Anais Vilchez}, Jorge and {Vink{\'o}}, J{\'o}zsef and {Westerling}, Aya L. and {Yang}, Yi and {Zheng}, WeiKang},
        title = "{JWST and Ground-based Observations of the Type Iax Supernovae SN 2024pxl and SN 2024vjm: Evidence for Weak Deflagration Explosions}",
      journal = {\apjl},
         year = 2025,
        month = aug,
       volume = {989},
       number = {2},
          eid = {L33},
        pages = {L33},
          doi = {10.3847/2041-8213/adf062},
archivePrefix = {arXiv},
       eprint = {2505.02944},
 primaryClass = {astro-ph.HE},
       adsurl = {https://ui.adsabs.harvard.edu/abs/2025ApJ...989L..33K}
}

@ARTICLE{Dimitriadis2025,
       author = {{Dimitriadis}, G. and {Burgaz}, U. and {Deckers}, M. and {Maguire}, K. and {Johansson}, J. and {Smith}, M. and {Rigault}, M. and {Frohmaier}, C. and {Sollerman}, J. and {Galbany}, L. and {Kim}, Y.-L. and {Liu}, C. and {Miller}, A.~A. and {Nugent}, P.~E. and {Alburai}, A. and {Chen}, P. and {Dhawan}, S. and {Ginolin}, M. and {Goobar}, A. and {Groom}, S.~L. and {Harvey}, L. and {Kenworthy}, W.~D. and {Kulkarni}, S.~R. and {Phan}, K. and {Popovic}, B. and {Riddle}, R.~L. and {Rusholme}, B. and {M{\"u}ller-Bravo}, T.~E. and {Nordin}, J. and {Terwel}, J.~H. and {Townsend}, A.},
        title = "{ZTF SN Ia DR2: The diversity and relative rates of the thermonuclear supernova population}",
      journal = {\aap},
         year = 2025,
        month = feb,
       volume = {694},
          eid = {A10},
        pages = {A10},
          doi = {10.1051/0004-6361/202451852},
archivePrefix = {arXiv},
       eprint = {2409.04200},
 primaryClass = {astro-ph.HE},
       adsurl = {https://ui.adsabs.harvard.edu/abs/2025A&A...694A..10D}
}

@article{Rodriguez_2023,
doi = {10.3847/1538-4357/ace2bd},
url = {https://doi.org/10.3847/1538-4357/ace2bd},
year = {2023},
month = {sep},
publisher = {The American Astronomical Society},
volume = {955},
number = {1},
pages = {71},
author = {{Rodr{\'\i}guez}, {\'O}smar and {Maoz}, Dan and {Nakar}, Ehud},
title = {The Iron Yield of Core-collapse Supernovae},
journal = {The Astrophysical Journal}
}

@ARTICLE{Banhidi2025,
       author = {{B{\'a}nhidi}, D. and {Barna}, B. and {Szalai}, T. and {Vink{\'o}}, J. and {B{\'\i}r{\'o}}, I.~B. and {Bostroem}, K.~A. and {Cs{\'a}nyi}, I. and {Davis}, K.~W. and {Foley}, R.~J. and {Galbany}, L. and {Jha}, S.~W. and {Howell}, D.~A. and {Kwok}, L.~A. and {P{\'a}l}, A. and {Pellegrino}, C. and {Rojas-Bravo}, C. and {Sz{\'e}kely}, P. and {Taggart}, K. and {Terreran}, G. and {Tinyanont}, S.},
        title = "{SN 2022xlp: The second-known well-observed, intermediate-luminosity Iax supernova}",
      journal = {\aap},
         year = 2025,
        month = nov,
       volume = {703},
          eid = {A64},
        pages = {A64},
          doi = {10.1051/0004-6361/202553922},
archivePrefix = {arXiv},
       eprint = {2509.07717},
 primaryClass = {astro-ph.SR},
       adsurl = {https://ui.adsabs.harvard.edu/abs/2025A&A...703A..64B}
}

@ARTICLE{cunningham_2024,
       author = {{Cunningham}, Tim and {Caiazzo}, Ilaria and {Prusinski}, Nikolaus Z. and {Fuller}, James and {Raymond}, John C. and {Kulkarni}, S.~R. and {Neill}, James D. and {Duffell}, Paul and {Martin}, Chris and {Toloza}, Odette and {Charbonneau}, David and {Kenyon}, Scott J. and {Lin}, Zeren and {Matuszewski}, Mateusz and {McGurk}, Rosalie and {Polin}, Abigail and {Yao}, Philippe Z.},
        title = "{Expansion Properties of the Young Supernova Type Iax Remnant Pa 30 Revealed}",
      journal = {\apjl},
         year = 2024,
        month = nov,
       volume = {975},
       number = {1},
          eid = {L7},
        pages = {L7},
          doi = {10.3847/2041-8213/ad713b},
archivePrefix = {arXiv},
       eprint = {2410.10940},
 primaryClass = {astro-ph.SR},
       adsurl = {https://ui.adsabs.harvard.edu/abs/2024ApJ...975L...7C}
}

@ARTICLE{singh_2025,
       author = {{Singh}, Mridweeka and {Kwok}, Lindsey A. and {Jha}, Saurabh W. and {Dastidar}, R. and {Larison}, Conor and {Filippenko}, Alexei V. and {Andrews}, Jennifer E. and {Andrews}, Moira and {Anupama}, G.~C. and {Arunachalam}, Prasiddha and {Auchettl}, Katie and {B{\'A}nhidi}, Dominik and {Barna}, Barnabas and {Bostroem}, K. Azalee and {Brink}, Thomas G. and {Cartier}, R{\'E}gis and {Chen}, Ping and {Christy}, Collin T. and {Coulter}, David A. and {Covarrubias}, Sofia and {Davis}, Kyle W. and {Dickinson}, Connor B. and {Dong}, Yize and {Farah}, Joseph and {Fl{\"O}rs}, Andreas and {Foley}, Ryan J. and {Franz}, Noah and {Fremling}, Christoffer and {Galbany}, Llu{\'I}s and {Gangopadhyay}, Anjasha and {Garg}, Aarna and {Gates}, Elinor L. and {Graur}, Or and {Gordon}, Alexa C. and {Hiramatsu}, Daichi and {Hoang}, Emily and {Howell}, D. Andrew and {Hsu}, Brian and {Johansson}, Joel and {Joshi}, Arti and {Kahinga}, Lordrick A. and {Kaur}, Ravjit and {Kumar}, Sahana and {Kumnurdmanee}, Piramon and {Kuncarayakti}, Hanindyo and {Lebaron}, Natalie and {Lidman}, C. and {Liu}, Chang and {Maeda}, Keiichi and {Maguire}, Kate and {Martin}, Bailey and {Mccully}, Curtis and {Mehta}, Darshana and {Menotti}, Luca M. and {Metevier}, Anne J. and {Miller}, A.~A. and {Misra}, Kuntal and {Tanner Murphey}, C. and {Newsome}, Megan and {Padilla Gonzalez}, Estefania and {Patra}, Kishore C. and {Pearson}, Jeniveve and {Piro}, Anthony L. and {Polin}, Abigail and {Ravi}, Aravind P. and {Rest}, Armin and {Rehemtulla}, Nabeel and {Meza Retamal}, Nicolas and {Robinson}, O.~M. and {Rojas-Bravo}, C{\'E}sar and {Sahu}, Devendra K. and {Sand}, David J. and {Schmidt}, Brian P. and {Schulze}, Steve and {Schwab}, Michaela and {Shrestha}, Manisha and {Siebert}, Matthew R. and {Simha}, Sunil and {Smith}, Nathan and {Sollerman}, Jesper and {Srivastav}, Shubham and {Subrayan}, Bhagya M. and {Szalai}, Tam{\'A}s and {Taggart}, Kirsty and {Teja}, Rishabh Singh and {Terwel}, Jacco H. and {Tinyanont}, Samaporn and {Valenti}, Stefano and {Vink{\'O}}, J{\'O}zsef and {Westerling}, Aya L. and {Yang}, Yi and {Zheng}, Weikang},
        title = "{Photometry and Spectroscopy of SN 2024pxl: A Luminosity Link Among Type Iax Supernovae}",
      journal = {arXiv e-prints},
         year = 2025,
        month = may,
          eid = {arXiv:2505.02943},
        pages = {arXiv:2505.02943},
          doi = {10.48550/arXiv.2505.02943},
archivePrefix = {arXiv},
       eprint = {2505.02943},
 primaryClass = {astro-ph.HE},
       adsurl = {https://ui.adsabs.harvard.edu/abs/2025arXiv250502943S}
}

@ARTICLE{wang_2024,
       author = {{Wang}, Lingzhi and {Hu}, Maokai and {Wang}, Lifan and {Yang}, Yi and {Yang}, Jiawen and {Gomez}, Haley and {Chen}, Sijie and {Hu}, Lei and {Chen}, Ting-Wan and {Mo}, Jun and {Wang}, Xiaofeng and {Baade}, Dietrich and {Hoeflich}, Peter and {Wheeler}, J. Craig and {Pignata}, Giuliano and {Burke}, Jamison and {Hiramatsu}, Daichi and {Howell}, D. Andrew and {McCully}, Curtis and {Pellegrino}, Craig and {Galbany}, Llu{\'\i}s and {Hsiao}, Eric Y. and {Sand}, David J. and {Zhang}, Jujia and {Uddin}, Syed A. and {Anderson}, J.~P. and {Ashall}, Chris and {Cheng}, Cheng and {Gromadzki}, Mariusz and {Inserra}, Cosimo and {Lin}, Han and {Morrell}, N. and {Morales-Garoffolo}, Antonia and {M{\"u}ller-Bravo}, T.~E. and {Nicholl}, Matt and {Gonzalez}, Estefania Padilla and {Phillips}, M.~M. and {Pineda-Garc{\'\i}a}, J. and {Sai}, Hanna and {Smith}, Mathew and {Shahbandeh}, M. and {Srivastav}, Shubham and {Stritzinger}, M.~D. and {Yang}, Sheng and {Young}, D.~R. and {Yu}, Lixin and {Zhang}, Xinghan},
        title = "{Newly formed dust within the circumstellar environment of SN Ia-CSM 2018evt}",
      journal = {Nature Astronomy},
         year = 2024,
        month = apr,
       volume = {8},
        pages = {504-519},
          doi = {10.1038/s41550-024-02197-9},
archivePrefix = {arXiv},
       eprint = {2310.14874},
 primaryClass = {astro-ph.HE},
       adsurl = {https://ui.adsabs.harvard.edu/abs/2024NatAs...8..504W}
}

@ARTICLE{clayton_2025,
       author = {{Clayton}, Geoffrey C. and {Wesson}, R. and {Fox}, Ori D. and {Shahbandeh}, Melissa and {Filippenko}, Alexei V. and {Nickson}, Bryony and {Engesser}, Michael and {Van Dyk}, Schuyler D. and {Zheng}, WeiKang and {Brink}, Thomas G. and {Yang}, Yi and {Temim}, Tea and {Smith}, Nathan and {Andrews}, Jennifer and {Ashall}, Chris and {De Looze}, Ilse and {Derkacy}, James M. and {Dessart}, Luc and {Dulude}, Michael and {Dwek}, Eli and {Foley}, Ryan J. and {Gezari}, Suvi and {Gomez}, Sebastian and {Gonzaga}, Shireen and {Indukuri}, Siva and {Jencson}, Jacob and {Johansson}, Joel and {Kasliwal}, Mansi and {Lane}, Zachary G. and {Lau}, Ryan and {Law}, David and {Marston}, Anthony and {Milisavljevic}, Dan and {O'Steen}, Richard and {Pierel}, Justin and {Rest}, Armin and {Sarangi}, Arkaprabha and {Siebert}, Matthew and {Skrutskie}, Michael and {Strolger}, Lou and {Szalai}, Tam{\'a}s and {Tinyanont}, Samaporn and {Wang}, Qinan and {Williams}, Brian and {Xiao}, Lin and {Zs{\'\i}ros}, Szanna},
        title = "{Very Late-time JWST and Keck Spectra of the Oxygen-rich Supernova 1995N}",
      journal = {\apj},
         year = 2025,
        month = oct,
       volume = {991},
       number = {2},
          eid = {133},
        pages = {133},
          doi = {10.3847/1538-4357/adfc72},
archivePrefix = {arXiv},
       eprint = {2505.01574},
 primaryClass = {astro-ph.SR},
       adsurl = {https://ui.adsabs.harvard.edu/abs/2025ApJ...991..133C}
}

@ARTICLE{lykou_2023,
       author = {{Lykou}, Foteini and {Parker}, Quentin A. and {Ritter}, Andreas and {Zijlstra}, Albert A. and {Hillier}, D. John and {Guerrero}, Mart{\'\i}n A. and {Le D{\^u}}, Pascal},
        title = "{A New Study on a Type Iax Stellar Remnant and its Probable Association with SN 1181}",
      journal = {\apj},
         year = 2023,
        month = feb,
       volume = {944},
       number = {2},
          eid = {120},
        pages = {120},
          doi = {10.3847/1538-4357/acb138},
archivePrefix = {arXiv},
       eprint = {2208.03946},
 primaryClass = {astro-ph.SR},
       adsurl = {https://ui.adsabs.harvard.edu/abs/2023ApJ...944..120L}
}

@ARTICLE{fox_14dt_2016,
       author = {{Fox}, Ori D. and {Johansson}, Joel and {Kasliwal}, Mansi and {Andrews}, Jennifer and {Bally}, John and {Bond}, Howard E. and {Boyer}, Martha L. and {Gehrz}, R.~D. and {Helou}, George and {Hsiao}, E.~Y. and {Masci}, Frank J. and {Parthasarathy}, M. and {Smith}, Nathan and {Tinyanont}, Samaporn and {Van Dyk}, Schuyler D.},
        title = "{An Excess of Mid-infrared Emission from the Type Iax SN 2014dt}",
      journal = {\apjl},
         year = 2016,
        month = jan,
       volume = {816},
       number = {1},
          eid = {L13},
        pages = {L13},
          doi = {10.3847/2041-8205/816/1/L13},
archivePrefix = {arXiv},
       eprint = {1510.08070},
 primaryClass = {astro-ph.HE},
       adsurl = {https://ui.adsabs.harvard.edu/abs/2016ApJ...816L..13F}
}

@article{matsuura_2015,
	author = {M. Matsuura and E. Dwek and M. J. Barlow and B. Babler and M. Baes and M. Meixner and Jos{\'{e}} Cernicharo and Geoff C. Clayton and L. Dunne and C. Fransson and Jacopo Fritz and Walter Gear and H. L. Gomez and M. A. T. Groenewegen and R. Indebetouw and R. J. Ivison and A. Jerkstrand and V. Lebouteiller and T. L. Lim and P. Lundqvist and C. P. Pearson and J. Roman-Duval and P. Royer and Lister Staveley-Smith and B. M. Swinyard and P. A. M. van Hoof and J. Th. van Loon and Joris Verstappen and Roger Wesson and Giovanna Zanardo and Joris A. D. L. Blommaert and Leen Decin and W. T. Reach and George Sonneborn and Griet C. Van de Steene and Jeremy A. Yates},
	doi = {10.1088/0004-637x/800/1/50},
	journal = {The Astrophysical Journal},
	month = {feb},
	number = {1},
	pages = {50},
	publisher = {American Astronomical Society},
	title = {A {STUBBORNLY} {LARGE} {MASS} {OF} {COLD} {DUST} {IN} {THE} {EJECTA} {OF} {SUPERNOVA} 1987A},
	url = {https://doi.org/10.1088%2F0004-637x%2F800%2F1%2F50},
	volume = {800},
	year = 2015}

@article{seitenzahl_2014,
	adsurl = {https://ui.adsabs.harvard.edu/abs/2014ApJ...792...10S},
	archiveprefix = {arXiv},
	author = {{Seitenzahl}, Ivo R. and {Timmes}, F.~X. and {Magkotsios}, Georgios},
	doi = {10.1088/0004-637X/792/1/10},
	eid = {10},
	eprint = {1408.5986},
	journal = {\apj},
	month = sep,
	number = {1},
	pages = {10},
	primaryclass = {astro-ph.SR},
	title = {{The Light Curve of SN 1987A Revisited: Constraining Production Masses of Radioactive Nuclides}},
	volume = {792},
	year = 2014}

@article{liu_2018,
	adsurl = {https://ui.adsabs.harvard.edu/abs/2018SciA....4.1054L},
	archiveprefix = {arXiv},
	author = {{Liu}, Nan and {Nittler}, Larry R. and {Alexander}, Conel M.~O. 'D. and {Wang}, Jianhua},
	doi = {10.1126/sciadv.aao1054},
	eprint = {1801.06463},
	journal = {Science Advances},
	month = jan,
	number = {1},
	pages = {eaao1054},
	primaryclass = {astro-ph.SR},
	title = {{Late formation of silicon carbide in type II supernovae}},
	volume = {4},
	year = 2018}

@article{ott_2019,
	adsurl = {https://ui.adsabs.harvard.edu/abs/2019ApJ...885..128O},
	author = {{Ott}, Ulrich and {Stephan}, Thomas and {Hoppe}, Peter and {Savina}, Michael R.},
	doi = {10.3847/1538-4357/ab41f3},
	eid = {128},
	journal = {\apj},
	month = nov,
	number = {2},
	pages = {128},
	title = {{Isotopes of Barium as a Chronometer for Supernova Dust Formation}},
	volume = {885},
	year = 2019}

@INPROCEEDINGS{hoppe_2010,
       author = {{Hoppe}, P.},
        title = "{Measurements of presolar grains}",
    booktitle = {Nuclei in the Cosmos},
         year = 2010,
        month = jan,
          eid = {21},
        pages = {21},
          doi = {10.22323/1.100.0021},
       adsurl = {https://ui.adsabs.harvard.edu/abs/2010nuco.confE..21H}
}

@ARTICLE{fok_2024,
       author = {{Fok}, Hung Kwan and {Pignatari}, Marco and {C{\^o}t{\'e}}, Beno{\^\i}t and {Trappitsch}, Reto},
        title = "{Silicon Isotopic Composition of Mainstream Presolar SiC Grains Revisited: The Impact of Nuclear Reaction Rate Uncertainties}",
      journal = {\apjl},
         year = 2024,
        month = dec,
       volume = {977},
       number = {1},
          eid = {L24},
        pages = {L24},
          doi = {10.3847/2041-8213/ad91ab},
archivePrefix = {arXiv},
       eprint = {2411.19935},
 primaryClass = {astro-ph.SR},
       adsurl = {https://ui.adsabs.harvard.edu/abs/2024ApJ...977L..24F}
}

@ARTICLE{hoppe_2019,
       author = {{Hoppe}, Peter and {Stancliffe}, Richard J. and {Pignatari}, Marco and {Amari}, Sachiko},
        title = "{Isotopic Signatures of Supernova Nucleosynthesis in Presolar Silicon Carbide Grains of Type AB with Supersolar $^{14}$N/$^{15}$N Ratios}",
      journal = {\apj},
         year = 2019,
        month = dec,
       volume = {887},
       number = {1},
          eid = {8},
        pages = {8},
          doi = {10.3847/1538-4357/ab521c},
       adsurl = {https://ui.adsabs.harvard.edu/abs/2019ApJ...887....8H}
}

@ARTICLE{hoppe_2024,
       author = {{Hoppe}, Peter and {Leitner}, Jan and {Pignatari}, Marco and {Amari}, Sachiko},
        title = "{Isotope studies of presolar silicon carbide grains from supernovae: new constraints for hydrogen-ingestion supernova models}",
      journal = {\mnras},
         year = 2024,
        month = jul,
       volume = {532},
       number = {1},
        pages = {211-222},
          doi = {10.1093/mnras/stae1523},
       adsurl = {https://ui.adsabs.harvard.edu/abs/2024MNRAS.532..211H}
}

@article{hoppe_2018,
	adsurl = {https://ui.adsabs.harvard.edu/abs/2018GeCoA.221..182H},
	author = {{Hoppe}, Peter and {Pignatari}, Marco and {Kodol{\'a}nyi}, J{\'a}nos and {Gr{\"o}ner}, Elmar and {Amari}, Sachiko},
	doi = {10.1016/j.gca.2017.01.051},
	journal = {\gca},
	month = jan,
	pages = {182-199},
	title = {{NanoSIMS isotope studies of rare types of presolar silicon carbide grains from the Murchison meteorite: Implications for supernova models and the role of $^{14}$C}},
	volume = {221},
	year = 2018}

@ARTICLE{shahbandeh_2023,
       author = {{Shahbandeh}, Melissa and {Sarangi}, Arkaprabha and {Temim}, Tea and {Szalai}, Tam{\'a}s and {Fox}, Ori D. and {Tinyanont}, Samaporn and {Dwek}, Eli and {Dessart}, Luc and {Filippenko}, Alexei V. and {Brink}, Thomas G. and {Foley}, Ryan J. and {Jencson}, Jacob and {Pierel}, Justin and {Zs{\'\i}ros}, Szanna and {Rest}, Armin and {Zheng}, WeiKang and {Andrews}, Jennifer and {Clayton}, Geoffrey C. and {De}, Kishalay and {Engesser}, Michael and {Gezari}, Suvi and {Gomez}, Sebastian and {Gonzaga}, Shireen and {Johansson}, Joel and {Kasliwal}, Mansi and {Lau}, Ryan and {De Looze}, Ilse and {Marston}, Anthony and {Milisavljevic}, Dan and {O'Steen}, Richard and {Siebert}, Matthew and {Skrutskie}, Michael and {Smith}, Nathan and {Strolger}, Lou and {Van Dyk}, Schuyler D. and {Wang}, Qinan and {Williams}, Brian and {Williams}, Robert and {Xiao}, Lin and {Yang}, Yi},
        title = "{JWST observations of dust reservoirs in type IIP supernovae 2004et and 2017eaw}",
      journal = {\mnras},
         year = 2023,
        month = aug,
       volume = {523},
       number = {4},
        pages = {6048-6060},
          doi = {10.1093/mnras/stad1681},
archivePrefix = {arXiv},
       eprint = {2301.10778},
 primaryClass = {astro-ph.HE},
       adsurl = {https://ui.adsabs.harvard.edu/abs/2023MNRAS.523.6048S}
}

@INPROCEEDINGS{amari_2014,
       author = {{Amari}, Sachiko and {Zinner}, Ernst and {Gallino}, Roberto},
        title = "{Abundances of presolar graphite and SiC from supernovae and AGB stars in the Murchison meteorite}",
    booktitle = {Origin of Matter and Evolution of Galaxies 2013},
         year = 2014,
       editor = {{Jeong}, Sunchan and {Imai}, Nobuaki and {Miyatake}, Hiroari and {Kajino}, Toshitaka},
       series = {American Institute of Physics Conference Series},
       volume = {1594},
        month = may,
    publisher = {AIP},
        pages = {307-312},
          doi = {10.1063/1.4874087},
       adsurl = {https://ui.adsabs.harvard.edu/abs/2014AIPC.1594..307A}
}

@incollection{zin07,
	address = {Oxford},
	author = {E. Zinner},
	booktitle = {Treatise on Geochemistry},
	doi = {http://dx.doi.org/10.1016/B0-08-043751-6/01144-0},
	pages = {1 - 33},
	publisher = {Pergamon},
	title = {1.02 - Presolar Grains},
	url = {http://www.sciencedirect.com/science/article/pii/B0080437516011440},
	year = {2007}}

@inproceedings{nit08,
	adsurl = {http://adsabs.harvard.edu/abs/2008nuco.confE..13N},
	author = {{Nittler}, L.~R.},
	booktitle = {Nuclei in the Cosmos (NIC X)},
	title = {{Presolar Stardust In The Solar System: Recent Advances for Nuclear Astrophysics}},
	year = 2008}

@article{don85,
	adsurl = {http://adsabs.harvard.edu/abs/1985ApJ...288..187D},
	author = {{Donn}, B. and {Nuth}, J.~A.},
	doi = {10.1086/162779},
	journal = {Astrophysical Journal},
	month = jan,
	pages = {187-190},
	title = {{Does nucleation theory apply to the formation of refractory circumstellar grains?}},
	volume = 288,
	year = 1985}

@article{dwe11,
	adsurl = {http://adsabs.harvard.edu/abs/2011ApJ...727...63D},
	archiveprefix = {arXiv},
	author = {{Dwek}, E. and {Cherchneff}, I.},
	doi = {10.1088/0004-637X/727/2/63},
	eid = {63},
	eprint = {1011.1303},
	journal = {\apj},
	month = feb,
	pages = {63},
	title = {{The Origin of Dust in the Early Universe: Probing the Star Formation History of Galaxies by Their Dust Content}},
	volume = 727,
	year = 2011}

\section*{Appendix}

\begin{longtable*}{cccccccc}
\caption{ New chemical network added to  current analysis.\\The rates mentioned are calculated using Chemical Kinetic approach (Either Theoretical, Experimental from database or estimated from similar reactions)  The reaction rates are expressed in Arrhenius form   $k = A_{ij} \times (T/300K)^\nu \times exp(-E_a/T )$ with $A_{ij}$ in $s^{-1}$ ,$cm^3 s^{-1}$, or $cm^6 s^{-1}$ for uni-, bi-and termolecular processes,respectively; $E_a$ is in Kelvin.\\SC13:\cite{sar13},SC13-\ce{SiO}: similar to SiO network in \cite{sar13}, NIST:\href{www.http://kinetics.nist.gov/kinetics/}{NIST/Kinetics} E: Estimated}
\label{tab: network}\\
\hline
Reaction & Reactants &  & Products & $A_{ij}$ & $\nu$ & E$_a$ & References \\ \hline
\endfirsthead
\multicolumn{8}{c}%
{{\bfseries Table \thetable\ (continue)}} \\
\hline
Reaction & Reactants &  & Products & $A_{ij}$ & $\nu$ & E$_a$ & References \\ \hline
\endhead
\hline
\endfoot

\endlastfoot
\multicolumn{8}{c}{(Fe$_n$O$_m$)  cluster} \\ \hline
1 & \ce{Fe}+\ce{O2} & $\rightarrow$ & \ce{FeO}+\ce{O} & \SI{2.0900e-9}{} & 0.0 & 10199.6 & NIST \\
2 & \ce{FeO}+\ce{O} & $\rightarrow$ & \ce{Fe}+\ce{O2} & \SI{4.6000E-10}{} & -0.4 & 350.0 & " \\
3 & \ce{Fe}+\ce{CO2} & $\rightarrow$ & \ce{FeO}+\ce{CO} & \SI{5.3800e-10}{} & 0.0 & 15033.1 & " \\
4 & \ce{Fe}+\ce{O2} & $\rightarrow$ & \ce{FeO2} & \SI{2.0500e-28}{} & -2.6 & 3170.2 & " \\
5 & \ce{FeO}+\ce{O2} & $\rightarrow$ & \ce{FeO2}+\ce{O} & \SI{1.0200e-11}{} & 0.4 & 8202.8 & " \\
6 & \ce{FeO}+\ce{CO2} & $\rightarrow$ & \ce{FeO2}+\ce{CO} & \SI{6.4800e-9}{} & 0.0 & 22979.5 & " \\
7 & \ce{FeO}+\ce{FeO2} & $\rightarrow$ & \ce{Fe2O3} & \SI{1.0000e-11}{} & 0.0 & 500.0 & E \\
8 & \ce{FeO2}+\ce{FeO2} & $\rightarrow$ & \ce{Fe2O3}+\ce{O} & \SI{1.0000e-11}{} & 0.0 & 500.0 & " \\
9 & \ce{FeO}+\ce{FeO} & $\rightarrow$ & \ce{Fe2O2} & \SI{4.6086e-17}{} & 0.0 & -2821.4 & SC13-\ce{SiO} \\
10 & \ce{Fe2O2} +\ce{FeO} & $\rightarrow$ & \ce{Fe3O3} & \SI{2.2388e-15}{} & 0.0 & -2878.9 & " \\
11 & \ce{Fe2O2} +\ce{Fe2O2} & $\rightarrow$ & \ce{Fe3O3}+\ce{FeO} & \SI{1.5265e-14}{} & 0.0 & -2386.8 & " \\
12 & \ce{Fe2O2} & $\rightarrow$ & \ce{FeO}+\ce{FeO} & \SI{7.7200e-7}{} & 0.0 & 0.0 & " \\
13 & \ce{Fe3O3} & $\rightarrow$ & \ce{Fe2O2}+\ce{FeO} & \SI{7.8300e-6}{} & 0.0 & 0.0 & " \\
14 & \ce{Fe2O2}+\ce{O2} & $\rightarrow$ & \ce{Fe2O3}+\ce{O} & \SI{1.0000e-11}{} & 0.0 & 500.0 & " \\
15 & \ce{Fe2O2}+\ce{SO} & $\rightarrow$ & \ce{Fe2O3}+\ce{S} & \SI{1.0000e-11}{} & 0.0 & 500.0 & " \\
16 & \ce{Fe2O3}+O & $\rightarrow$ & \ce{Fe2O2}+\ce{O2} & \SI{1.0000e-12}{} & 0.0 & 0.0 & " \\
17 & \ce{Fe2O3}+S & $\rightarrow$ & \ce{Fe2O2}+\ce{SO} & \SI{1.0000e-12}{} & 0.0 & 0.0 & " \\
18 & \ce{Fe3O3}+\ce{O2} & $\rightarrow$ & \ce{Fe3O4}+\ce{O} & \SI{1.0000e-13}{} & 0.0 & 500.0 & " \\
19 & \ce{Fe3O3}+\ce{SO} & $\rightarrow$ & \ce{Fe3O4}+\ce{S} & \SI{1.0000e-13}{} & 0.0 & 500.0 & " \\
20 & \ce{Fe2O3}+\ce{FeO} & $\rightarrow$ & \ce{Fe3O4} & \SI{7.4626e-16}{} & 0.0 & -2878.9 & " \\
21 & \ce{Fe2O2}+\ce{FeO} & $\rightarrow$ & \ce{Fe2O3}+\ce{Fe} & \SI{7.4626e-16}{} & 0.0 & -2878.9 & " \\
22 & \ce{Fe3O3}+\ce{FeO} & $\rightarrow$ & \ce{Fe3O4}+\ce{Fe} & \SI{5.0884e-15}{} & 0.0 & -2386.8 & " \\
23 & \ce{Fe}+\ce{Fe2O3} & $\rightarrow$ & \ce{Fe2O2}+\ce{FeO} & \SI{1.0000e-15}{} & 0.0 & 4000.0 & " \\
24 & \ce{Fe}+\ce{Fe3O4} & $\rightarrow$ & \ce{Fe3O3}+\ce{FeO} & \SI{1.0000e-15}{} & 0.0 & 8000.0 & " \\ \hline
\multicolumn{8}{c}{[Mg$_{1-x}$Fe$_x$SiO$_3$]   and [Mg$_{1-x}$Fe$_x$SiO$_4$] clusters} \\ \hline
25 & \ce{Mg}+\ce{Si2O3} & $\rightarrow$ & \ce{MgSi2O3} & \SI{1.0000e-12}{} & 0.0 & 0.0 & SC13 \\
26 & \ce{MgSi2O3}+\ce{O2} & $\rightarrow$ & \ce{MgSi2O4}+\ce{O} & \SI{1.0000e-12}{} & 0.0 & 0.0 & " \\
27 & \ce{MgSi2O4}+\ce{Mg} & $\rightarrow$ & \ce{Mg2Si2O4} & \SI{1.0000e-12}{} & 0.0 & 0.0 & " \\
28 & \ce{Mg2Si2O4}+\ce{O2} & $\rightarrow$ & \ce{Mg2Si2O5}+\ce{O} & \SI{1.0000e-12}{} & 0.0 & 0.0 & " \\
29 & \ce{Mg2Si2O5}+\ce{O2} & $\rightarrow$ & \ce{Mg2Si2O6}+\ce{O} & \SI{1.0000e-12}{} & 0.0 & 0.0 & " \\
30 & \ce{Mg2Si2O6}+\ce{Mg} & $\rightarrow$ & \ce{Mg3Si2O6} & \SI{1.0000e-12}{} & 0.0 & 0.0 & " \\
31 & \ce{Mg3Si2O6}+\ce{O2} & $\rightarrow$ & \ce{Mg3Si2O7}+\ce{O} & \SI{1.0000e-12}{} & 0.0 & 0.0 & " \\
32 & \ce{Mg3Si2O7}+\ce{Mg} & $\rightarrow$ & \ce{Mg4Si2O7} & \SI{1.0000e-12}{} & 0.0 & 0.0 & " \\
33 & \ce{Mg4Si2O7}+\ce{O2} & $\rightarrow$ & \ce{Mg4Si2O8}+\ce{O} & \SI{1.0000e-12}{} & 0.0 & 0.0 & " \\
34 & \ce{MgSi2O3}+\ce{SO} & $\rightarrow$ & \ce{MgSi2O4}+\ce{S} & \SI{1.0000e-12}{} & 0.0 & 0.0 & " \\
35 & \ce{Mg2Si2O4}+\ce{SO} & $\rightarrow$ & \ce{Mg2Si2O5}+\ce{S} & \SI{1.0000e-12}{} & 0.0 & 0.0 & " \\
36 & \ce{Mg2Si2O5}+\ce{SO} & $\rightarrow$ & \ce{Mg2Si2O6}+\ce{S} & \SI{1.0000e-12}{} & 0.0 & 0.0 & " \\
37 & \ce{Mg3Si2O6}+\ce{SO} & $\rightarrow$ & \ce{Mg3Si2O7}+\ce{S} & \SI{1.0000e-12}{} & 0.0 & 0.0 & " \\
38 & \ce{Mg4Si2O7}+\ce{SO} & $\rightarrow$ & \ce{Mg4Si2O8}+\ce{S} & \SI{1.0000e-12}{} & 0.0 & 0.0 & " \\
39 & \ce{Fe}+\ce{Si2O3} & $\rightarrow$ & \ce{FeSi2O3} & \SI{1.0000e-12}{} & 0.0 & 0.0 & as 25 \\
40 & \ce{FeSi2O3}+\ce{O2} & $\rightarrow$ & \ce{FeSi2O4}+\ce{O} & \SI{1.0000e-12}{} & 0.0 & 0.0 & as 26 \\
41 & \ce{FeSi2O4}+\ce{Fe} & $\rightarrow$ & \ce{Fe2Si2O4} & \SI{1.0000e-12}{} & 0.0 & 0.0 & as 27 \\
42 & \ce{Fe2Si2O4}+\ce{O2} & $\rightarrow$ & \ce{Fe2Si2O5}+\ce{O} & \SI{1.0000e-12}{} & 0.0 & 0.0 & as 28 \\
43 & \ce{Fe2Si2O5}+\ce{O2} & $\rightarrow$ & \ce{Fe2Si2O6}+\ce{O} & \SI{1.0000e-12}{} & 0.0 & 0.0 & as 29 \\
44 & \ce{Fe2Si2O6}+\ce{Fe} & $\rightarrow$ & \ce{Fe3Si2O6} & \SI{1.0000e-12}{} & 0.0 & 0.0 & as 30 \\
45 & \ce{Fe3Si2O6}+\ce{O2} & $\rightarrow$ & \ce{Fe3Si2O7}+\ce{O} & \SI{1.0000e-12}{} & 0.0 & 0.0 & as 31 \\
46 & \ce{Fe3Si2O7}+\ce{Fe} & $\rightarrow$ & \ce{Fe4Si2O7} & \SI{1.0000e-12}{} & 0.0 & 0.0 & as 32 \\
47 & \ce{Fe4Si2O7}+\ce{O2} & $\rightarrow$ & \ce{Fe4Si2O8}+\ce{O} & \SI{1.0000e-12}{} & 0.0 & 0.0 & as 33 \\
48 & \ce{FeSi2O3}+\ce{SO} & $\rightarrow$ & \ce{FeSi2O4}+\ce{S} & \SI{1.0000e-12}{} & 0.0 & 0.0 & as 34 \\
49 & \ce{Fe2Si2O4}+\ce{SO} & $\rightarrow$ & \ce{Fe2Si2O5}+\ce{S} & \SI{1.0000e-12}{} & 0.0 & 0.0 & as 35 \\
50 & \ce{Fe2Si2O5}+\ce{SO} & $\rightarrow$ & \ce{Fe2Si2O6}+\ce{S} & \SI{1.0000e-12}{} & 0.0 & 0.0 & as 36 \\
51 & \ce{Fe3Si2O6}+\ce{SO} & $\rightarrow$ & \ce{Fe3Si2O7}+\ce{S} & \SI{1.0000e-12}{} & 0.0 & 0.0 & as 37 \\
52 & \ce{Fe4Si2O7}+\ce{SO} & $\rightarrow$ & \ce{Fe4Si2O8}+\ce{S} & \SI{1.0000e-12}{} & 0.0 & 0.0 & as 38 \\
53 & \ce{MgSi2O4}+\ce{Fe} & $\rightarrow$ & \ce{MgFeSi2O4} & \SI{1.0000e-12}{} & 0.0 & 0.0 & as 27 \\
54 & \ce{FeSi2O4}+\ce{Mg} & $\rightarrow$ & \ce{MgFeSi2O4} & \SI{1.0000e-12}{} & 0.0 & 0.0 & " \\
55 & \ce{MgFeSi2O4}+\ce{O2} & $\rightarrow$ & \ce{MgFeSi2O5}+\ce{O} & \SI{1.0000e-12}{} & 0.0 & 0.0 & as 28 \\
56 & \ce{MgFeSi2O5}+\ce{O2} & $\rightarrow$ & \ce{MgFeSi2O6}+\ce{O} & \SI{1.0000e-12}{} & 0.0 & 0.0 & as 29 \\
57 & \ce{Fe2Si2O6}+\ce{Mg} & $\rightarrow$ & \ce{MgFe2Si2O6} & \SI{1.0000e-12}{} & 0.0 & 0.0 & as 29 \\
58 & \ce{MgFeSi2O6}+\ce{Mg} & $\rightarrow$ & \ce{Mg2FeSi2O6} & \SI{1.0000e-12}{} & 0.0 & 0.0 & as 30 \\
59 & \ce{MgFeSi2O6}+\ce{Fe} & $\rightarrow$ & \ce{MgFe2Si2O6} & \SI{1.0000e-12}{} & 0.0 & 0.0 & " \\
60 & \ce{MgFe2Si2O6}+\ce{O2} & $\rightarrow$ & \ce{MgFe2Si2O7}+\ce{O} & \SI{1.0000e-12}{} & 0.0 & 0.0 & as 31 \\
61 & \ce{Mg2FeSi2O6}+\ce{O2} & $\rightarrow$ & \ce{Mg2FeSi2O7}+\ce{O} & \SI{1.0000e-12}{} & 0.0 & 0.0 & " \\
62 & \ce{Fe3Si2O7}+\ce{Mg} & $\rightarrow$ & \ce{MgFe3Si2O7} & \SI{1.0000e-12}{} & 0.0 & 0.0 & as 32 \\
63 & \ce{MgFe2Si2O7}+\ce{Mg} & $\rightarrow$ & \ce{Mg2Fe2Si2O7} & \SI{1.0000e-12}{} & 0.0 & 0.0 & " \\
64 & \ce{MgFe2Si2O7}+\ce{Fe} & $\rightarrow$ & \ce{MgFe3Si2O7} & \SI{1.0000e-12}{} & 0.0 & 0.0 & " \\
65 & \ce{Mg2FeSi2O7}+\ce{Mg} & $\rightarrow$ & \ce{Mg3FeSi2O7} & \SI{1.0000e-12}{} & 0.0 & 0.0 & " \\
66 & \ce{Mg2FeSi2O7}+\ce{Fe} & $\rightarrow$ & \ce{Mg2Fe2Si2O7} & \SI{1.0000e-12}{} & 0.0 & 0.0 & " \\
67 & \ce{MgFe3Si2O7}+\ce{O2} & $\rightarrow$ & \ce{MgFe3Si2O8}+\ce{O} & \SI{1.0000e-12}{} & 0.0 & 0.0 & as 33 \\
68 & \ce{Mg2Fe2Si2O7}+\ce{O2} & $\rightarrow$ & \ce{Mg2Fe2Si2O8}+\ce{O} & \SI{1.0000e-12}{} & 0.0 & 0.0 & " \\
69 & \ce{Mg3FeSi2O7}+\ce{O2} & $\rightarrow$ & \ce{Mg3FeSi2O8}+\ce{O} & \SI{1.0000e-12}{} & 0.0 & 0.0 & as 34 \\
70 & \ce{MgFeSi2O4}+\ce{SO} & $\rightarrow$ & \ce{MgFeSi2O5}+\ce{S} & \SI{1.0000e-12}{} & 0.0 & 0.0 & as 35 \\
71 & \ce{MgFeSi2O5}+\ce{SO} & $\rightarrow$ & \ce{MgFeSi2O6}+\ce{S} & \SI{1.0000e-12}{} & 0.0 & 0.0 & as 36 \\
72 & \ce{MgFe2Si2O6}+\ce{SO} & $\rightarrow$ & \ce{MgFe2Si2O7}+\ce{S} & \SI{1.0000e-12}{} & 0.0 & 0.0 & as 37 \\
73 & \ce{Mg2FeSi2O6}+\ce{SO} & $\rightarrow$ & \ce{Mg2FeSi2O7}+\ce{S} & \SI{1.0000e-12}{} & 0.0 & 0.0 & " \\
74 & \ce{MgFe3Si2O7}+\ce{SO} & $\rightarrow$ & \ce{MgFe3Si2O8}+\ce{S} & \SI{1.0000e-12}{} & 0.0 & 0.0 & as 38 \\
75 & \ce{Mg2Fe2Si2O7}+\ce{SO} & $\rightarrow$ & \ce{Mg2Fe2Si2O8}+\ce{S} & \SI{1.0000e-12}{} & 0.0 & 0.0 & " \\
76 & \ce{Mg3FeSi2O7}+\ce{SO} & $\rightarrow$ & \ce{Mg3FeSi2O8}+\ce{S} & \SI{1.0000e-12}{} & 0.0 & 0.0 & " \\
77 & \ce{FeSi2O3}+\ce{O} & $\rightarrow$ & \ce{FeSiO3}+\ce{SiO} & \SI{1.0000e-12}{} & 0.0 & 0.0 & E \\
78 & \ce{Fe2Si2O4}+\ce{O} & $\rightarrow$ & \ce{Fe2SiO4}+\ce{SiO} & \SI{1.0000e-12}{} & 0.0 & 0.0 & " \\
79 & \ce{MgFeSi2O4}+\ce{O} & $\rightarrow$ & \ce{MgFeSiO4}+\ce{SiO} & \SI{1.0000e-12}{} & 0.0 & 0.0 & " \\
80 & \ce{MgSi2O3}+\ce{O} & $\rightarrow$ & \ce{MgSiO3}+\ce{SiO} & \SI{1.0000e-12}{} & 0.0 & 0.0 & " \\
81 & \ce{Mg2Si2O4}+\ce{O} & $\rightarrow$ & \ce{Mg2SiO4}+\ce{SiO} & \SI{1.0000e-12}{} & 0.0 & 0.0 & " \\ \hline
\multicolumn{8}{c}{cluster   fregmentation} \\ \hline
82 & \ce{FeO}+M & $\rightarrow$ & \ce{Fe}+O+M & \SI{4.4000e-10}{} & 0.0 & 98600.0 & E\\
83 & \ce{Fe2O2}+M & $\rightarrow$ & \ce{FeO}+\ce{FeO}+M & \SI{4.4000e-10}{} & 0.0 & 98600.0 & as 82\\
84 & \ce{FeO2}+M & $\rightarrow$ & \ce{FeO}+O+M & \SI{8.0180e-11}{} & 0.0 & 26900.0 & SC13-\ce{SiO} \\
85 & \ce{Fe2O3}+M & $\rightarrow$ & \ce{Fe2O2}+O+M & \SI{5.0000E-10}{} & 0.0 & 55000.0 & " \\
86 & \ce{Fe3O3}+M & $\rightarrow$ & \ce{Fe2O2}+\ce{FeO}+M & \SI{4.4000e-10}{} & 0.0 & 98600.0 & " \\
87 & \ce{Fe3O4}+M & $\rightarrow$ & \ce{Fe3O3}+O+M & \SI{5.0000E-10}{} & 0.0 & 55000.0 & " \\
88 & \ce{MgSi2O3}+M & $\rightarrow$ & \ce{Si2O3}+\ce{Mg}+M & \SI{1.0000e-10}{} & 0.0 & 98600.0 & " \\
89 & \ce{MgSi2O3}+M & $\rightarrow$ & \ce{MgSiO3}+\ce{Si}+M & \SI{1.0000e-10}{} & 0.0 & 98600.0 & E \\
90 & \ce{Mg2Si2O4}+M & $\rightarrow$ & \ce{Mg2SiO4}+\ce{Si}+M & \SI{1.0000e-10}{} & 0.0 & 98600.0 & " \\
91 & \ce{MgSi2O4}+M & $\rightarrow$ & \ce{MgSi2O3}+O+M & \SI{1.0000e-10}{} & 0.0 & 98600.0 & SC13 \\
92 & \ce{Mg2Si2O4}+M & $\rightarrow$ & \ce{MgSi2O4}+\ce{Mg}+M & \SI{1.0000e-10}{} & 0.0 & 98600.0 & " \\
93 & \ce{Mg2Si2O5}+M & $\rightarrow$ & \ce{Mg2Si2O4}+\ce{O}+M & \SI{1.0000e-10}{} & 0.0 & 98600.0 & " \\
94 & \ce{Mg2Si2O6}+M & $\rightarrow$ & \ce{Mg2Si2O5}+\ce{O}+M & \SI{1.0000e-10}{} & 0.0 & 98600.0 & " \\
95 & \ce{Mg3Si2O6}+M & $\rightarrow$ & \ce{Mg2Si2O6}+\ce{Mg}+M & \SI{1.0000e-10}{} & 0.0 & 98600.0 & " \\
96 & \ce{Mg3Si2O7}+M & $\rightarrow$ & \ce{Mg3Si2O6}+\ce{O}+M & \SI{1.0000e-10}{} & 0.0 & 98600.0 & " \\
97 & \ce{Mg4Si2O7}+M & $\rightarrow$ & \ce{Mg3Si2O7}+\ce{Mg}+M & \SI{1.0000e-10}{} & 0.0 & 98600.0 & " \\
98 & \ce{Mg4Si2O8}+M & $\rightarrow$ & \ce{Mg4Si2O7}+\ce{O}+M & \SI{1.0000e-10}{} & 0.0 & 98600.0 & " \\
99 & \ce{FeSi2O3}+M & $\rightarrow$ & \ce{Si2O3}+\ce{Fe}+M & \SI{1.0000e-10}{} & 0.0 & 98600.0 & as 88 \\
100 & \ce{FeSi2O3}+M & $\rightarrow$ & \ce{FeSiO3}+\ce{SiO}+M & \SI{1.0000e-10}{} & 0.0 & 98600.0 & as 89 \\
101 & \ce{Fe2Si2O4}+M & $\rightarrow$ & \ce{Fe2SiO4}+\ce{SiO}+M & \SI{1.0000e-10}{} & 0.0 & 98600.0 & as 90 \\
102 & \ce{FeSi2O4}+M & $\rightarrow$ & \ce{FeSi2O3}+O+M & \SI{1.0000e-10}{} & 0.0 & 98600.0 & as 91 \\
103 & \ce{Fe2Si2O4}+M & $\rightarrow$ & \ce{FeSi2O4}+\ce{Fe}+M & \SI{1.0000e-10}{} & 0.0 & 98600.0 & as 92 \\
104 & \ce{Fe2Si2O5}+M & $\rightarrow$ & \ce{Fe2Si2O4}+\ce{O}+M & \SI{1.0000e-10}{} & 0.0 & 98600.0 & as 93 \\
105 & \ce{Fe2Si2O6}+M & $\rightarrow$ & \ce{Fe2Si2O5}+\ce{O}+M & \SI{1.0000e-10}{} & 0.0 & 98600.0 & as 94 \\
106 & \ce{Fe3Si2O6}+M & $\rightarrow$ & \ce{Fe2Si2O6}+\ce{Fe}+M & \SI{1.0000e-10}{} & 0.0 & 98600.0 & as 95 \\
107 & \ce{Fe3Si2O7}+M & $\rightarrow$ & \ce{Fe3Si2O6}+\ce{O}+M & \SI{1.0000e-10}{} & 0.0 & 98600.0 & as 96 \\
108 & \ce{Fe4Si2O7}+M & $\rightarrow$ & \ce{Fe3Si2O7}+\ce{Fe}+M & \SI{1.0000e-10}{} & 0.0 & 98600.0 & as 97 \\
109 & \ce{Fe4Si2O8}+M & $\rightarrow$ & \ce{Fe4Si2O7}+\ce{O}+M & \SI{1.0000e-10}{} & 0.0 & 98600.0 & as 98 \\
110 & \ce{MgFeSi2O4}+M & $\rightarrow$ & \ce{FeSi2O4}+\ce{Mg}+M & \SI{1.0000e-10}{} & 0.0 & 98600.0 & as 90 \\
111 & \ce{MgFeSi2O4}+M & $\rightarrow$ & \ce{MgSi2O4}+\ce{Fe}+M & \SI{1.0000e-10}{} & 0.0 & 98600.0 & " \\
112 & \ce{MgFeSi2O4}+M & $\rightarrow$ & \ce{MgFeSiO4}+\ce{Si}+M & \SI{1.0000e-10}{} & 0.0 & 98600.0 & " \\
113 & \ce{MgFeSi2O5}+M & $\rightarrow$ & \ce{MgFeSi2O4}+O+M & \SI{1.0000e-10}{} & 0.0 & 98600.0 & as 93 \\
114 & \ce{MgFeSi2O6}+M & $\rightarrow$ & \ce{MgFeSi2O5}+O+M & \SI{1.0000e-10}{} & 0.0 & 98600.0 & as 94 \\
115 & \ce{MgFe2Si2O6}+M & $\rightarrow$ & \ce{Fe2Si2O6}+\ce{Mg}+M & \SI{1.0000e-10}{} & 0.0 & 98600.0 & as 95 \\
116 & \ce{MgFe2Si2O6}+M & $\rightarrow$ & \ce{MgFeSi2O6}+\ce{Fe}+M & \SI{1.0000e-10}{} & 0.0 & 98600.0 & " \\
117 & \ce{Mg2FeSi2O6}+M & $\rightarrow$ & \ce{MgFeSi2O6}+\ce{Mg}+M & \SI{1.0000e-10}{} & 0.0 & 98600.0 & " \\
118 & \ce{Mg2FeSi2O6}+M & $\rightarrow$ & \ce{Mg2Si2O6}+\ce{Fe}+M & \SI{1.0000e-10}{} & 0.0 & 98600.0 & " \\
119 & \ce{MgFe2Si2O7}+M & $\rightarrow$ & \ce{MgFe2Si2O6}+O+M & \SI{1.0000e-10}{} & 0.0 & 98600.0 & as 96 \\
120 & \ce{Mg2FeSi2O7}+M & $\rightarrow$ & \ce{Mg2FeSi2O6}+O+M & \SI{1.0000e-10}{} & 0.0 & 98600.0 & " \\
121 & \ce{MgFe3Si2O7}+M & $\rightarrow$ & \ce{Fe3Si2O7}+\ce{Mg}+M & \SI{1.0000e-10}{} & 0.0 & 98600.0 & as 97 \\
122 & \ce{MgFe3Si2O7}+M & $\rightarrow$ & \ce{MgFe2Si2O7}+\ce{Fe}+M & \SI{1.0000e-10}{} & 0.0 & 98600.0 & " \\
123 & \ce{Mg2Fe2Si2O7}+M & $\rightarrow$ & \ce{MgFe2Si2O7}+\ce{Mg}+M & \SI{1.0000e-10}{} & 0.0 & 98600.0 & " \\
124 & \ce{Mg2Fe2Si2O7}+M & $\rightarrow$ & \ce{Mg2FeSi2O7}+\ce{Fe}+M & \SI{1.0000e-10}{} & 0.0 & 98600.0 & " \\
125 & \ce{Mg3FeSi2O7}+M & $\rightarrow$ & \ce{Mg2FeSi2O7}+\ce{Mg}+M & \SI{1.0000e-10}{} & 0.0 & 98600.0 & " \\
126 & \ce{Mg3FeSi2O7}+M & $\rightarrow$ & \ce{Mg3Si2O7}+\ce{Fe}+M & \SI{1.0000e-10}{} & 0.0 & 98600.0 & " \\
127 & \ce{MgFe3Si2O8}+M & $\rightarrow$ & \ce{MgFe3Si2O7}+O+M & \SI{1.0000e-10}{} & 0.0 & 98600.0 & as 98 \\
128 & \ce{Mg2Fe2Si2O8}+M & $\rightarrow$ & \ce{Mg2Fe2Si2O7}+O+M & \SI{1.0000e-10}{} & 0.0 & 98600.0 & " \\
129 & \ce{Mg3FeSi2O8}+M & $\rightarrow$ & \ce{Mg3FeSi2O7}+O+M & \SI{1.0000e-10}{} & 0.0 & 98600.0 & " \\
130 & \ce{MgSiO3}+M & $\rightarrow$ & \ce{MgO2}+\ce{SiO}+M & \SI{1.0000e-10}{} & 0.0 & 98600.0 & E \\
131 & \ce{Mg2SiO4}+M & $\rightarrow$ & \ce{MgSiO3}+\ce{MgO}+M & \SI{1.0000e-10}{} & 0.0 & 98600.0 &  \\
132 & \ce{FeSiO3}+M & $\rightarrow$ & \ce{FeO2}+\ce{SiO}+M & \SI{1.0000e-10}{} & 0.0 & 98600.0 & " \\
133 & \ce{Fe2SiO4}+M & $\rightarrow$ & \ce{FeSiO3}+\ce{FeO}+M & \SI{1.0000e-10}{} & 0.0 & 98600.0 & " \\
134 & \ce{MgFeSiO4}+M & $\rightarrow$ & \ce{FeSiO3}+\ce{MgO}+M & \SI{1.0000e-10}{} & 0.0 & 98600.0 & " \\
135 & \ce{MgFeSiO4}+M & $\rightarrow$ & \ce{MgSiO3}+\ce{FeO}+M & \SI{1.0000e-10}{} & 0.0 & 98600.0 & " \\ 
\hline
\end{longtable*}

\begin{table*}[h]
\caption{Compton Electron-induced Reactions, Corresponding Mean Energy per Ion Pair $W_i$ from \cite{sar13}, and Arrhenius Coefficient A as a Function of Ejecta Model. The rates are expressed in an Arrhenius form. See \cite{cherchneff2009} for details.  $k_C^a: A \times(T/T_0)^{0.43}exp(-2324.88/T )$\qquad $k_C^b: A \times(T/T_0)^{0.08}exp(-2082.99/T )$\quad$ k_C^c: A \times(T/T_0)^{0.11}exp(-1335.44/T )$\qquad  $k_C^d: A\times(T/T_0)^{0.06}exp(-1259.58/T )$}
\label{tab: rates}
\begin{tabular}{ccccccc}
\hline
        &                                    &           & \texttt{N1def}     & \texttt{N10def}        & \texttt{N100def}    & \texttt{N100ddt}   \\
Species & Reactions                          & $W_i$(eV) & \Ni=0.034\Ms       & \Ni=0.183\Ms & \Ni=0.355\Ms        & \Ni=0.604\Ms       \\
        &                                    &           & A- 0.08 \Ms$^a$ & A- 0.48 \Ms$^b$     & A- 1.31 \Ms$^c$ & A- 1.4 \Ms$^d$ \\ \hline
CO      & $\rightarrow$ \ce{O+} + C          & 768       & \SI{2.33e-8}{}  & \SI{1.58e-7}{}     & \SI{1.94e-7}{} & \SI{1.73e-7}{} \\
        & $\rightarrow$ O + \ce{C+}          & 247       & \SI{7.25e-8}{}  & \SI{4.92e-7}{}     & \SI{6.05e-7}{} & \SI{5.37e-7}{} \\
        & $\rightarrow$ C + O                & 125       & \SI{1.43e-7}{}  & \SI{9.72e-7}{}     & \SI{1.19e-6}{} & \SI{1.06e-6}{}  \\
        & $\rightarrow$ \ce{CO+} + \ce{e-}   & 34        & \SI{5.27e-7}{}  & \SI{3.57e-6}{}     & \SI{4.39e-6}{} & \SI{3.90e-6}{}  \\ \hline
O       & $\rightarrow$ \ce{O+} + \ce{e-}    & 46.2      & \SI{3.88e-7}{}  & \SI{2.63e-6}{}     & \SI{3.23e-6}{} & \SI{2.87e-6}{} \\
C       & $\rightarrow$ \ce{C+} +\ce{ e-}    & 36.4      & \SI{4.92e-7}{}  & \SI{3.34e-6}{}     & \SI{4.10e-6}{} & \SI{3.64e-6}{} \\ \hline
SiO     & $\rightarrow$ \ce{O+} + Si         & 678       & \SI{2.64e-8}{}  & \SI{1.79e-7}{}     & \SI{2.20e-7}{} & \SI{1.96e-7}{}  \\
        & $\rightarrow$ O +\ce{ Si+}         & 218       & \SI{8.21e-8}{}  & \SI{5.57e-7}{}     & \SI{6.85e-7}{} & \SI{6.08e-7}{}  \\
        & $\rightarrow$ O + Si               & 110       & \SI{1.63e-7}{}  & \SI{1.10e-6}{}     & \SI{1.36e-6}{} & \SI{1.21e-6}{}  \\
        & $\rightarrow$ \ce{SiO+ }+\ce{ e-}  & 30        & \SI{5.97e-6}{}  & \SI{4.05e-6}{}     & \SI{4.98e-6}{} & \SI{4.42e-6}{} \\ \hline
Ne      & $\rightarrow$ \ce{Ne+}   + \ce{e-} & 36.4      & \SI{4.92e-7}{}  & \SI{3.34e-6}{}     & \SI{4.10e-6}{} & \SI{3.64e-6}{}  \\ \hline
\end{tabular}

\end{table*}
\end{document}